\documentclass[useAMS,usenatbib]{mn2e}
\usepackage{graphics,epsfig,txfonts}
\bibpunct{(}{)}{;}{a}{}{,}

\usepackage{graphicx}
\usepackage{longtable}
\usepackage{amssymb}
\usepackage{xcolor}
\usepackage{color}
\usepackage{lipsum}
\usepackage{textcomp}
\usepackage{threeparttable}
\usepackage[colorlinks=true,linkcolor=blue,citecolor=blue,
bookmarks=true,bookmarksopen=true,bookmarksnumbered=true,draft=false]{hyperref}
\usepackage{enumerate}

\usepackage{placeins}
\usepackage{float}

\usepackage[normalem]{ulem}
\usepackage{soul}
\usepackage{blindtext}
\usepackage[hyphenbreaks]{breakurl}

\newcommand{\ltsima}{$\buildrel < \over \sim$}
\newcommand{\lsim}{\lower.5ex\hbox{\ltsima}}
\newcommand{\gtsima}{$\buildrel > \over \sim$}
\newcommand{\gsim}{\lower.5ex\hbox{\gtsima}}

\newcommand{\src}{V404\,Cyg\xspace}

\newcommand{\swift}{{\it Swift}\xspace}

\newcommand{\eqb}{\begin{eqnarray}}
\newcommand{\eqe}{\end{eqnarray}}

\begin{document}

%\linenumbers

% \title{The X-ray dust scattered rings of the black hole low mass binary \src
% % \thanks{Based on observations with \xmm, an ESA Science Mission with instruments and contributions directly funded by ESA Member states and the USA (NASA); with Swift, a NASA mission with international participation.}
% }

% \author{G.~Vasilopoulos\inst{1}\and F.~Haberl\inst{1} \and K.~Dennerl\inst{1} \and R.~Sturm\inst{1}   \and A.~Udalski\inst{2}}
% \author{Georgios~Vasilopoulos\inst{1}\and Maria~Petropoulou\inst{2} et al.}

\title[The X-ray rings of LMXB \src]{The X-ray dust scattered rings of the black hole low mass binary \src}
\author[G. Vasilopoulos \& M. Petropoulou]{G.~Vasilopoulos$^1$\thanks{Email: gevas@mpe.mpg.de} \& M. Petropoulou$^{2}$\thanks{Einstein Postdoctoral Fellow}\\
$^{1}$Max-Planck-Institut f\"ur extraterrestrische Physik,Giessenbachstra{\ss}e, 85748 Garching, Germany \\
$^2$ Department of Physics and Astronomy, Purdue University, 525 Northwestern
Avenue, West Lafayette, IN 47907, USA}

% \titlerunning{\src}
 
% \institute{Max-Planck-Institut f\"ur extraterrestrische Physik,
%            Giessenbachstra{\ss}e, 85748 Garching, Germany\\
% 	   \email{gevas@mpe.mpg.de}
% 	   \and
% 	   Department of Physics and Astronomy, 
% 	   Purdue University, 
% 	   West Lafayette, IN, USA}
%  
\date{Received ?? ??? 2015 / Accepted ?? ??? 2015}

% \begin{document}
\maketitle
% 
% \abstract{aims}
% {methods}
% {results}
% {}
% {}
\begin{abstract}
We report on the first detection of X-ray dust scattered rings from the Galactic low mass X-ray binary \src.
The observation of the system with \swift/XRT on June 30 2015 revealed the presence
of five concentric ring-like structures centred at the position of \src. Follow-up \swift/XRT observations
allowed a time-dependent study of the  X-ray rings. Assuming
that these are the result of small-angle, single X-ray scattering by dust grains along the line of sight,
we find that their angular size scales as $\theta \propto\sqrt{t}$ in agreement with theoretical predictions. The dust grains are concentrated in five dust layers located at about $2.12, 2.05, 1.63, 1.50$ and $1.18$~kpc from the observer. These coincide roughly with locations of enhanced extinction as determined by infrared photometry. Assuming that the grain size distribution is described by a generalized Mathis-Rumpl-Nordsieck model, we find that the power-law index  of the most distant cloud is $q\sim 4.4$, while  $q \sim 3.5-3.7$ in all other clouds.  We constrain at a $3\sigma$ level the maximum grain  size
of the intermediate dust layers  in the range $0.16-0.20\,\mu$m and set a lower limit  of $\sim 0.2\,\mu$m in the other clouds. Hints of an exponential cutoff at the angular intensity profile of the outermost X-ray ring suggest that the smallest grains have sizes $0.01 \mu{\rm m}\le \alpha_{\min} \lesssim 0.03\,\mu$m.  Based on the relative ratios of dust column densities we find the highest dust concentration at $\sim 1.6$~kpc. Our results indicate a gradient in the dust properties within 1~kpc from \src.
% The derived dust properties of the nearest to Earth cloud are ,thus, compatible with a MNR model with a silicate grain composition.
\end{abstract}

\begin{keywords}
X-ray: binaries -- individual: V404 Cygni -- X-ray: ISM -- (ISM:) dust
\end{keywords}

\maketitle

\section{Introduction}
\label{sec-intro}
V404 Cyg, also known as GS 2023+338, is a nearby \citep[d = 2.39 $\pm$ 0.14 kpc][]{Miller-Jones2009} low mass X-ray binary
(LMXB) with an orbital period of $\sim 6.47$~d \citep[see e.g.][]{Casares1992, CasaresCharles1994}.
It consists of a $\sim 1\ M_{\odot}$ late type G or early type K 
companion star \citep{Casares1992, Wagner1992, CasaresCharles1994} and a compact object.
The latter is black hole (BH) with an estimated mass of $12^{+3}_{-2} M_{\odot}$ \citep{Shahbaz1996} 
\citep[for other BH mass estimates, see][]{Casares1992, CasaresCharles1994}.
\src is an X-ray transient source with three confirmed
historical outbursts in 1936, 1958 and 1989 \citep{Richter1989}. It 
was in the 1989 outburst \citep[for a detailed analysis, see ][]{Zycki1999} that \src was 
discovered in  X-rays by the Geminga satellite \citep{Makino1989}.

After a long period of quiescence (26 yr), the Swift Burst Alert Telescope (BAT) was triggered on June 15 2015 
(18:32 UT) following the systems re-brightening (Barthelmy et al. GCN \#17929). Soon after the BAT trigger, the system
was detected in X-rays by MAXI on board of the International Space Station \citep{MAXI2015}, which initiated multi-wavelength observations 
of the  source \citep[see e.g.][]{Garner2015, Gazeas2015, Mooley2015,Tetarenko2015}. The subsequent, still ongoing, monitoring of the 
renewed activity of \src  across the electromagnetic spectrum resulted in observations that can be used not only for studying the properties of the LMXB itself \citep[e.g.][]{King2015, Rodriguez2015} but also  those of its surrounding medium. For the latter, the 
detection, for the first time, of ring-like structures centered at the position of \src having
a typical angular size of a few arcmin \citep{Beardmore2015}, is of particular importance. The formation of the X-ray rings
can be understood in terms of scattering of the source's X-ray photons by  dust grains of the interstellar medium (ISM), and in this
regard, their study can provide information for the properties of the ISM and the dust distribution in the direction
of the source.

It was originally proposed by \cite{Overbeck1965} that a bright X-ray point source
would appear surrounded by a diffuse X-ray emission (halo) due to the small angle scattering from dust grains in the ISM.
First observational evidence for the X-ray scattering from interstellar grains were presented
by \cite{Rolf1983} for the source GX339–4. As the scattering cross section depends on the  properties of the dust grains, such as size and composition,
as well as on the distribution of dust along the line of sight (LOS) \citep{vandeHulst1957}, analysis of X-ray halos
from Galactic and extragalactic sources can be used for determining the properties of the interstellar dust
{ \citep[e.g.][]{Mauche1986, GreinerPredehl1995, PredehlSchmitt1995, GreinerDennerl1996, PredehlKlose1996, Draine2003, Costantini2005, SmithDame2006,Corrales2015}.}
If the spatial distribution of dust in the direction of a variable X-ray source is known, 
constraints to the source distance can be then set by using the time delay between the direct and scattered signals 
\citep{Truemper1973}. This method has been applied to several systems, including the X-ray binaries Cyg X-3 \citep{Predehl2000} and 
Cyg X-1 \citep{Xiang2011}, as well as 
and the soft gamma-ray repeater (SGR) SGR 1806-20 \citep{Svirski2011}. 

Although the scattering from bright X-ray point sources manifests itself usually as an
X-ray halo \citep[e.g.][]{Rolf1983}, the combination of a short duration outburst
from a variable X-ray source with a series of discrete dust slabs
along the LOS would result in the appearance of multiple (typically, one or two) X-ray rings.
Analysis of dust scattered rings have been applied to a variety of systems, such as
gamma-ray bursts (GRBs) \citep[e.g.][and references therein]{Klose1994, Vaughan2004, Vianello2007},
anomalous X-ray pulsars \citep[see][for 1E1547.0–5408]{Tiengo2010}, and SGR \citep[e.g.][]{Kouveliotou2001}. 

Here we report on the recent \swift/XRT observations of the first ever observed dust scattered X-ray rings from
\src. These observations offer a unique opportunity to study the properties of dust along the LOS in the direction of \src for a number of reasons:
(i) at least five dust scattered  X-ray rings can be identified in the X-ray image, while in most cases the maximum number of observed rings is 
two to three; (ii) the expansion of the five rings can be followed for  a minimum period  of  $\sim$23~d for the outer ring, up to 
a maximum period of $\sim$36~d for the innermost rings; 
(iii) \src is one of the nearest,  most luminous LMXB in quiescence \citep{Miller-Jones2009}; and (iv) its distance is well constrained. 
Aim of our study is to derive the basic properties of the dust, such as their spatial distribution and 
typical size, by applying a simple, yet well proven theoretical framework, to the recent \swift/XRT observations. 

The present paper is structured as follows. In \S\ref{sec-theory} we outline the basic formalism used in our analysis of the X-ray dust scattering.
We report on the observations and the data reduction  in \S\ref{sec-observations}. We continue in \S\ref{sec-results} with a presentation
of our analysis and results. In \S\ref{sec-discussion} we discuss several aspects
of our analysis and conclude in \S\ref{sec-summary} with a summary of our results.

\section{Theoretical framework}
\label{sec-theory}
The dust grains in the ISM are responsible for X-ray scattering and absorption. The latter becomes important especially
at energies below 1~keV, although the dust grains may be composed by elements with K-shell edges above 1 keV \citep[see e.g.][]{Smith1998}.
Moreover, the analytic Rayleigh-Gans (RG) approximation for the scattering cross section $d\sigma/d\Omega$ breaks down 
at energies $E<1$~keV, where one should use the solution of Mie theory \citep{vandeHulst1957}. {
We note that a detailed treatment
of the X-ray scattering goes even beyond the Mie solution, as this is also derived
under certain simplifying assumptions, e.g. spherical shape, that may not be satisfied for the dust grains of the ISM}. 

Here, we limit our analysis at $E\ge 1$~keV where X-ray absorption is not as important as in lower energies and
make use of the RG approximation, which roughly speaking, is valid
for photon energies in keV  significantly larger  than the grain size in $\,\mu$m. Furthermore, we 
neglect the effects of multiple X-ray scatterings, which become 
important if the scattered intensity is comparable to the intensity of the point source. This translates to
$\tau_{\rm sc} >1.3$ at 1~keV \citep[][]{MathisLee1991}, where $\tau_{\rm sc}$ is the optical depth for scattering by dust.
We can estimate $\tau_{\rm sc}$ for \src using the relation $\tau_{\rm sc} \approx 0.15 \ A_{\rm V} (1\ {\rm keV}/E)^{-1.8}$ \citep{Draine2004},
where $A_{\rm V}$ is the visual extinction in the direction of the source. By adopting $A_{\rm V}\sim 4$ \citep[][]{Casares1993},
we find $\tau_{\rm sc}\sim 0.6$ at 1~keV; {this value would be lower by a factor of $\sim 2$,  if the relation of \cite{PredehlSchmitt1995} was
used.} It is, therefore, reasonable to assume that multiple X-ray scatterings are not important for our study.
Our results should be considered in the light of these assumptions.
\begin{figure}
 \resizebox{\hsize}{!}{\includegraphics[angle=0,clip=]{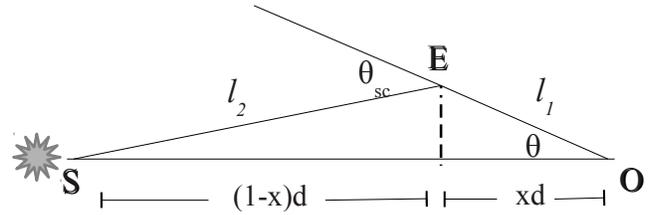}}
  \caption{Sketch of the dust scattering geometry.} 
  \label{fig:dust_sc}
\end{figure}

The observed time delay $\Delta t$ between the X-ray outburst from the source and the { detection of the} X-ray dust scattered rings 
is the result of different path lengths (see e.g. Fig.~\ref{fig:dust_sc}), and is given by 
\eqb
\Delta t = \frac{\ell_1+\ell_2}{c}-\frac{d}{c},
\label{eq0}
\eqe
where $\ell_1 = x d/\cos \theta$, $\ell_2=\sqrt{ (1-x)^2 d^2+ (xd)^2 \tan\theta^2}$, $\theta$ is the observed angular radius
of the ring and $x d$ is the distance of the dust screen from the observer. { Expression (\ref{eq0}) 
is derived under the assumption that the dust scattered rings observed at time $t$ are
the result of a single, short duration outburst occurring at $t_0$; this is reasonable as long as $\delta t \ll \Delta t$, where $\delta t$ is the outburst duration.} 
In the limit of small angles ($\theta \ll 1$), where
$\tan \theta \approx \theta + \mathcal{O}(\theta^3)$ and $\cos\theta \approx 1-\theta^2/2$, the time delay for each ring
is approximately given by
\eqb
\Delta t \approx \frac{d}{2c}\frac{x}{1-x}\theta^2 + \mathcal{O}(\theta^3).
\label{delay}
\eqe
The above expression may be rewritten in a more useful form as 
\eqb
\theta(t)=1 \ {\rm arcmin} \left(\frac{19.84 (1-x)}{x}\right)^{1/2} \left(\frac{1\ {\rm kpc}}{d}\right)^{1/2}\left(\frac{\Delta t}{1 \ {\rm d}} \right)^{1/2}.
\label{theta}
\eqe

In the RG approximation of small scattering angles ($\theta_{\rm sc}$) the cross section is written as \citep[e.g][]{Mauche1986}:
\eqb
\frac{d\sigma \left(E, \bar{\alpha},\theta_{\rm sc} \right)}{d\Omega} = C_{\rm d}{(E)}\ \bar{\alpha}^6 e^{-\theta_{\rm sc}^2/2\Theta^2( {E,\bar{\alpha}})},
\label{cross}
\eqe
where 
\eqb
\Theta{(E,\bar{\alpha})} =\frac{1.04 \ {\rm arcmin}} {E_{\rm keV} \bar{\alpha}},
\label{theta-big}
\eqe
with $E_{\rm keV}\equiv E/1$~keV and $\bar{\alpha}\equiv \alpha/ 1 \ \,\mu$m.  The normalization of the scattering
cross section is given by \citep[][]{Mauche1986}
\eqb
C_{\rm d}{(E)}=1.1\ \left(\frac{2Z}{M}\right)^2\left(\frac{\rho}{3 \ {\rm gr \ cm}^{-3}}\right)^2\left(\frac{F(E)}{Z} \right)^2 \ {\rm cm}^2 \ {\rm sr}^{-1}.
\eqe
In the above, $Z$ is the atomic number, $M$ is the molecular mass (in amu), $\rho$ is the mass density, $E$ is the X-ray energy in keV and 
$F(E)$ is the form factor \citep{Henke1981}. The factor $F(E)/Z$ is close to unity when considering the scattering
of X-rays with energies 1-6~keV, since most of the K-edges of light elements ($Z \le 10$) lie
below this range\footnote{A table for X-ray absorption edges and lines can be found
at \url{http://www.kayelaby.npl.co.uk/atomic_and_nuclear_physics/4_2/4_2_1.html}} \citep[][]{Dewey1969}. 
Moreover, the factor $2Z/M$ is of order unity for all the abundant elements in grains \citep[see e.g.][]{MathisLee1991}, and $3$~gr cm$^{-3}$
is the typical mass density of dust grains \citep[see e.g.][]{MRN1977, Draine2011}. 

For the general case of a continuous dust distribution between the source and the observer, the scattered intensity 
(in units of erg cm$^{-2}$ s$^{-1}$ keV$^{-1}$ sr$^{-1}$) is given by \citep[e.g.][]{MathisLee1991}:
\eqb
I_{\rm sc}\left(\theta, E\right) = F_{\rm X}(E)N_{\rm H}\int_{\alpha_{\min}}^{\alpha_{\max}}\!\!d\alpha \ n(\alpha)\int_{0}^{1-\epsilon}\!\!\! 
dx \ \frac{f(x)}{(1-x)^2} \frac{d\sigma \left(E, \alpha,\theta_{\rm sc} \right)}{d\Omega}
\label{Isc}
\eqe
where $\theta$ is the observed angle, $F_{\rm X}(E)$ is the observed (i.e.,  not corrected for absorption) 
differential energy flux of the {source} (in erg cm$^{-2}$ s$^{-1}$ keV$^{-1}$), 
$N_{\rm H}$ is the total hydrogen column density along the LOS, $n(\alpha)$ is 
the number of grains per hydrogen atom with radii between $\alpha$ and $\alpha+d{\alpha}$,
$f(x)$ is the spatial distribution of grains, 
normalized to unity for the interval
(0,1) and $\epsilon\approx \theta$ for small $\theta$, i.e. up to several arcmin \citep{Smith1998}.
In the case where a spatially discrete distribution of dust is implied by the observations (i.e. rings vs. diffuse halo),
we may write  $f(x)= \sum_{i=1}^{N}\delta(x-x_i)$ \citep[see also][]{Mauche1986} and eq.~(\ref{Isc}) becomes 
$I_{\rm sc}\left(\theta, E\right) = \sum_{i=1}^{N}I_{\rm sc, i}\left(\theta, E\right)$, where 
\eqb
I_{\rm sc, i}\left(\theta, E\right) = F_{\rm X}(E)N_{\rm H,i}\frac{1}{\left(1-x_{\rm i}\right)^2}\int_{{\alpha}_{\min, \rm i}}^{{\alpha}_{\max, \rm i}}\!\!d{\alpha} \ n_{\rm i}({\alpha})
\ \frac{d\sigma \left(E, {\alpha},\theta_{\rm sc} \right)}{d\Omega},
\label{Isc-delta}
\eqe
{where  $N_{\rm H,i}$ is the hydrogen column density of the cloud.
For an X-ray burst of fluence $\Phi_{\rm X}(E)$  (in erg cm$^{-2}$ keV$^{-1}$) that leads to the production of ring-like structures instead
of an X-ray halo, eq.~(\ref{Isc-delta}) becomes 
\eqb
F_{\rm sc, i}\left(\theta, E\right) = \Phi_{\rm X}(E)N_{\rm H,i}\frac{g(x)}{\left(1-x_{\rm i}\right)^2}\int_{{\alpha}_{\min, \rm i}}^{{\alpha}_{\max, \rm i}}\!\!d{\alpha} \ n_{\rm i}({\alpha})
\ \frac{d\sigma \left(E, {\alpha},\theta_{\rm sc} \right)}{d\Omega},
\label{Fsc-delta}
\eqe
where $g(x)\equiv 2\pi \theta d\theta/dt =  2\pi c (1-x)/x d$ (see eq.~(\ref{delay}))  and 
$F_{\rm sc, i}$ is the scattered flux integrated over the angular extend of the X-ray ring (in erg cm$^{-2}$ s$^{-1}$ keV$^{-1}$).}

For a power-law size distribution, i.e. $n(\alpha) \propto \alpha^{-q}$, 
the radial profile of the scattered intensity (or flux) from one dust layer
should exhibit two power-law segments ($I_{\rm sc} \propto\,$const for $\theta \lesssim \theta_1$, where $\theta_1$ is given by eq.~(\ref{theta-big}) for
$\alpha=\alpha_{\max}$, and $I_{\rm sc}\propto \theta ^{-7+q}$) followed by an exponential cutoff
at an angle determined by the smallest size grains \citep[see e.g.][]{Svirski2011}.

% Note that besides the normalization constanteqs.~(\ref{Isc-delta}) and (\ref{Fsc-delta}) differ only in their normalization constants

Although a  power-law distribution is scale-free, in the case 
of X-ray scattering by a power-law grain distribution, an average grain size can still be defined \citep[][]{Mauche1986},
since the scattering cross section introduces a characteristic grain size (see e.g. eqs.~(\ref{cross}) and (\ref{theta-big})).
Using the integrated cross section over scattering angles, i.e.
\eqb
\sigma \left(E, \bar{\alpha}\right) =6.3\times 10^{-7}\ \bar{\alpha}^{4}\left(\frac{2Z}{M}\right)^2\left(\frac{\rho}{3 \ {\rm gr \ cm}^{-3}}\right)^2\left(\frac{F(E)}{Z} \right)^2 \ {\rm cm}^2
\eqe
the average grain size is calculated by
\eqb
\langle \alpha \rangle = \frac{\int d\alpha \ \sigma(E,\alpha) \alpha  n(\alpha)}{\int d\alpha \sigma(E,\alpha) n(\alpha)} 
\approx \frac{5-q}{6-q}\alpha_{\max},
\label{average}
\eqe
where the approximation holds for $q<5$.  Equation (\ref{average}) will be used to derive the average dust grain size using the best-fit results for $q$ and $\alpha_{\max}$ (see \S\ref{sec-results}).

%%%%%%%%%%%%%%%%%%%%%%%%%%%%%%%%%%%%%%%%%%%%%%%%%%%%%%%%%%%%%%%%%
%%%%%%%%%%%%%%%%%%%%%%%%%%%%%%%%%%%%%%%%%%%%%%%%%%%%%%%%%%%%%%%%%
%%%%%%%%%%%%%%%%%%%X-ray%%%%%%%%%%%%%%%%%%%%%%%%%%%%%%%%%%%%%%%%%
%%%%%%%%%%%%%%%%%%%%%%%%%%%%%%%%%%%%%%%%%%%%%%%%%%%%%%%%%%%%%%%%%
%%%%%%%%%%%%%%%%%%%%%%%%%%%%%%%%%%%%%%%%%%%%%%%%%%%%%%%%%%%%%%%%%

\section{Observations and data reduction}
\label{sec-observations}

% \begin{figure}
%  \resizebox{\hsize}{!}{\includegraphics[angle=0,clip=]{./plots_bu/image_71_rgb_v2.ps}}
% \caption{RGB image of the X-ray dust scattered rings observed by \swift/XRT at MJD~57203.45 (obsid: 00031403071). The color coding
% used is R: 0.5-1.5~keV, G: 1.5-2.5~keV, and  B: 2.5-10~keV.  The  \swift/XRT field-of-view is enclosed
% by the green circle, while a white circle of radius $7.5\arcmin$ is overplotted for guiding the eye.}
% \label{fig:image_rgb}
% \end{figure}
For the study of the X-ray rings we used \swift/XRT data which are publicly available from the {\tt Swift Data Center}\footnote{http://swift.gsfc.nasa.gov/sdc/}.
We ran the standard XRT pipeline to derive the cleaned event lists from the level one products. 
The observations were performed in two instrument modes, photon counting (PC) and window timing (WT) mode. For the WT data we kept single and double pixel events (grades 0-2), while for the PC data analysis we used single to quadruple events (grades 0-12). 

The \swift/XRT data reduction and the analysis of the radial profiles of the X-ray rings was performed with the use of standardized procedures from the {\tt ftools}\footnote{https://heasarc.gsfc.nasa.gov/ftools/} software packages.
For the creation of the \swift/XRT light curve %(plotted on the bottom panel of Fig. \ref{fig:expansion_rings}) 
we used  the {\tt HEASoft} program {\tt xselect}. The source events were extracted from a 40\arcsec region, while a similar size region was used for the background. 
The background correction was made by using the standard {\tt FTool} {\tt lcmath}. 

The  position of \src on each of the images was calculated using the {\tt xrtcentroid} tool, and 
the derived value was adopted as the center for all the X-ray rings displayed on the image under study. 
Finally, the  radial profiles of the rings count rate per unit area were extracted by using the {\tt ximage} program.

For the event selection/extraction of the individual rings we used the {\tt HEASoft} program {\tt xselect}. 
The X-ray spectra of the rings ware extracted from an annulus with inner and outer radii determined
from the best fit radial profile of the ring (details about the fitting procedure can be found at \S\ref{sec-results}). 
The {\tt xrtmkarf} task was used to create the ancillary response function (ARF) used for the spectral analysis. 
The ARF was created for an extended source without PSF corrections from the vignetting-corrected exposure map.

For those \swift/XRT snapshots where the rings were partially outside the field-of-view (FOV) of the telescope, appropriate correction factors were estimated taking into account the extraction region and the exposure map of the instrument. The corrections were finally used as a scaling factor for the intensities derived after the fitting of the { X-ray} spectra.    

Throughout the monitoring of the system (MJD 57203.5-57239.5) the X-ray rings could always be easily distinguished. {To be exact, this holds
for the two innermost rings, while the two intermediate rings were clearly visible until MJD~57231.3 and the outermost
ring could not be detected after MJD~57226.5}.
For the four inner rings we chose to perform the analysis in pairs.
This approach was adopted since their positions were separated by less than the full width at half maximum (FWHM) of their radial profiles, thus their properties could not be disentangled. Additionally by summing the detected photons of two adjacent rings we improved the signal-to-noise of our data. In order to test that this merging did not affect the results of our analysis, we 
extracted partially the X-ray spectrum from different regions of the double peaked radial profile and repeated the same analysis; in both cases,
we found similar results.
%we performed several tests, namely for each pair of rings we extracted partially the X-ray spectrum from different regions of the double peaked radial profile and repeated the same analysis.

The extracted X-ray spectra were then analyzed with {\tt xspec} \citep{Arnaud1996} version 12.8.0. 
All the spectra were fitted with a phenomenological model of an absorbed power-law ({\tt tbnew*powerlaw}) and 
the elemental abundances of the ISM were set after \citet{wilms2000}. Due to the 
lack of photons {of} energies higher than 4~keV, the spectra were fitted in the energy range of 0.7-4~keV.
The regions for determining the background contamination of the spectra were selected from annuli at the edges of the rings in study.
The size of each annulus was based on the radial profile of the dust scattered rings. Details about the background selection can be found 
in \S\ref{sec-activity}.     
{By fitting first all the X-ray spectra with $\chi^2$-statistics, we obtained reduced $\chi^2$ values between 0.8 and 1.4 that justify the absorbed power-law model selection. For those observations with sufficient statistics {($\gtrsim 400$~counts)} the binned spectra were then used and fitted with $\chi^2$-statistics. In all other cases, the un-binned spectra were used and fitted with C-statistics \citep{Cash1979}. }
From the best-fit model we then derived the intensity values at 1.0, 1.5, 2.0 and 2.5 keV.
The observation log of the {23} \swift/XRT observations used for the analysis of the dust scattered rings can be found in 
Table \ref{tab:xray-obs} of appendix~\ref{appenA} {where they are listed in ascending order with respect to their starting time of observation.} 

For the study of the overall activity of the system we additionally used all available INTEGRAL data.
The satellite started monitoring the source from revolution 1554 till revolution 1563 with data made immediately public. Observations made during Revolutions 1555 and 1556 were done under an AO-12 approved Target-of-Opportunity program (PI: Rodriguez). The PI has kindly allowed public access to the data in the consolidated version\footnote{These are available in the following link: \url{http://www.isdc.unige.ch/integral/Operations/Shift/QLAsources/V404_Cygni/V404_Cygni.php}} \citep[][]{Rodriguez2015,Kuulkers2015}.

% The \swift/XRT observations used for extracting the light curve are not included in the table, 

\section{Analysis and Results}
\label{sec-results}
\subsection{Ring expansion}
\label{sec-expansion}
\begin{figure}
 \resizebox{\hsize}{!}{\includegraphics[angle=0,clip=]{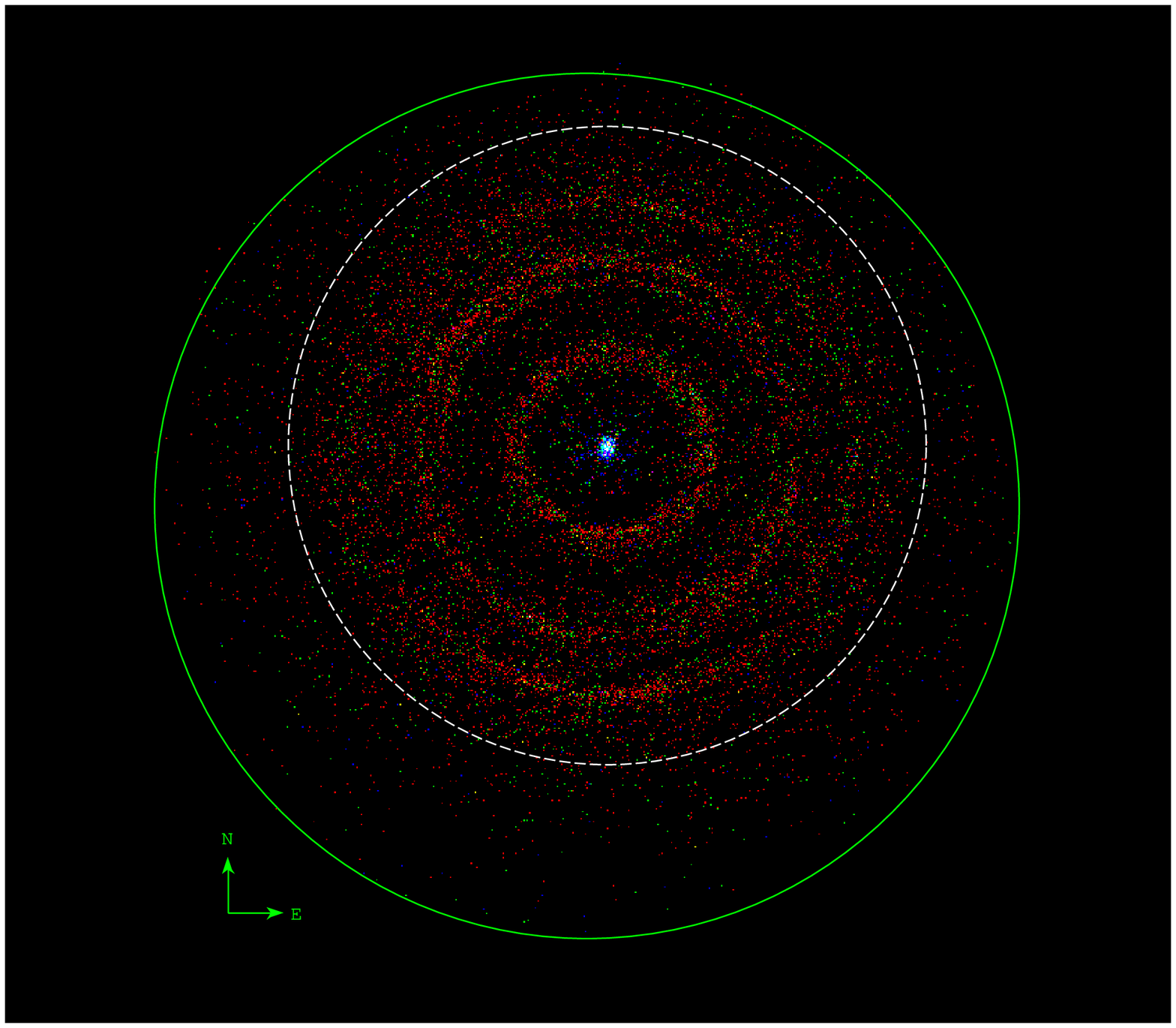}}
\caption{RGB image of the X-ray dust scattered rings observed by \swift/XRT at MJD~57205.5 {(\swift OBS-ID 00033861006)}. The color coding
used is R: 0.5-1.5~keV, G: 1.5-2.5~keV, and  B: 2.5-10~keV.  The  \swift/XRT field-of-view is enclosed
by the green circle, while a white circle of radius $9\arcmin$ is overplotted for guiding the eye.}
\label{fig:image_rgb}
\end{figure}
The radial expansion of the dust rings was modeled using the 1-10~keV \swift/XRT data, noting, however, that 
the dust scattered emission falls off rapidly at energies  $\gtrsim 2.5$~keV.
This is also illustrated
in Fig.~\ref{fig:image_rgb} where only the central source is detected in { the energy band 2.5-10~keV}.
Although we limited the analysis of XRT data above 1~keV, where approximate expressions for the scattering cross section
from dust are applicable  (for more details, see \S\ref{sec-theory}), 
we verified that our results on the  location and expansion of the rings
are not affected by the inclusion of lower energy X-ray data.
 
To model the angular profile of the dust scattered X-ray emission
and to pinpoint the angular sizes of the rings, we assumed a uniform 
background that was optimized and then subtracted from the data. {As we discuss in more detail in \S\ref{sec-activity}, 
the evolution of the angular profiles with time reveals a a variable
background emission, which is, most probably, related to the \src activity since its trigger.} 
% Thus for each observation we used a uniform background that 
The background subtracted angular profiles of the dust scattered emission  exhibit five major peaks that are
related to the five ring-like structures observed with \swift/XRT.
From this point on, we will refer to them as follows: the two innermost rings or ``rings 1 and 2''; the two intermediate
rings or ``rings 3 and 4''; and the outermost ring or ``ring 5''. 

We fitted the background subtracted angular profile using a model that consists
of a King profile and five Lorentzian functions ($f_L$) with three free parameters each:
\eqb
f_L^{(i)}= \frac{1}{4\pi}\frac{f_{0i} \gamma_i}{\sqrt{(r-r_{0i})^2 + \left(\frac{\gamma_i}{2}\right)^2}}, \ i=1,\dots,5.
\eqe
In the above expression, $f_L^{(i)}$ is in units of ct s$^{-1}$ arcmin$^{-2}$ and $r_{0i}, \gamma_i$ are in arcsec. 
The use of Lorentzian functions over-imposed on a King profile for modeling the observed radial profile of the dust rings
is  motivated by the following: 
\begin{enumerate}
 \item the point spread function (PSF) of \swift/XRT is well modeled
by a King profile \citep{Moretti2005}.
\item simulations performed by \cite{Tiengo2010} for the radial profile of a thin ring
that is broadened by a PSF with King profile showed that a Lorentzian gives a better description than a Gaussian function.
We verified that the replacement of the Lorentzian with Gaussian functions leads to poor fits, except for observations with poor signal to noise ratio between the ring profile and the local background or in cases where the X-ray ring is projected near the edge of the FOV of \swift/XRT, {such as observation \#19} (MJD 57223.4). In this case, the outermost ring was better described by a Gaussian function.
\end{enumerate}

In short, in all cases but one we applied the two-component model described above. 
The observed radial profiles (symbols) and the best-fit models (solid lines) for an indicative subset of the analyzed observations
are shown in Fig.~\ref{fig:radial}, while the rest are presented in Fig.~\ref{fig:radial-all} of Appendix~\ref{appenB}. 
The position of the peaks and their errors, which are used in our subsequent analysis, are summarized
in Table~\ref{Tab:modelfit}.
\begin{figure*}
\includegraphics[width=0.32\textwidth,angle=0,clip=]{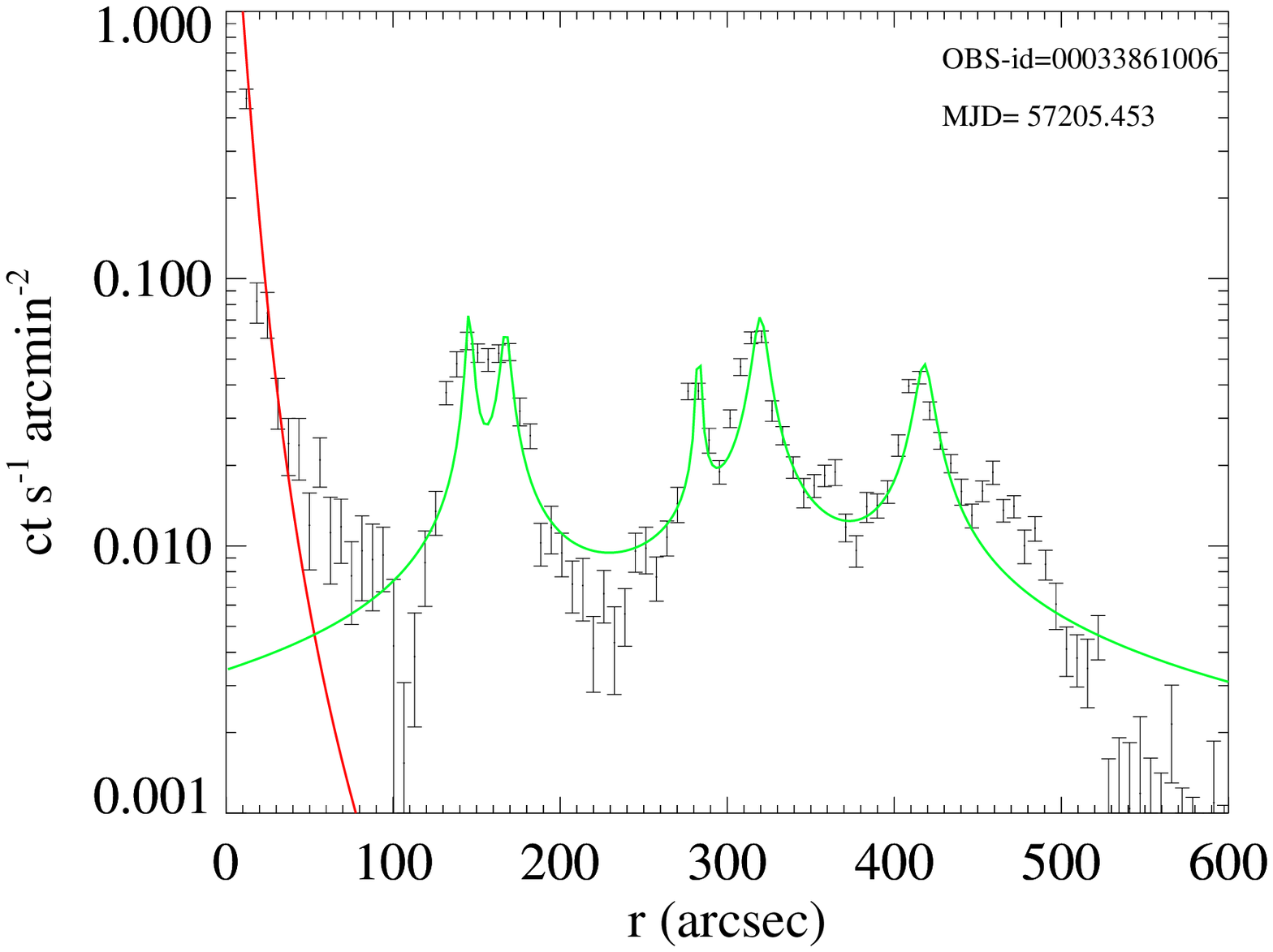}
\includegraphics[width=0.32\textwidth,angle=0,clip=]{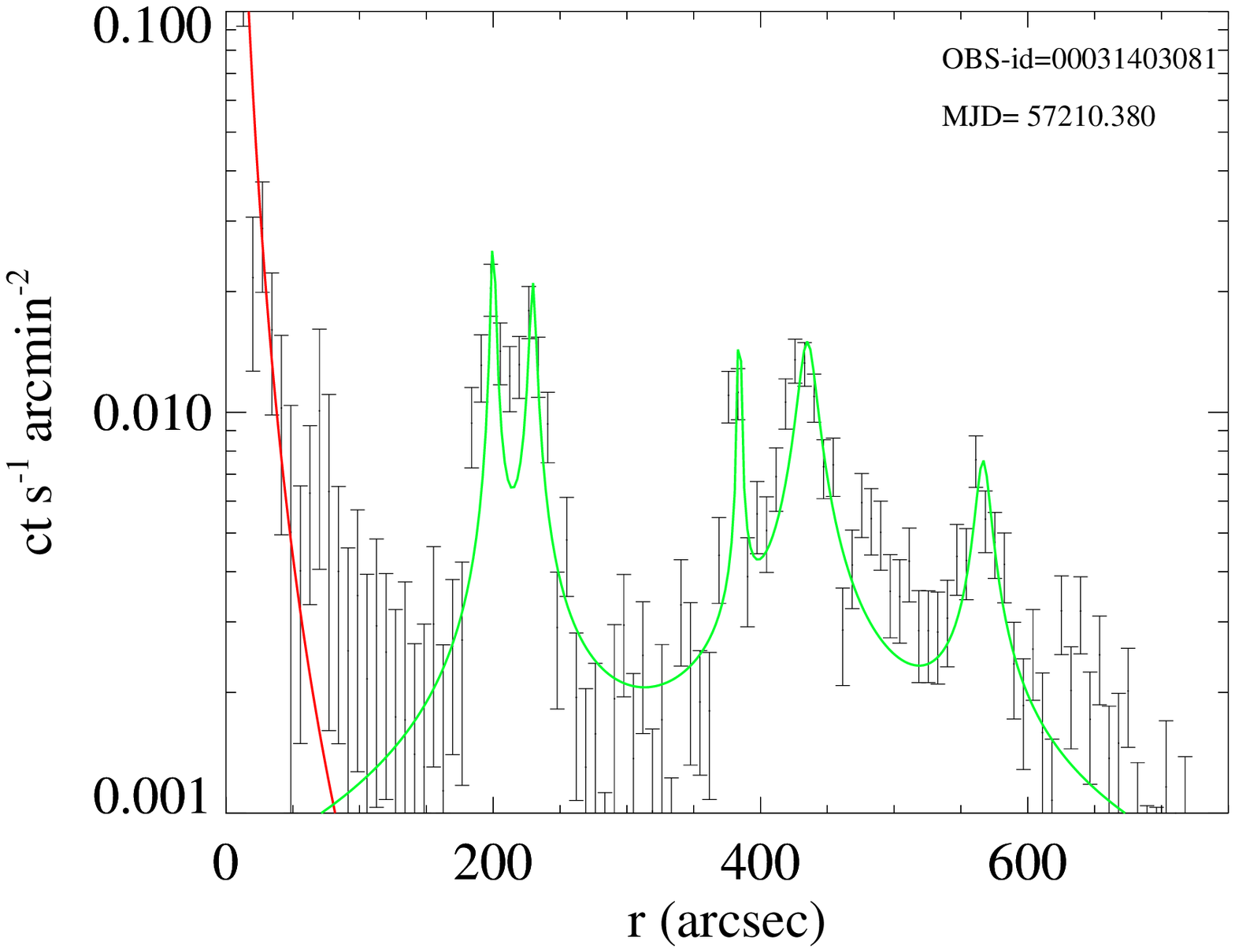}
\includegraphics[width=0.32\textwidth,angle=0,clip=]{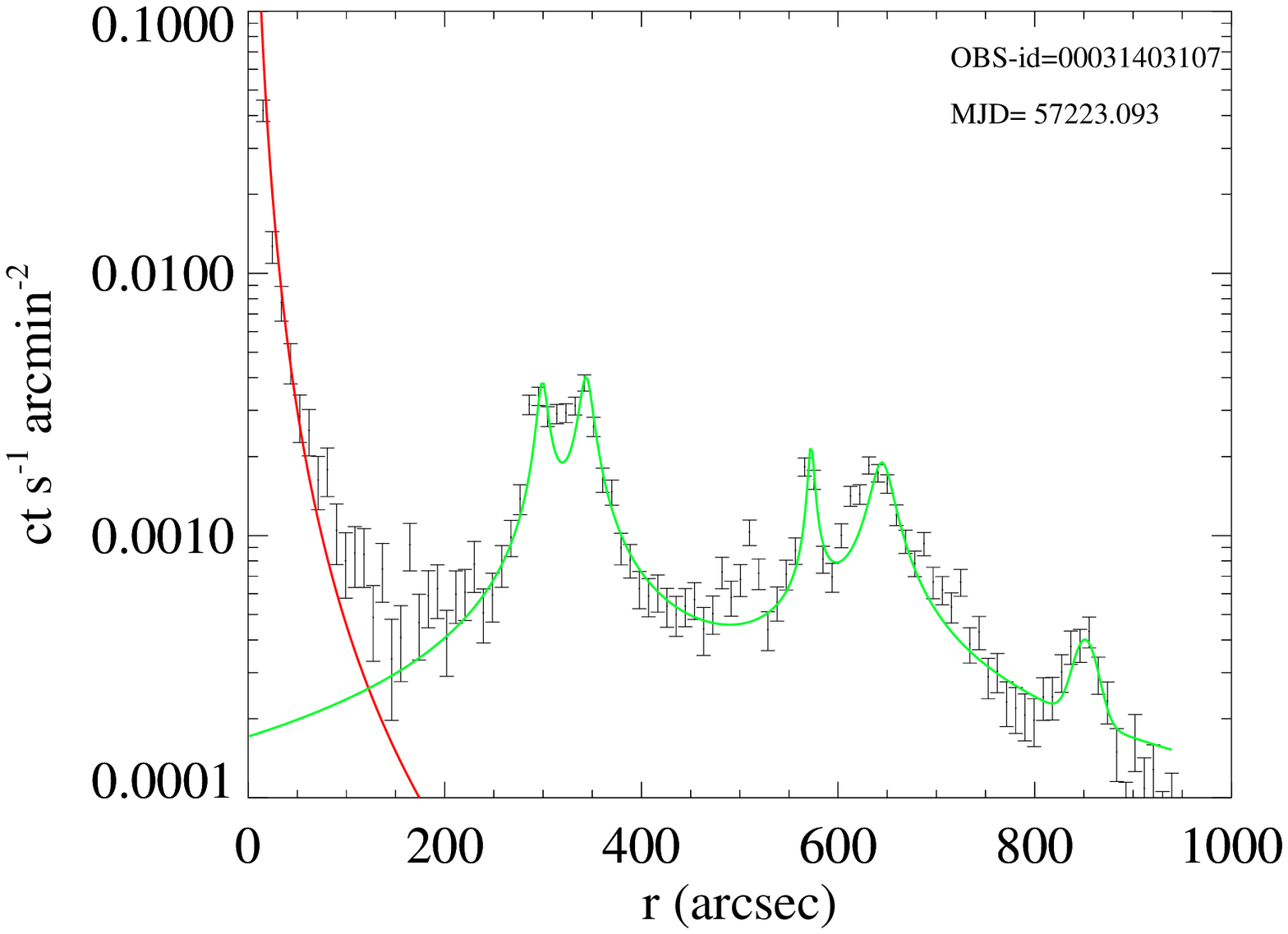}
  \caption{The background subtracted radial profile of the dust rings for three indicative \swift/XRT observations obtained in PC mode.
  The fitted model is composed by a King profile (red line) and five Lorentzian functions (green line).
  In some observations, such as { \# 19},  the outermost ring (ring 5) was best modeled by a Gaussian function.}
  \label{fig:radial}
\end{figure*}
Inspection of Fig.~\ref{fig:radial} shows that the best-fit model is particularly successful in capturing the peaks of the angular profile,
and in most cases, it provides a good description of the data. However, for some observations, {e.g. \#1 and \#2}, the reduced $\chi^2$ value of the global fit is high. This suggests that the two-component model cannot account for all the details of the respective radial profile. For example,
the poor global fits of observations {\#1 and \#2 (see Figs.~\ref{fig:radial} and \ref{fig:radial-all}, respectively)} are mainly caused by the difference between the  data and the model at large angular distances (i.e., $\theta \gtrsim 350\arcsec$ and $450\arcsec$, respectively). The observed excess of emission, 
 with respect to the predicted one by the fifth Lorentzian function  (see  Fig.~\ref{fig:radial}), can be understood in  terms of a diffuse X-ray scattered emission,
which, if the photon statistics allow, can be observed as an X-ray halo (see also Fig.~\ref{fig:image_rgb}). {This may be the result of dust layers having a  finite width instead of being infinitely thin, as we assumed in our analysis for simplifying reasons. Other sources of the observed excess may be the X-ray scattered emission due to prior outburst activity (see \S\ref{sec-activity})  and/or multiple scatterings, although the latter are expected to be the least important source as discussed in \S\ref{sec-theory}}.

\begin{figure}
 \centering
% \resizebox{\hsize}{!}{\includegraphics[angle=0,clip=]{./plots_bu/expansion_rings0508.ps}}
\resizebox{\hsize}{!}{\includegraphics[angle=0,clip=]{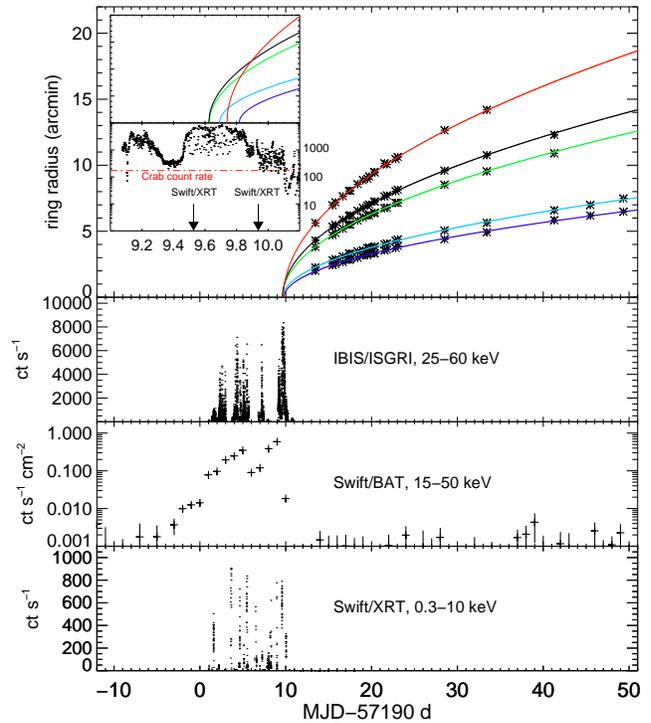}}
  \caption{The bottom three panels show (from bottom to top): the 60~s binned \swift/XRT (0.3-10~keV) light curve, 
  the daily binned \swift/BAT X-ray light curve at 
  15-50~keV based on the results from the BAT transient monitor \citep[][]{Krimm2013}, and the 
  INTEGRAL/ISGRI light-curve at 25-60~keV \citep[see text,][]{Rodriguez2015,Kuulkers2015}.
  Top panel: Expansion of the five dust rings detected by \swift/XRT (symbols) along with the best-fit model (solid lines)  given by eq.~(\ref{theta}).
  The positional uncertainties of the rings are smaller than the size of the asterisk symbol used in this graphics and thus are not plotted. 
  A close-up at the period of a major outburst
  detected by INTEGRAL/ISGRI  ($\sim$ MJD 57199.7) is shown in the
  inset plot. The arrows denote the dates of  simultaneous observations with \swift/XRT, while the horizontal dashed
  line corresponds to the count rate of Crab.}
  \label{fig:expansion_rings}
\end{figure}

Although the two-component model does not provide an overall good fit to some  observations, it still 
allows for an accurate determination of the angular size of the rings. 
This argument is also {\sl a posteriori} supported  by the fitting results of the ring expansion, which 
are summarized in Tab.~\ref{tab:expansion} and Fig.~\ref{fig:expansion_rings}. 
\begin{table}
\caption{Fitting parameters for the ring expansion using eq.~(\ref{theta}).}
\begin{center}
\scalebox{0.9}{
\begin{threeparttable}
\begin{tabular}{cccc}
\hline\hline
Ring &  $t_{0}$ & $x_i$\tnote{a} & $d_i$  \\ \noalign{\smallskip}
 \#  & [MJD-57190] &    &[kpc]   \\  \noalign{\smallskip}
\hline\noalign{\smallskip}
1 & 9.81$\pm$0.12& 0.8866$\pm$0.0006  &2.12$\pm$0.12(0.002)\tnote{b} \\  \noalign{\smallskip}
2 & 9.71$\pm$0.10& 0.8576$\pm$0.0006  &2.05$\pm$0.12(0.001) \\  \noalign{\smallskip}
3 & 9.62$\pm$0.06& 0.6839$\pm$0.0007  &1.63$\pm$0.10(0.001) \\  \noalign{\smallskip}
4 & 9.64$\pm$0.03& 0.6300$\pm$0.0003  &1.50$\pm$0.09(0.001) \\  \noalign{\smallskip}
5 & 9.73$\pm$0.05&0.4958$\pm$0.0006  &1.18$\pm$0.07(0.001) \\  \noalign{\smallskip}
\hline
\end{tabular}
\tnote{a} The values are calculated by fixing $t_0=57199.75$~d and performing a fit with only one free parameter. \\
\tnote{b} The error listed outside the parenthesis is the total one, i.e. calculated using error propagation, whereas
the one enclosed in the parenthesis is the $1\sigma$ statistical error {derived from the fit of the ring expansion.}
\end{threeparttable}
}
\end{center}
\label{tab:expansion}
\end{table}

The results in Tab.~\ref{tab:expansion} support the hypothesis of a single outburst of short duration occurring
at $t_0$ that is being scattered by dust clouds of small width, i.e. dust layers or screens, located at different locations between the observer and the source. Henceforth, we adopt the following convention: the dust layer responsible for the detection of the ring $i$ will be labeled as the ``dust cloud $i$''. 
The best-fit value for $t_0$ suggests that an X-ray outburst of short duration should have taken place at MJD~$=57190+t_0 \sim 57199.75$, namely
$\sim 3.7$~days before the first detection of the X-ray rings. 
Figure~\ref{fig:expansion_rings} (top panel) shows that the expansion
of the rings follows the theoretical prediction, i.e. $\theta \propto \sqrt{\Delta t}$, where $\Delta t$ is the time delay since the outburst. 
At late times the angular size of the outer rings (3-5) is larger than the \swift/XRT FOV, which also explains the lack of data points in the respective model curves in Fig.~\ref{fig:expansion_rings}. 
Interestingly, a broad X-ray flare with its peak centered at $\sim$MJD~57199.7
has been detected in both soft\footnote{Henceforth, we will refer to the 0.3-10~keV energy band of \swift/XRT
as the ``soft'' X-ray band, unless stated otherwise.} and hard X-rays, i.e. in 0.3-10~keV with \swift/XRT, in 15-50~keV with 
\swift/BAT \citep[BAT transient monitor,][]{Krimm2013},  and 25-60~keV with INTEGRAL IBIS/ISGRI. The respective light curves
are also shown in Fig.~\ref{fig:expansion_rings}. 

A close-up to the time-interval of the X-ray outburst as detected by INTEGRAL IBIS/ISGRI 
is shown in the inset of the top panel in Fig.~\ref{fig:expansion_rings}. The arrows denote the dates were simultaneous \swift/XRT
observations are available, while the horizontal dashed line corresponds to the Crab count rate (172.1$\pm$5.6 ct s$^{-1}$) as measured by INTEGRAL IBIS/ISGRI during satellite revolution 1528; it is noteworthy that the X-ray flux during this burst corresponds to the highest flux observed by INTEGRAL
after the trigger of the \src activity.  Although the outburst has a duration of $\delta t \sim 0.9$~days, 
this is still short enough compared to the time delay of the first X-ray ring detection, as to make the $\delta$-function approximation
of the X-ray outburst valid (see also \S\ref{sec-theory}). In any case, the derived value for $t_0$  coincides (within the errors) with the INTEGRAL/ISGRI outburst period.

Fixing $t_0=57199.75$~d and performing a one-parameter fit to the ring expansion using eq.~(\ref{theta}), 
we derived the distances of the dust layers. These fall in the range $\sim$1.2-2.1~kpc from the observer with the most distant one being located
close to the source and producing the innermost X-ray rings.  The calculated errors to the distances 
are dominated by the  error of the source distance itself ($d=2.39\pm0.14$), while  the $1\sigma$ statistical errors { derived from the fit of the ring expansion}
are approximately one order of magnitude smaller (see Tab.~\ref{tab:expansion}). We found that the spatial separation  of the dust clouds that produce the two innermost X-ray rings is small, i.e. $\sim 0.07$~kpc. This is in agreement with their observed angular proximity (see e.g. Figs.~\ref{fig:image_rgb} and  \ref{fig:radial}) and points towards a common origin of their dust content. 
If we were to relax the assumption of infinitely thin dust layers, our results  would suggest
the presence of one dust cloud at a distance $\sim 2$~kpc, with a finite width of $\sim 100$~pc and a non-uniform density that peaks roughly at the positions $x_1$ and $x_2$. {For the width estimation, we used the FWHM of the first two Lorentzian functions, as derived from modeling the radial profile of the first \swift/XRT observation}. Similar conclusions  can be drawn for the clouds 3 and 4.

\begin{figure}
 \resizebox{\hsize}{!}{\includegraphics[angle=0,clip=]{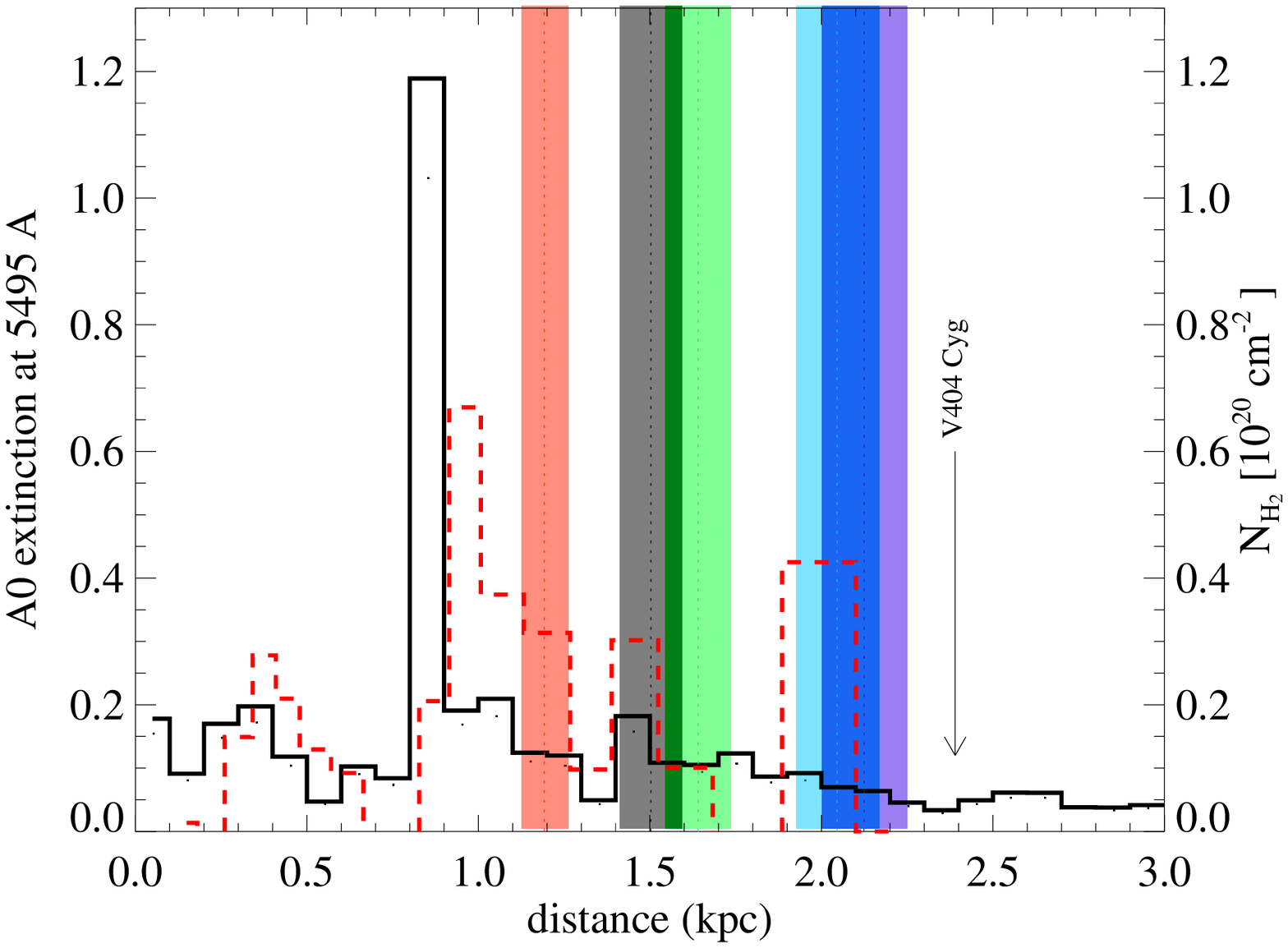}}
  \caption{The extinction $A_{\lambda}$ as a function of distance from the observer
  in the direction of \src (black lines). The data are obtained from three-dimensional extinction maps
  based on IPHAS photometry \citep{Sale2014}. 
  The distribution of $N_{\rm H_2}$ (red dashed lines)  as derived from CO measurements in the direction of \src \citep{Dame2001},
  using the Galactic rotation curve by \citep{Clemens1985} for  $R_{\odot}=8.5$~kpc and $V_{\odot}=220$~km/s, is also shown.
  The inferred distances of the dust layers are marked
  with dotted vertical lines, while the width of the colored bars denotes the total error. Same color coding as in Fig.~\ref{fig:expansion_rings} is used.} 
  \label{fig:dust_extinction}
\end{figure}

The distances that we derived can be then compared with other indirect
measurements of the dust distribution in the direction of \src, such 
as measurements of the extinction  $A_{\lambda}$. We choose, in particular, 
the three-dimensional extinction maps based on IPHAS photometry \citep{Sale2014} and plot in Fig.~\ref{fig:dust_extinction} 
the extinction  $A_{\lambda}$  vs. the distance from the observer in the direction of \src (black histogram). 
The vertical dotted lines denote the position of the dust layers as derived by modeling the ring expansion
and the width corresponds to the total error. Similar color coding as in Fig.~\ref{fig:expansion_rings} has been used.

In the direction of \src the extinction is relatively low, except for a peak at $\sim 0.85$~kpc, where $A_{\lambda} \simeq 1.2$.
The positions of the dust layers fall into regions where $A_{\lambda}\sim 0.1-0.2$. 
However, we find no dust layer located at the distance
of maximum extinction. An X-ray ring produced by scattering on a hypothetical dust layer
located at $\sim 0.85$~kpc and detected simultaneously with the other five rings
on MJD~57205.5 should have an angular size of $\sim 9\arcmin$ (see eq.~(\ref{theta})). 
The \swift/XRT image, however, shows no signs of an additional ring with that angular size 
(see white circle in Fig.~\ref{fig:image_rgb}). Since the 
hypothetical ring should fall within the FOV of \swift/XRT, the absence of a dust-scattered ring in combination with 
the high extinction value might imply a very small scattering cross section (see e.g. eq.~(\ref{cross})).
Since we observe no ring structure even at $E=1$~keV, we may estimate a lower limit of the average
grain size of a hypothetical dust cloud located at $0.85$~kpc. Substitution of $\theta=9\arcmin$ and $E=1$~keV in eq.~(\ref{theta-big})
results in $\alpha \gg 0.11\,\mu$m, which translates to $\alpha_{\max}\gg 0.18\,\mu$m for $q=3.5$ (see eq.~(\ref{average})).

In addition, we compared the derived distances with the distribution of molecular hydrogen (H$_2$) in the direction
of \src as derived by CO measurements. For this, we used
the data from the Galaxy CO survey \citep{Dame2001}, and, in particular, the velocity-resolved, moment-masked (i.e., noise-suppresed) 
\citep{Dame2011} radio maps of the Individual Survey DHT18 (Second Quadrant, $1/4^{\rm o}$ sampled).
For the de-projection of the observed velocities we 
used the rotation curve of \cite{Clemens1985} for $V_{\odot}=220$~km/s and  $R_{\odot}=8.5$~kpc. {The latter
is close to the value derived by recent  dynamic studies of the Galactic center \citep{Chatzopoulos2015}, namely 
$R_{\odot}=8.33\pm0.11$.} The distribution 
of $N_{\rm H_2}$ (in cm$^{-2}$) is over plotted in Fig.~\ref{fig:dust_extinction} with a red dashed line. 
The derived distances of the five dust layers
seem to coincide with those where the column density appears enhanced. The maximum column density ($\sim 6\times 10^{19}$~cm$^{-2}$)
is found at a distance close to the maximum extinction  $A_{\lambda}$, where no dust cloud is inferred from the \swift/XRT observations.

% The abrupt cutoff of the $N_{H_2}$ histogram  at 2~kpc reflects the fact that the measured radial velocities
% cannot be explained due to motions around the Galactic center for the adopted Galactic rotation curve. 
% This is a result of the geometry of the problem and the resolution of the CO maps; 
% let us consider the measured LSR velocities of the CO map at a longitude $l$.
% % let us assume a distance $R_{\odot}$ from the galactic center and a line of sight along a direction of longitude $l$.
% Then, the minimum distance between the LOS and the galactic center is $R_{\odot}\sin(l)$, while the CO map observation measures LSR velocities along the same LOS that yield galactocentric distances smaller than $R_{\odot}sin(l)$, these velocities cannot be explained from the Galactic rotation curve. Given a galactocentric distance of $R_{\odot}$=8.5, for the direction of \src we calculate that $R_{\odot}\sin(l)$=8.12 kpc, which translates to a distance from the observer of 2.48 kpc ($R_{\odot}\cos(l)$). It is a unfortunate  coincidence that \src is locates at a distance of $\sim$2.39 kpc, thus at a location that is poorly resolved by by the CO map for the given Galactic rotation curve.

\begin{figure}
 \resizebox{\hsize}{!}{\includegraphics[angle=0,clip=]{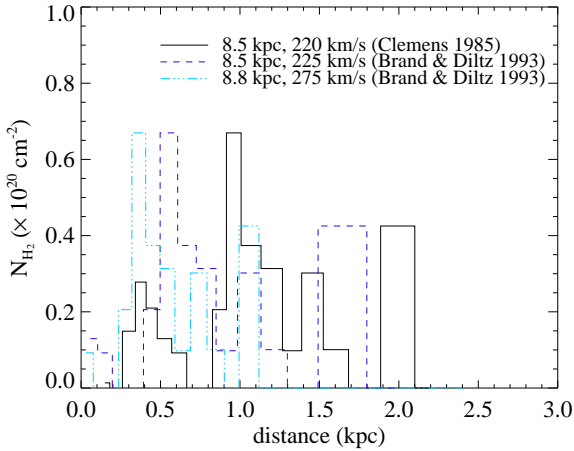}}
  \caption{The distribution of molecular hydrogen in the direction of \src as inferred from the same CO radio maps
  but for different values of the Galactic constants and Galactic rotation curves (see inset legend).} 
  \label{fig:CO}
\end{figure}

It is noteworthy that the  distribution of $N_{\rm H_2}$ is very sensitive to the adopted values of
the Galactic constants ($R_{\odot}$,$V_{\odot}$) and of the Galactic rotation curve itself 
\citep[see e.g.][]{Clemens1985, Sofue2009,Bhattacharjee2014}. This is illustrated in 
Fig.~\ref{fig:CO}, where we plot the derived distribution of $N_{\rm H_{2}}$ using
three different values of Galactic constants, namely ($R_{\odot}$~kpc,$V_{\odot}$~km/s)=$\left[(8.5, 220), (8.5, 225), (8.8, 275)\right]$. 
For the first pair of values (black line), the Galactic rotation curve by 
\cite{Clemens1985}  was used, while the rotation curve by \cite{BrandBlitz1993} was adopted for the other two (cyan dashed-dotted and blue dashed lines).

In summary, we have showed that the angular structure of the \swift/XRT images can be adequately explained
by scattering of a short duration X-ray outburst on dust layers located at distances $\gtrsim 1.2$~kpc from the observer. 
These conclusions are independent of the composition and properties of the dust, which will be 
determined in the following section.

\subsection{Properties of the dust clouds}
\label{sec-dustprop}
% Here, we generate the energy-dependent radial intensity profiles of the rings ($I^{(i)}_{\rm sc}(\theta, E)$) in order to derive
% the basic properties of the dust grains. 
To model the evolution of the specific { scattered flux} of the X-ray rings,
we assumed that the grain size distribution in all five clouds is described by a generalized 
Mathis-Rumpl-Nordsieck (MRN) model \citep[][]{MRN1977}:
\eqb
n(\bar{\alpha}) = A \bar{\alpha}^{-q}H[\bar{\alpha}-\bar{\alpha}_{\min}]H[\bar{\alpha}_{\max}-\bar{\alpha}],
\eqe
where $\bar{\alpha}\equiv \alpha/1 \ \,\mu$m {and $H[x]$ is the Heaviside function defined to be equal to 1 for $x>0$ and to 0, otherwise.} The model has four free parameters, namely
the slope $q$, the normalization $A$, the minimum $\bar{\alpha}_{\min}$ and maximum
$\bar{\alpha}_{\max}$ grain sizes. The normalization $A$ (dimensionless) is related to the 
total number of grains in a given cloud through $N_{\rm tot}=\int d\bar{\alpha}n(\bar{\alpha})= A \left(\bar{\alpha}_{\max}^{1-q}-
\bar{\alpha}_{\min}^{1-q}\right)/\left(1-q\right)$.
In the following, we assume that the model parameters are different among the dust clouds.
Substitution of the power-law distribution in eq.~(\ref{Fsc-delta}) results in
\eqb
F_{\rm sc, i}\left(\theta, E\right) = \mathcal{F}^{(i)}_{\rm X}(E)\int_{\bar{\alpha}_{\min}}^{\bar{\alpha}_{\max}}\!\!d\bar{\alpha} \ \bar{\alpha}^{6-q}
\exp\left(-\frac{\theta^2}{2 (1-x_i)^2\Theta^2(E,\bar{\alpha})}\right),
\label{Fsc-final}
\eqe
where we used the approximation $\theta_{\rm sc} \approx \theta / (1-x_i)$ and the normalization
is given by
\eqb
\mathcal{F}_{\rm X}^{(i)}(E) = 2.54\times10^{-11}\frac{{\Phi_{\rm X}(E)} N_{\rm H, i} A_{\rm i} C_{\rm d,i}(E)}{x_{\rm i} \left(1-x_{\rm i}\right)},
\label{norm}
\eqe
{where $\mathcal{F}_{\rm X}^{(i)}$ is in units of erg cm$^{-2}$ s$^{-1}$  keV$^{-1}$ and $N_{\rm d}\equiv N_{\rm H,i}A_{\rm i}$ is the column density of dust,{ that is the number of grains per cm$^2$.} In principle, by fitting the scattered {flux} of the individual rings as given by eq.~(\ref{Fsc-final}),
the following parameters of the dust clouds can be determined: $q$, $\bar{\alpha}_{\min}$,$\bar{\alpha}_{\min}$ and $\mathcal{F}_{\rm X}(E)$.
 \begin{figure*}
\centering
 \includegraphics[width=0.33\textwidth,angle=0,clip=]{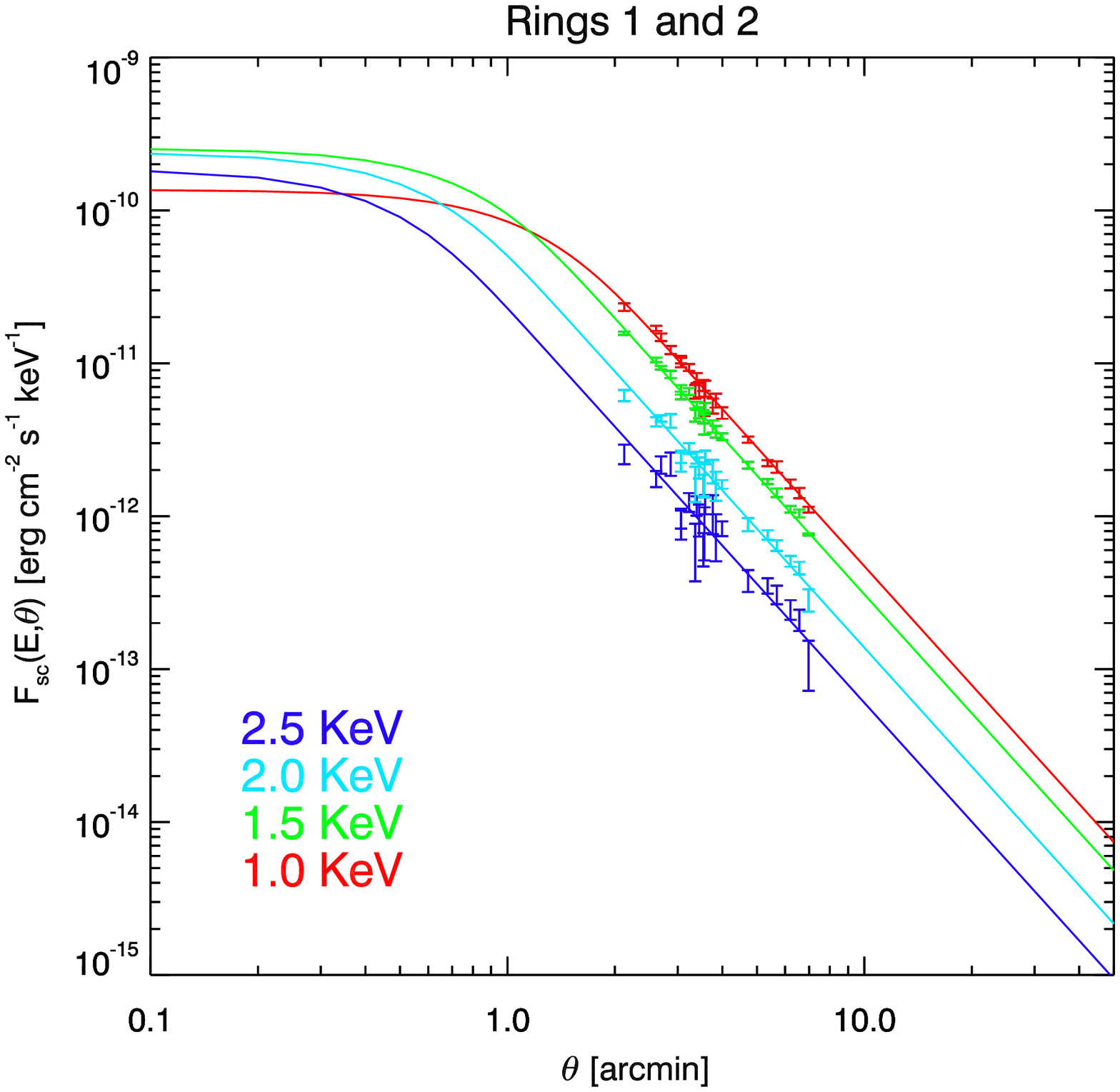}
 \includegraphics[width=0.33\textwidth,angle=0,clip=]{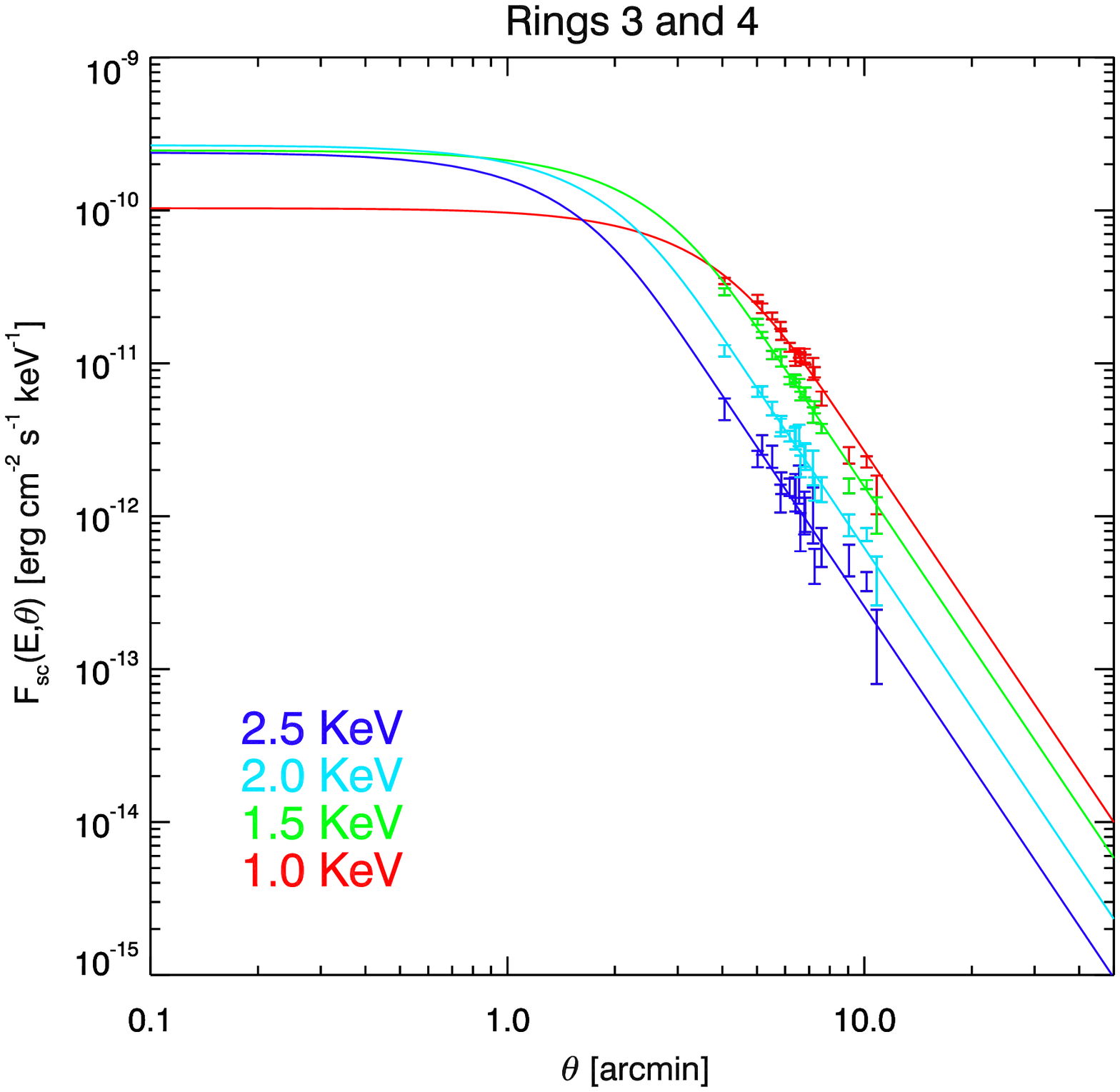}
 \includegraphics[width=0.33\textwidth,angle=0,clip=]{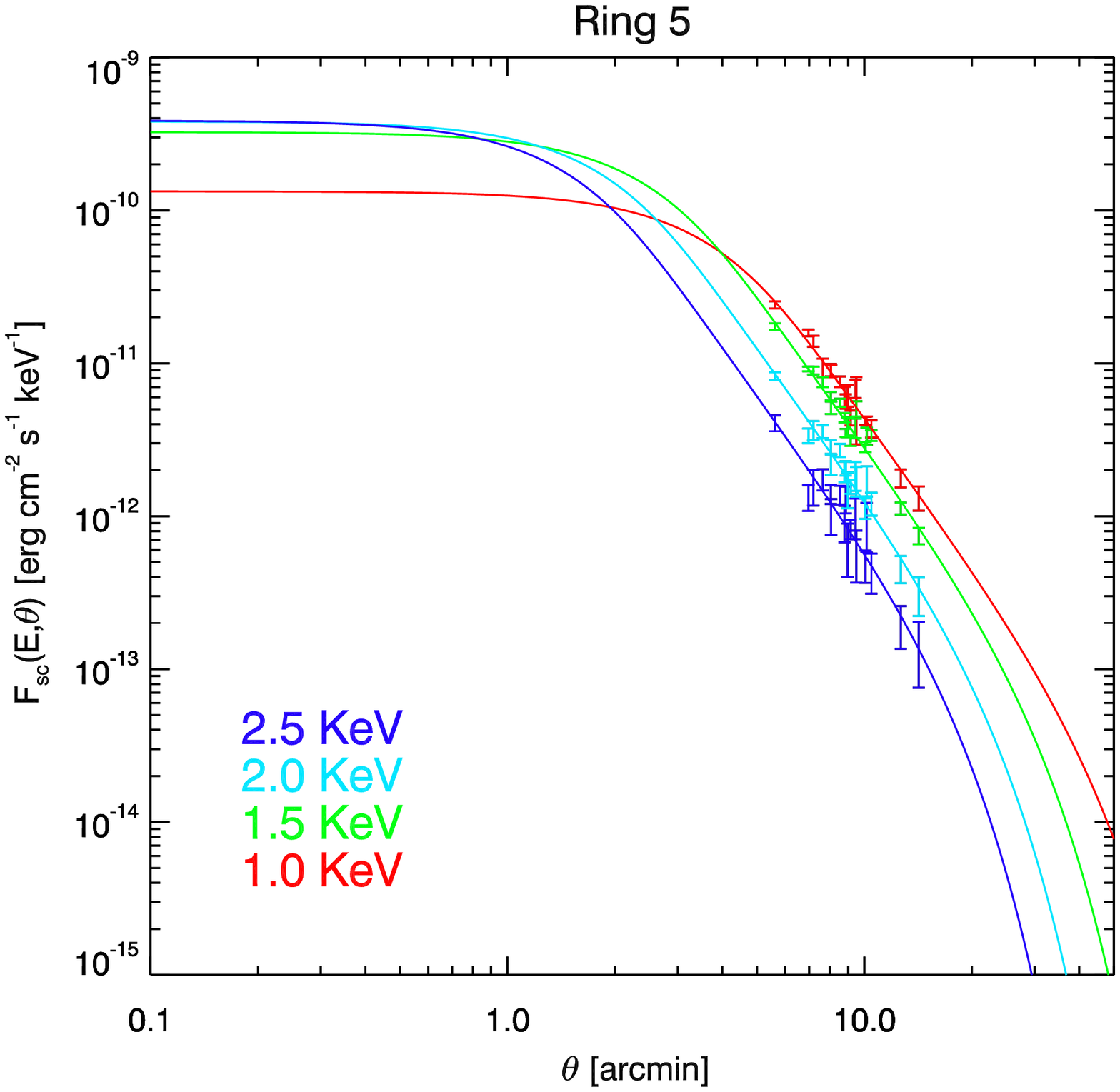}
  \caption{Scattered {flux} of the X-ray rings as a function of the observed angular size $\theta$ for 
  different energies marked on the plot. The cumulative profiles of the two innermost rings and rings 3-4 are shown in the left and
  middle panels, respectively. The {flux} profile of the outermost ring is plotted in the right panel.
  For the cumulative  profiles, we used the average angular size of the two adjacent rings. The best-fit model 
  is plotted in all panels with a solid line.}  
  \label{fig:intensity}
\end{figure*}
\begin{table*}
\caption{Best-fit values derived by modelling the radial profile of the X-ray scattered { flux} at four different energies.}
\begin{center}
\scalebox{0.9}{
\begin{threeparttable}
\begin{tabular}{cccc c cccc c}
\hline
Ring(s) & $\alpha_{\min}$  & $\alpha_{\max}$ & $q$ & $\langle \alpha \rangle$ & $\mathcal{F}_{\rm X}(1 {\rm keV})$ & $\mathcal{F}_{\rm X}(1.5 {\rm keV})$ & $\mathcal{F}_{\rm X}(2 {\rm keV})$ & $\mathcal{F}_{\rm X}(2.5 {\rm keV})$ & $\chi^2_{\rm red}$ (dof)\\
        & [$\,\mu$m]         &  [$\,\mu$m]    &   &    [$\,\mu$m]             & \multicolumn{4}{c}{[erg cm$^{-2}$ s$^{-1}$ keV$^{-1}$]} & \\
\hline 
1-2& $10^{-5}\pm$N/A\tnote{a} & $0.18\pm0.03$ & $4.42\pm0.02$& $0.07\pm0.01$ & $(3.0\pm0.2)\times 10^{-8} $ & $(5.5\pm0.3)\times 10^{-8} $ & $(5.2\pm0.3)\times 10^{-8} $  & $(4.0\pm0.3)\times 10^{-8}$ &  1.90(81) \\
3-4&$10^{-5}\pm$N/A\tnote{a} & $0.17\pm0.01$ & $3.53\pm0.06$ & $0.10\pm0.01$ & $(1.8\pm0.3)\times 10^{-7} $ & $(4.3\pm0.8)\times 10^{-7} $ & $(4.7\pm0.9)\times 10^{-7} $  & $(4.2\pm0.9)\times 10^{-7}$ & 1.91(70)\\
5  & $0.02\pm0.01$ & $0.24\pm0.03$ & $3.77\pm0.11$ & $0.13\pm 0.02$ & $(4.2\pm1.1)\times 10^{-8} $ & $(1.0\pm0.3)\times 10^{-7} $ & $(1.2\pm0.3)\times 10^{-7} $  & $(1.2\pm0.4)\times 10^{-7}$ & 1.89(65)\\
\hline
\end{tabular}
\item{a} As the best-fit is obtained for a value equal to the imposed lower bound, the error calculation is not applicable in this case.
\end{threeparttable}
}
\end{center}
\label{tab:intensity}
\end{table*}
Due to the angular proximity of the rings 1-2 and 3-4 (see also \S\ref{sec-observations}), the use of
the {flux} profile extracted from individual rings is less reliable \citep[see e.g.][]{Tiengo2010}. 
For the purposes of the fit, we therefore
used the cumulative {scattered flux} of rings 1-2 and 3-4, assuming that the dust properties of the
the clouds that are responsible for the  formation of adjacent X-ray rings are the same. This assumption
is also justified by the proximity of the dust clouds 1-2 and 3-4, as it was derived in \S\ref{sec-expansion} by modelling the ring expansion (see also Tab.~\ref{tab:expansion}). From this point on, we will treat the innermost rings as one, with an angular size given by the mean of the angular
sizes of the rings 1 and 2. In addition, in eq.~(\ref{Fsc-final})  we substitute 
$x_{\rm i}$ by the mean of the distances, i.e. $\langle x\rangle_{12}=(x_1+x_2)/2$. The same applies for the third and fourth rings.

For each individual ring, we jointly fitted $F_{\rm sc,i}(\theta,E)$ at four different energies 
using the expression (\ref{Fsc-final}). To account for the energy dependence of $\mathcal{F}_{\rm X}$, we allowed for four
different normalizations. In total, the number of free parameters required for modelling one ring are seven: four 
related to the energy dependent-normalization plus three related to the grain size distribution (the fourth free parameter
of the grain size distribution, $A$, is absorbed in the normalization).

For each ring, we first  searched for the best-fit model after leaving free all the model parameters. 
We imposed, however, limits on both $\alpha_{\min}$ and $\alpha_{\max}$. These were respectively bounded in the ranges
 $10^{-5}-0.1\,\mu$m and $0.1-0.5\,\mu$m. {We caution the reader that our results regarding the largest grains
 should be trusted up to $\sim 0.3\,\mu$m, given that for larger grains the RG approximation overestimates by at least a factor of $\sim 3-4$ \citep[e.g.][]{Smith1998}
 the scattered flux of X-ray photons with energy $E \sim 1$~keV (see also \S\ref{sec-discussion} for a relevant discussion).}
 
 Figure~\ref{fig:intensity} shows the angular profile of the 
 scattered {flux} for rings 1-2 (left panel), rings 3-4 (middle panel) and ring 5 (right panel)
 at four different energies (coloured symbols) along with the best-fit model (solid lines). 
The respective parameter values are listed in Table~\ref{tab:intensity}. 
To facilitate a comparison among the rings, the specific scattered flux of the rings is also shown in Fig.~\ref{fig:best-fit}.
The fact that the profiles of rings 3-4 and ring 5 are almost overlapping does not
necessarily mean that the corresponding dust layers have the same hydrogen column density, since the dust cloud 5 is located closer
to the observer (see also below). The intensity profiles of all rings, except for the outermost, show no evidence of an exponential
cutoff, which could strongly constrain the minimum size of the grains (see \S\ref{sec-theory}). 
\swift/XRT observations of the innermost rings  after  MJD~57239.3 ({observation \#21}) could shed light on the actual size of the smallest
grains. However, the signal-to-noise ratio of follow-up observations  was not high enough (time exposure $\lesssim 1$ks) 
for a reliable spectral extraction and fitting.
We note that the data at 2.5~keV show systematically a larger scatter, which  is due to a deficit of photons
at the higher energy part of the spectrum. 
 \begin{figure}
\centering
 \resizebox{\hsize}{!}{\includegraphics{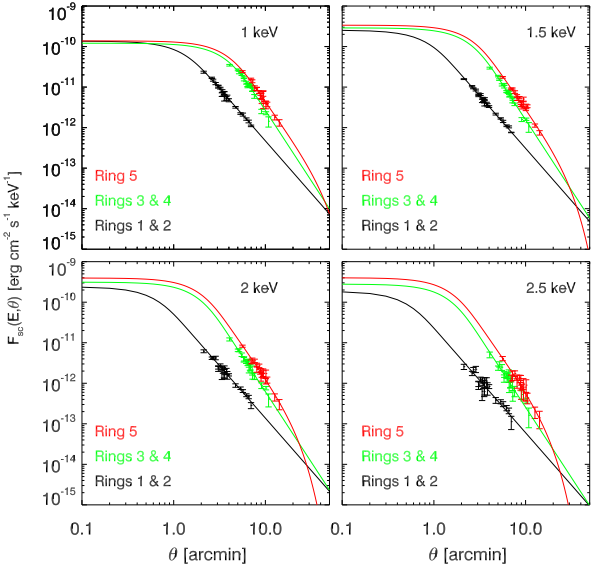}}
  \caption{Comparison of the intensity angular profiles  of the X-ray rings  at fixed energies.}  
  \label{fig:best-fit}
\end{figure}
 \begin{figure*}
 \centering
 \includegraphics[width=0.32\textwidth,angle=0,clip=]{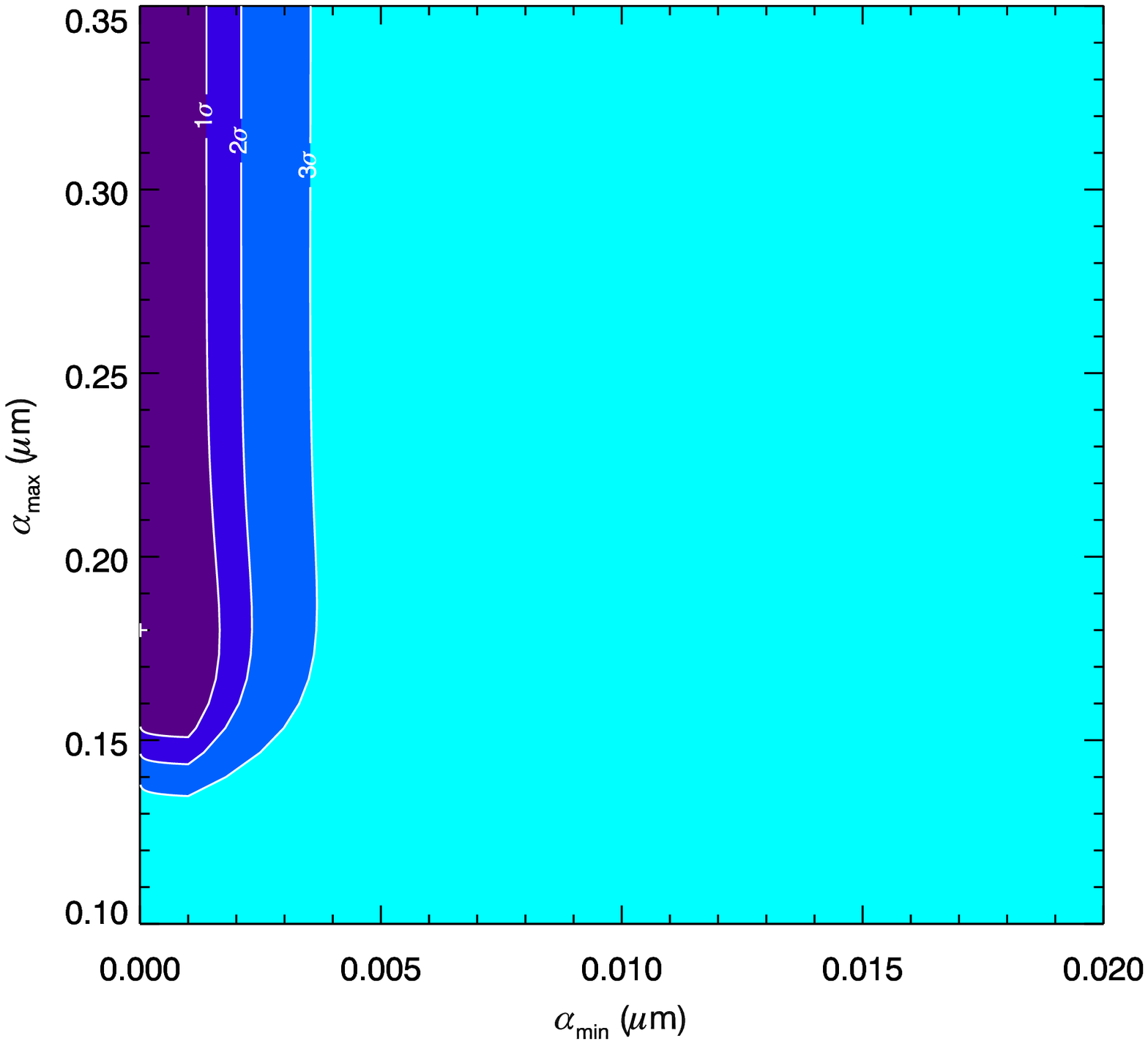} 
 \includegraphics[width=0.32\textwidth,angle=0,clip=]{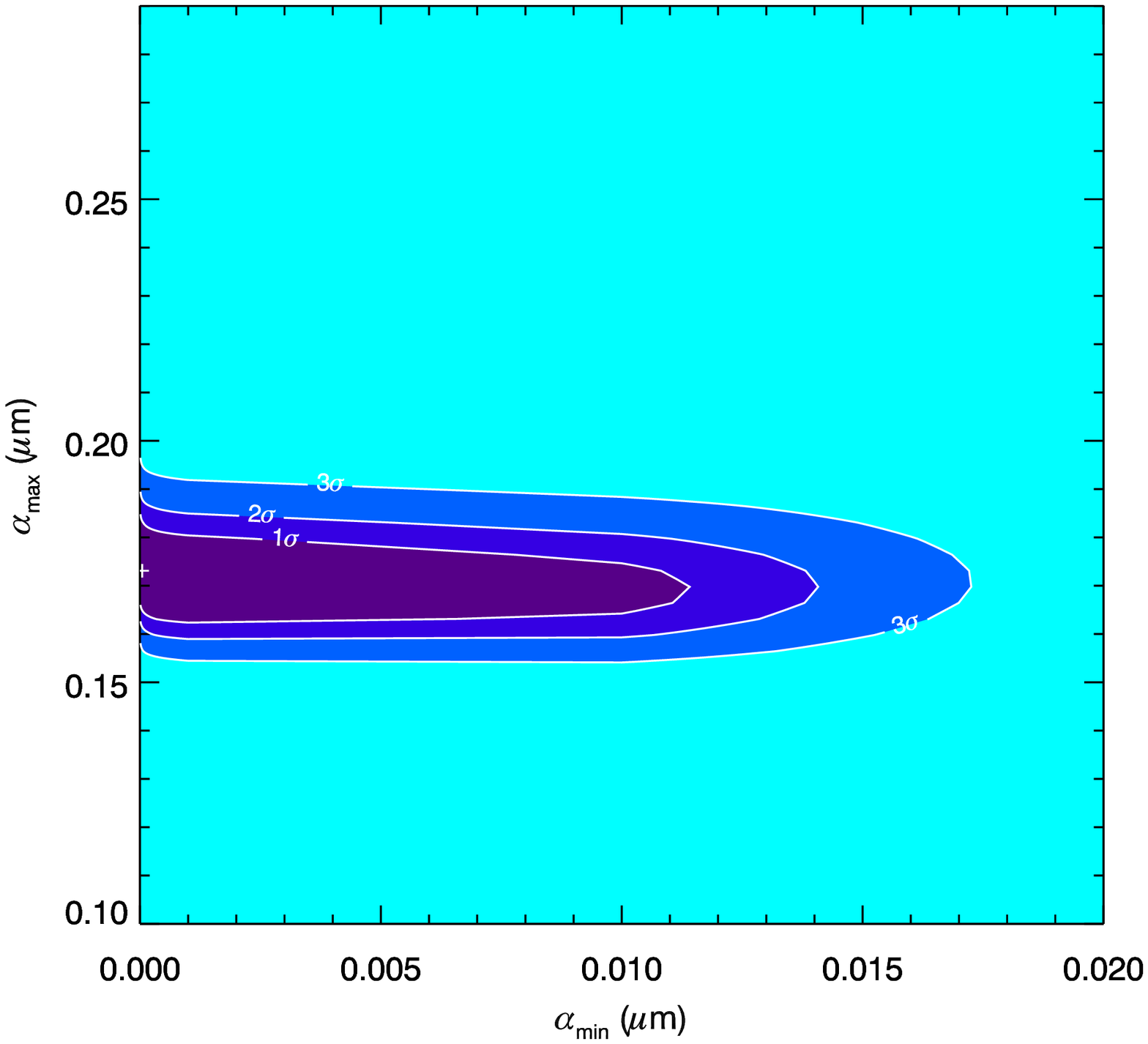}
 \includegraphics[width=0.32\textwidth,angle=0,clip=]{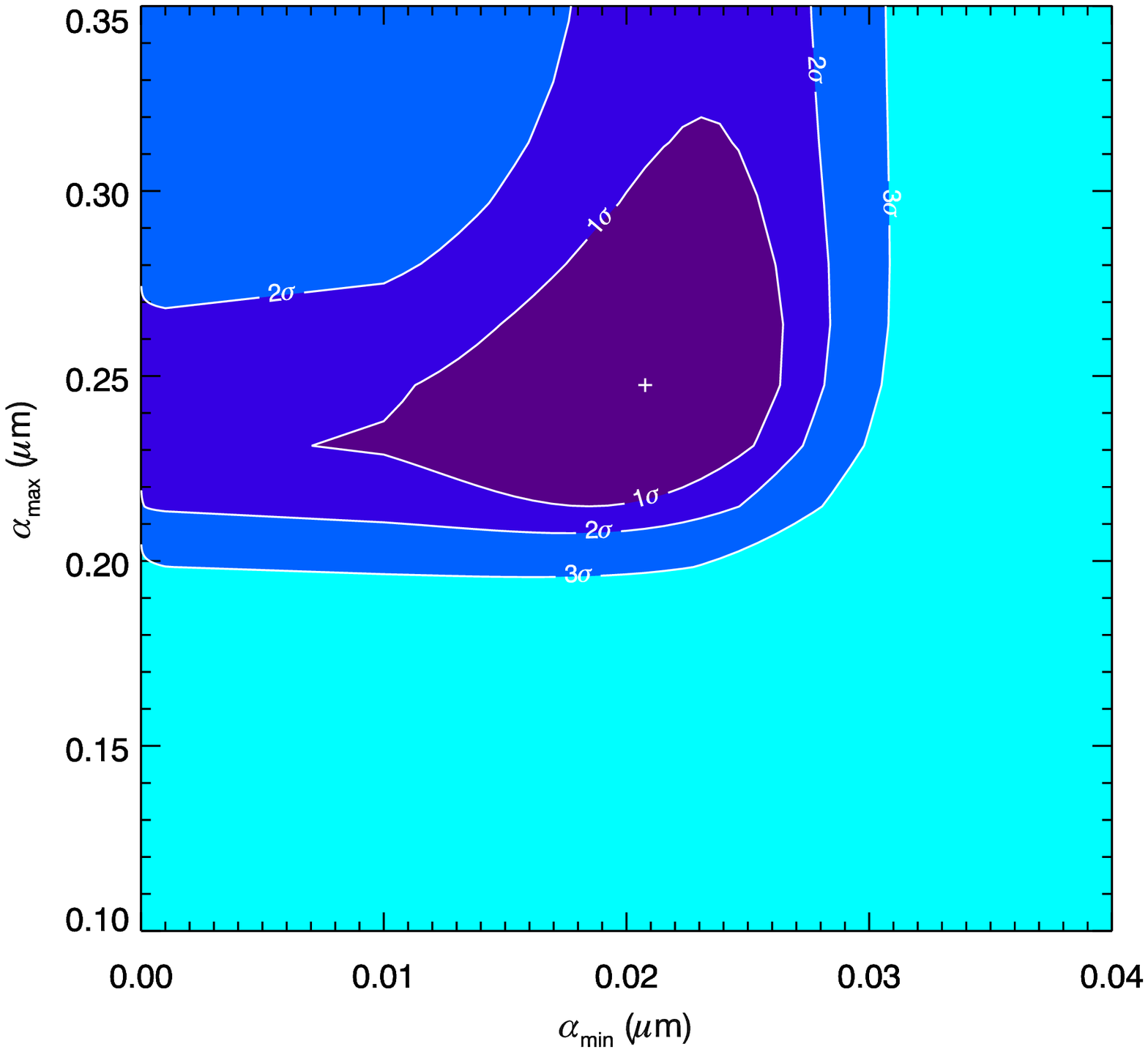}  \\
 \includegraphics[width=0.32\textwidth,angle=0,clip=]{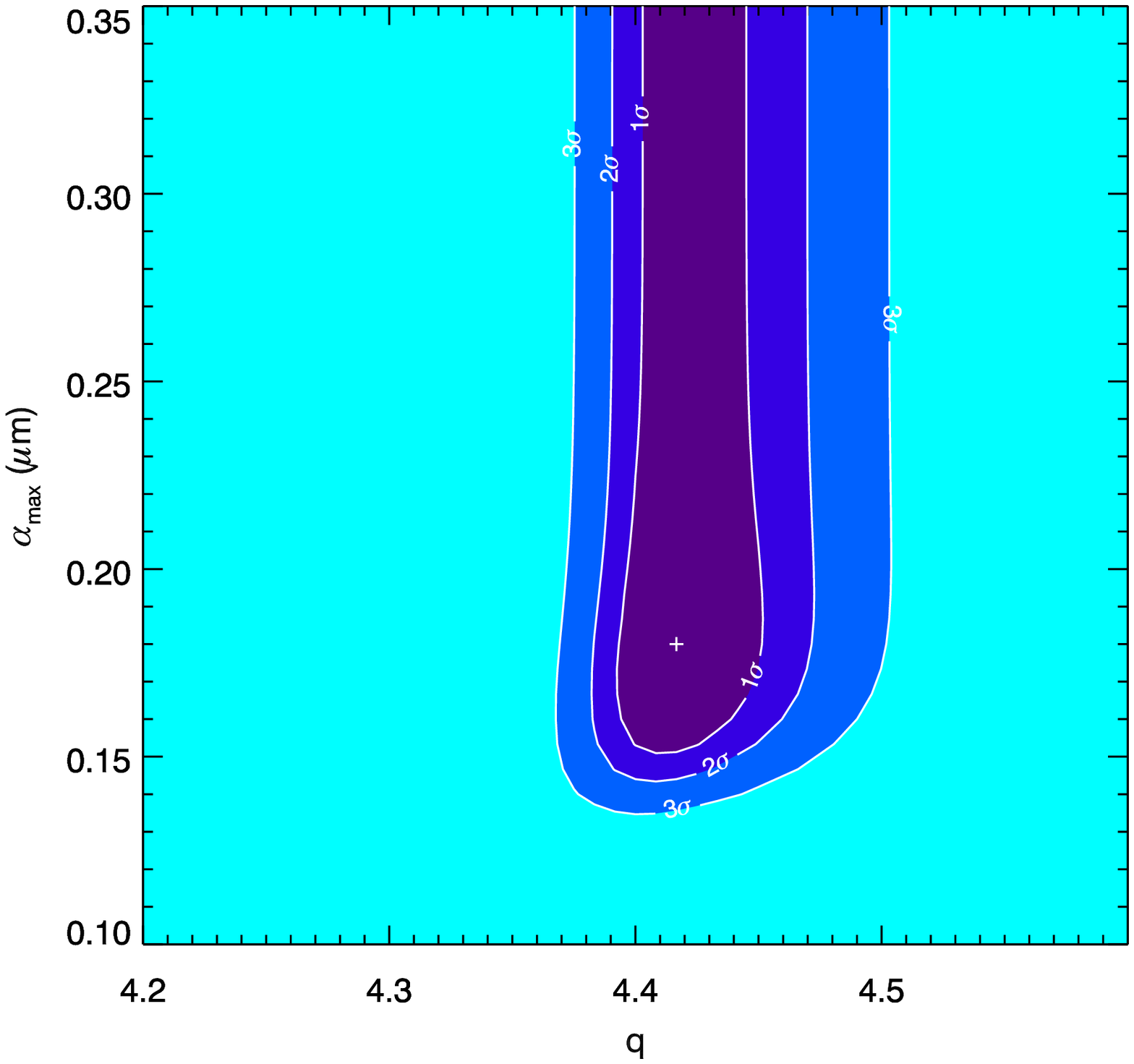} 
 \includegraphics[width=0.32\textwidth,angle=0,clip=]{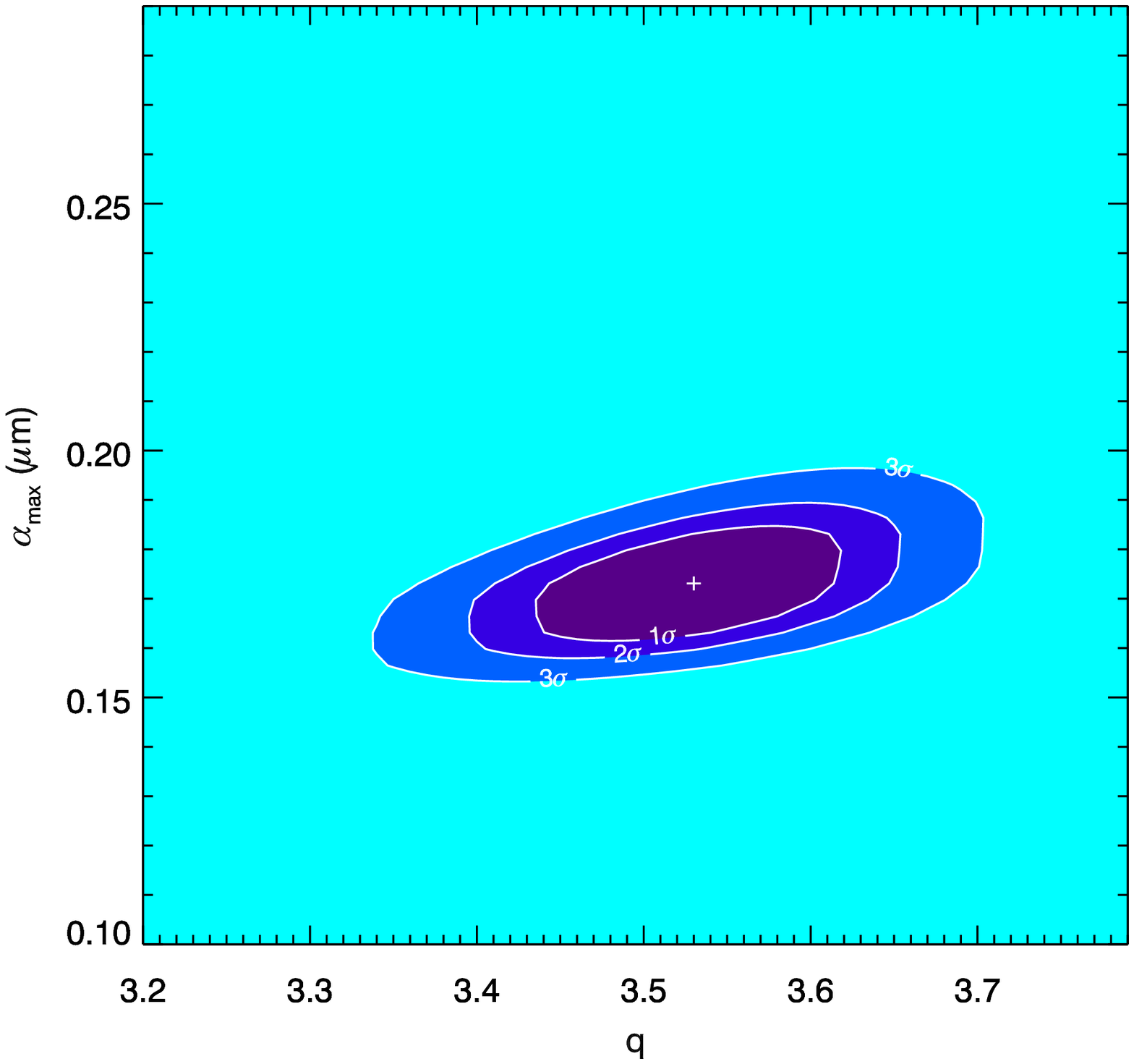}  
 \includegraphics[width=0.32\textwidth,angle=0,clip=]{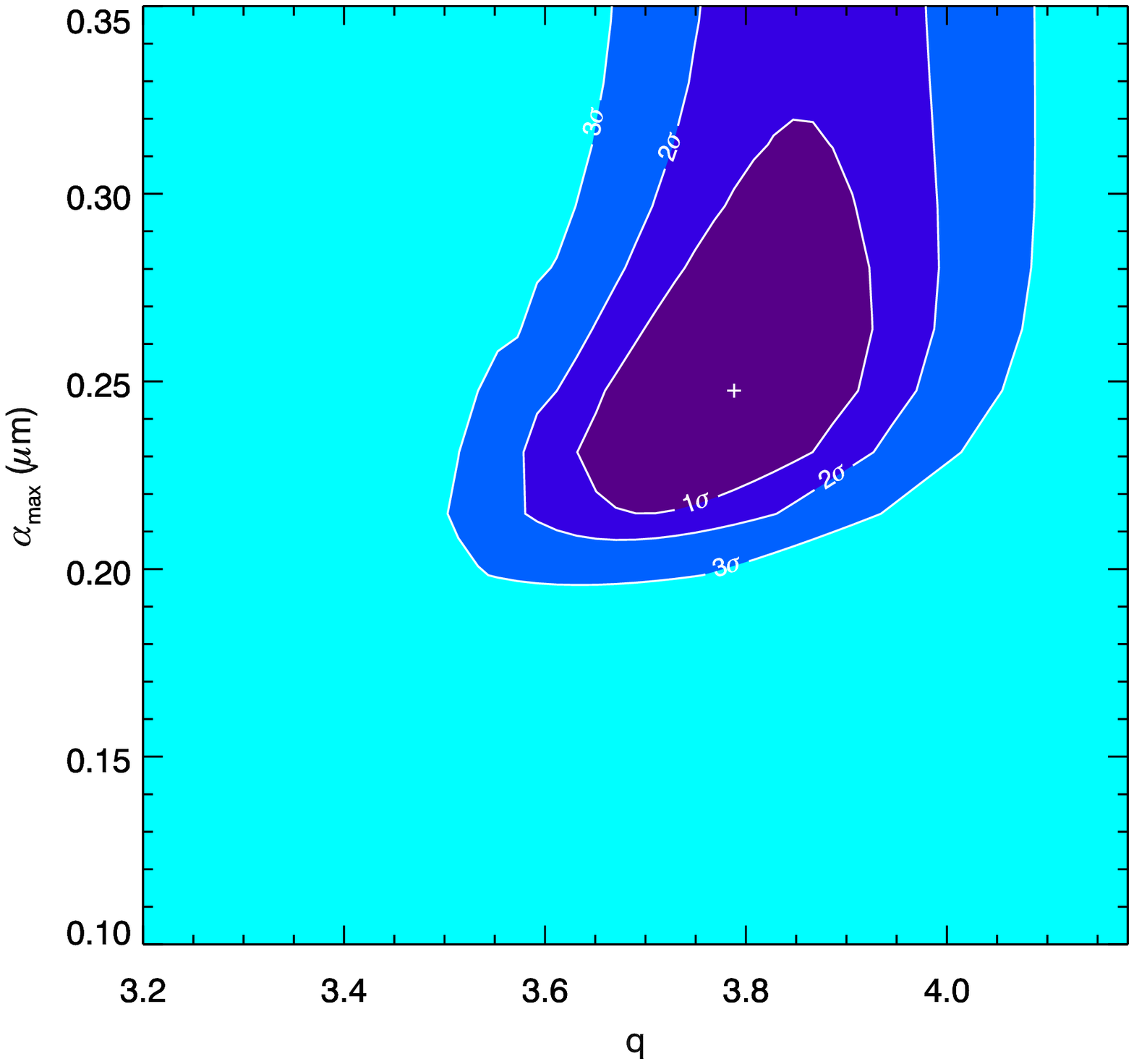}
  \caption{Contour plots for two model parameters (top row: $\alpha_{\min}-\alpha_{\max}$; bottom row $\alpha_{\max}-q$) 
  as derived by jointly fitting at four different energies the scattered {flux} of the rings 1-2 (left column), rings 3-4 (middle column)
  and ring 5 (right column). The lines mark the 1$\sigma$, 2$\sigma$ and 3$\sigma$ confidence levels.
  In all cases, the best-fit value is marked with a cross.}
  \label{fig:contours}
\end{figure*}

As a  second step, we created a grid of values for two model parameters, namely $\alpha_{\min}-\alpha_{\max}$ and $q-\alpha_{\max}$. Based
on the $\chi^2$ value of the fits, the 1$\sigma$, 2$\sigma$ and 3$\sigma$  significance levels  were then computed and the 
contour plots for $\alpha_{\min}-\alpha_{\max}$ and $\alpha_{\max}-q$ were produced. These are presented in Fig.~\ref{fig:contours} for rings 1-2
(left column), rings 3-4 (middle column), and ring 5 (right column). 
A few things worth mentioning follow:
\begin{itemize}
 \item Rings 1-2:
only a lower limit on the maximum grain size ($\alpha_{\max} \gtrsim {0.15} \,\mu$m) can be determined.
This is to be expected, since the \swift/XRT data show no evidence of a flat angular profile at $\theta \lesssim 2\arcmin$
(see also \S\ref{sec-theory}). Moreover, a minimum grain size larger than $\sim 4\times 10^{-3}\,\mu$m can be already excluded at a $3 \sigma$
level. The power-law index of the size distribution can be constrained though, with a best-fit value $q=4.42\pm0.02$.
\item Rings 3-4: in contrast to the most distant dust clouds, for the dust clouds 3 and 4 
we can constrain both $\alpha_{\max}$ and $q$ at a 3$\sigma$  level. 
The best-fit value of the former is compatible with the lower limit of $\alpha_{\max}$ derived from rings 1-2, whereas
the grain size distribution in clouds 3 and 4 is best described by a shallower
power-law with $q \sim 3.5$. No lower limit can be found for the minimum grain size, yet
the data of rings 3-4 exclude values larger than $\sim 0.017\,\mu$m at a 3$\sigma$ level.
\item Ring 5: the angular profile 
of the specific intensity shows hints of an exponential steepening at large angles ($\gtrsim 10\arcmin$), which
is imprinted on the contour plot of 
$\alpha_{\min}$ and $\alpha_{\max}$ (top panel on the right). At the 1$\sigma$ level we find a narrow range
of values for $\alpha_{\min}$ (0.01-0.025$\,\mu$m). At the 3$\sigma$ level, however, only an upper limit on the size
of small grains can be imposed, i.e. $\alpha_{\min} \lesssim 0.03\,\mu$m. Similar conclusions can be drawn for $\alpha_{\max}$.
The largest grains cannot be smaller than $0.2\,\mu$m, in agreement with the other dust clouds. Finally, the power-law index
of the size distribution is constrained at the $3\sigma$ level in the range $3.5-4.0$.
\end{itemize}

From the best-fit values of the normalization at a fixed energy, we can obtain the ratios of the hydrogen column densities for different
dust layers, such as:
\eqb
 \frac{N_{\rm H,12}}{N_{\rm H, 5}} \approx \frac{\mathcal{F}_{\rm X}^{(12)}(E)}{\mathcal{F}^{(5)}_{\rm X}(E)} 
 \frac{\left(1-\langle x \rangle_{12} \right){\langle x \rangle_{12}}}{\left(1-x_5\right){x_5}},
 \label{ratio}
\eqe
where we {assumed that the number of grains per hydrogen atom is the same in all clouds} and dropped the factor $C_{\rm d, i}$, since this 
is not sensitive to the grain composition \citep[][]{Mauche1986} (see also \S\ref{sec-theory}). In the above,
$N_{\rm H, 12}$ represents the average column density of dust clouds 1 and 2. As already mentioned in \S\ref{sec-expansion}
and later in \S\ref{sec-activity}, these are most probably one dust cloud of finite width, whose average column density we are probing.
Having derived the ratios {at} four different energies, we calculated the weighted means, namely
% $N_{\rm H,12}/N_{\rm H, 5}=0.028\pm0.004$, $N_{\rm H,12}/N_{\rm H, 34}=0.016\pm0.002$, 
% and $N_{\rm H,34}/N_{\rm H, 5}=\mathbf{1.810\pm0.298}$.     3.5259199      0.58075171
$N_{\rm H,12}/N_{\rm H, 5}={0.19\pm0.03}$, $N_{\rm H,12}/N_{\rm H, 34}={0.059\pm0.006}$, 
and $N_{\rm H,34}/N_{\rm H, 5}={3.5\pm0.6}$. Thus, the intermediate dust clouds have the highest column density. This is { 3.5} times higher than that of the  fifth dust cloud, which may also explain the approximately equal intensities of rings 3-4 and 5, given their different distances from the observer. {Interestingly,} the column density of the clouds located closest to \src is $\sim 6\% N_{\rm H, 34}$. 
These findings are  in rough qualitative agreement with 
the extinction histogram shown in Fig.~\ref{fig:dust_extinction}, where
the lowest extinction values are found at distances $\gtrsim 2$~kpc, i.e. where
the dust layers 1 and 2 are located. 
{Quantitatively, we find that the ratios $N_{\rm H,34}/N_{\rm H, 5}$ and $N_{\rm H,34}/N_{\rm H, 12}$ as derived from the extinction maps \citep{Sale2014} are by a factor of $\sim3$ smaller than our estimates, while the $N_{\rm H,12}/N_{\rm H, 5}$ ratio is the same (within uncertainty values). This implies that our analysis is overestimating the $N_{\rm H,34}$ column density by a factor of $\sim3$. A simple explanation for this difference is that the dust-to-hydrogen ratio is not the same for all clouds as originally assumed in our calculations (the $A_{\rm i}/A_{\rm j}$ term has dropped from the l.h.s. of eq. (\ref{ratio})). In any case, the ratio of the normalizations $\mathcal{F}_{\rm X}^{(i)}$ can be directly related to the ratio of dust column densities $N_{\rm H,i}A_{\rm i}$ without resorting to any additional assumptions.}

\subsection{Effects of the X-ray activity since MJD~57188.5}
 \label{sec-activity}
In contrast to other studies of dust scattered X-ray emission, where the source may exhibit one
major outburst \citep[see e.g.][]{Vaughan2004, Vianello2007}, the LMXB \src 
entered a new period of enhanced activity on MJD~57188.5 and, since then,
it  exhibited multiple X-ray flares in both soft and hard energy bands \citep[][]{MottaAtel2015, FerrignoAtel2015, RodriguezAtel2015, SegretoAtel2015}. 
In summary, 
\begin{itemize}
 \item   during MJD~57188.5-57191.0 no significant flares were detected by either \swift/BAT or \swift/XRT.
 \item multiple flares have been detected in the period of MJD~57191-57198 with the count rate reaching a maximum 
 on MJD~57194 (see e.g. Fig.~\ref{fig:expansion_rings} and \cite{Rodriguez2015}).
 \item finally, on $\sim$MJD~57199.7, a major outburst that lasted approximately $\sim 0.9$~d, was detected by all observing instruments. 
 Interestingly, its fluence in 25-60~keV is approximately equal to the fluence measured by INTEGRAL/ISGRI over
  the whole period between MJD~57191 and MJD~57198.
%  \item in the data of ISGRI and for the period MJD~57191-57198, we identified two outbursts (composed by ... flares)
%   of duration $\sim 1$~d . Their fluxes, which ... lower than the flux of the 57199 outburst, are ...
%   \item 
\end{itemize}
This preceding X-ray outburst activity may be imprinted on the angular profiles shown in Fig.~\ref{fig:radial}, which 
can be understood as follows. The simultaneous observation of X-ray rings due to the 
scattering of a single outburst from dust layers located at different distances from the observer
implies that the scattered photons should travel along paths of the same length, i.e. the scattering
points on the dust layers should belong {to} an ellipse. 
This is illustrated in Fig.~\ref{fig:dust_ecl}, where two ellipses corresponding
to different time-delays are shown. The colored regions denote the dust layers where the X-ray scattering
takes place, with the color gradient mimicking the gradient of the column density. Here, the inclination of
the dust layers with respect to the LOS has been arbitrarily chosen to be 90\textdegree.
Notice that the scattered emission of a prior outburst at $t'_0 < t_0$ (blue line)
by the dust layer at distance $x_1$ could  ``contaminate'' the scattered emission observed, at the same angle, but
caused by the scattering of an outburst at $t_0$ on the dust
layer at distance $x_2$ (green line). Thus, the X-ray scattering from bursts occurring before  
$t_0=$57199.75 could be imprinted on the angular profile of the rings shown in Fig.~\ref{fig:radial}. 
 \begin{figure}
%   \resizebox{\hsize}{!}{\includegraphics[angle=-90,clip=]{./plots_bu/dust_rings67.ps}}
   \resizebox{\hsize}{!}{\includegraphics[angle=0,clip=]{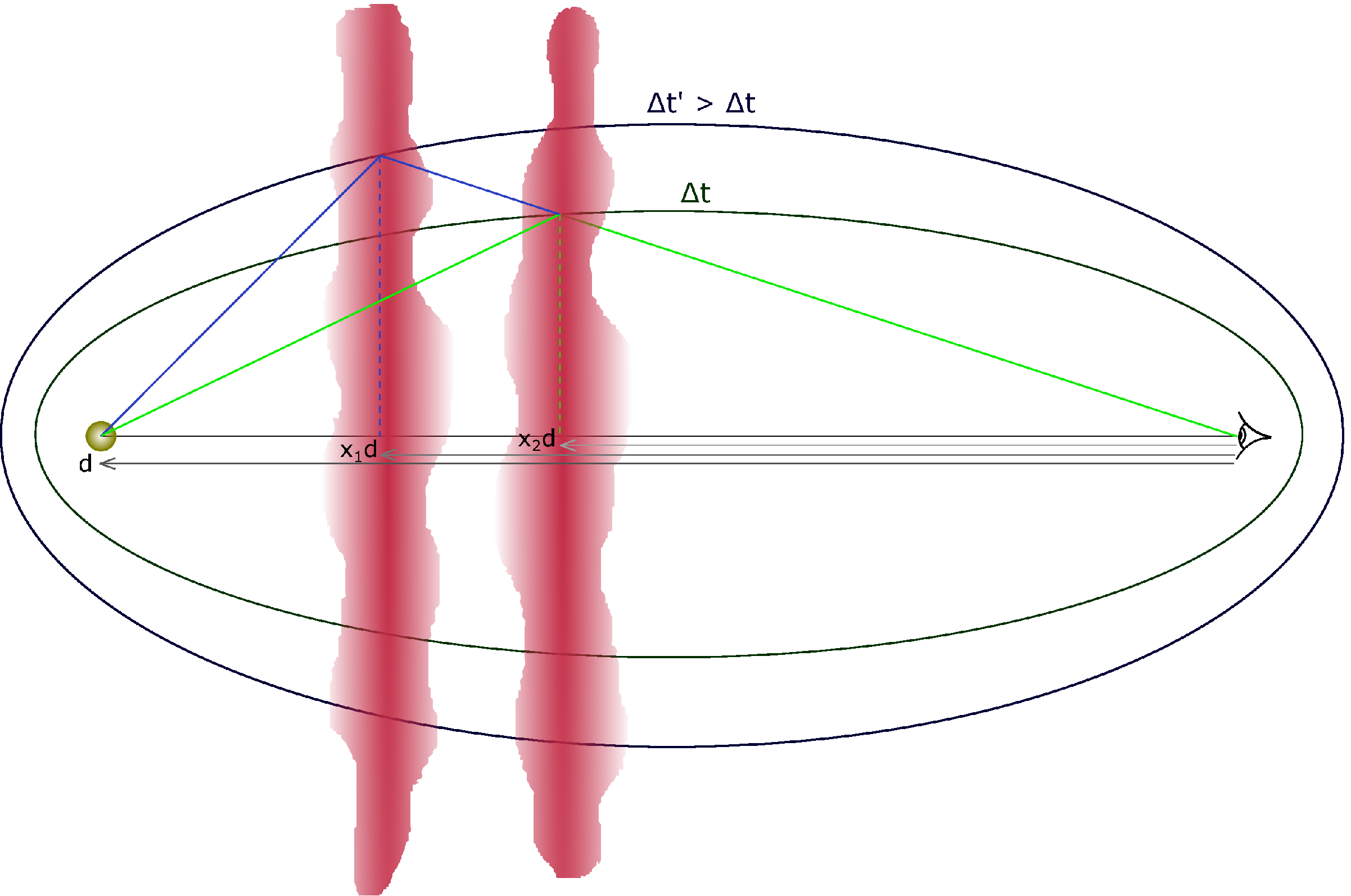}}
  \caption{An illustration of the locus of points determined by scattering events on dust screens (colored regions)
  located at different distances from the observer. The inner and outer ellipses correspond to different time delays marked on the plot.
  The colored solid lines show the travel paths of photons that are: (i) scattered by different dust layers, (ii) emitted 
  during source outbursts at $t'_0<t_0$ (blue line) and $t_0$ (green line) and  (iii) detected simultaneously with the same observed angle.} 
  \label{fig:dust_ecl}
\end{figure}

Having derived the distances of the dust layers by modeling
the ring expansion due to the latest, major outburst on $t_0=$~57199.75~d, we may estimate the expansion
of ring-like structures caused by the earlier activity of \src described above.
This is illustrated in  Fig.~\ref{fig:overlap_rings}, where 
the expansion of dust scattered rings due to outbursts on $t_0=$MJD~57194.5 and 57192.0 are shown
with dashed and dotted lines, respectively. In all cases,
the angular region enclosed by the solid and dotted lines corresponds to a maximum ``region of influence''
from bursts taking place after the onset of the \src activity. 
All the \swift/XRT observations of the innermost rings (blue and cyan lines) are affected by the previously
scattered emission on the most distant clouds. Similarly, the third and fourth rings (green and black lines)
are predominantly affected by previous bursts that were scattered on the dust clouds 3 and 4. The contamination becomes
important after MJD~57210. Finally, the outermost ring (red lines) is mostly contaminated by the scattered emission on the dust clouds 3 and 4.
Interestingly, the last two \swift/XRT detections of the outermost ring are the least affected by prior scattered emission.

A question that naturally arises is whether the earlier X-ray activity of the source could
give rise to distinct ring-like structures similar to the detected ones. To answer this question one has
to estimate the contamination of the previous bursts in the dust scattered rings of the final burst (MJD~57199-57200). We proceed with some simple estimations for the ratio of the fluence in soft X-rays (0.3-10 keV) during and before the burst ($R_{\rm fl}$). 
% The contamination is proportional to the total counts emitted by the source for a given period of time, thus strong but short bursts will have the same contribution as weaker but longer bursts.
Since the system is highly variable an accurate measurement of the fluence in soft X-rays would require a continuous monitoring of the system, 
while the \swift/XRT observations cover the burst for small time intervals (typically $\sim$1-2 ks).
However, we may estimate $R_{\rm fl}$ using the
fluence in the 25-60~keV band as measured by INTEGRAL IBIS/ISGRI during the same intervals, assuming a correlation between soft and hard X-rays.   
% By directly comparing the ratio between the total photon measured by IBIS/ISGRI before and after MJD~57199, we estimate that $\sim$48\% of the total fluence was emitted after MJD~57199, and during the major burst that produced the dust scattered rings.
Between MJD~57191 and MJD~57198 the system was observed with INTEGRAL IBIS/ISGRI for $\sim$5.1~d, while during
the major burst the system was continuously monitored ($\sim$1 day exposure time).
Assuming that the average flux of \src during the observing gaps was similar to the observed one,
we estimated that $\sim$39\% of the exposure corrected fluence of the system was emitted after MJD~57199. 
In addition, we found that the average flux 
during the major burst increased by a factor of $\sim 5$ compared to the average flux prior to it.
In this first period of enhanced X-ray activity (MJD~57191-57198), multiple strong bursts
spanning over one-day intervals, i.e. MJD~57193.6-57194.5 and MJD~57194.6-57195.7, can be identified (see also Fig.~\ref{fig:expansion_rings}).
Yet, the respective fluences
were estimated to be $\sim$31\% of the fluence during the major burst; the same applies for the fluxes. 
These estimates can be extended to the soft X-ray band under the former assumptions { (i.e., correlated soft and hard X-ray emission)}.
 \begin{figure}
\resizebox{\hsize}{!}{\includegraphics[angle=0,clip=]{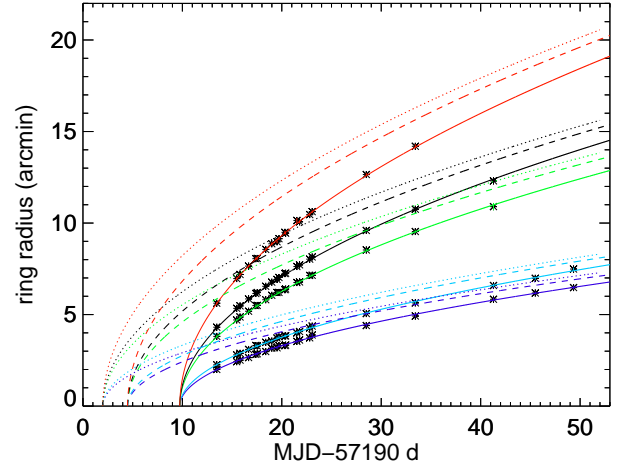}}
\caption{Expansion of dust scattered rings for three bursts occurring at different times: $t_0$=MJD~57199.75 (solid lines), MJD~57194.5 (dashed lines), and MJD~57192.0 (dotted lines). 
  The \swift/XRT observations are plotted with asterisks. 
  The earlier burst represents the start of the enhanced activity of \src, while
  the latest major burst is the one responsible for  the \swift/XRT detection of the rings.} 
  \label{fig:overlap_rings}
\end{figure}

%  A question that naturally arises is whether the earlier X-ray activity of the source could
%  give rise to distinct ring-like structures similar to the detected ones.
%  The answer is most probably not, 
 As the estimated fluence of the strongest
 earlier outbursts was by  a factor of $\sim 3$  less than that of the major outburst, {these would probably not  give rise to distinct ring-like structures similar to the detected ones.}  What is observed as ``bridge'' emission between the peak positions of successive 
 rings (see Fig.~\ref{fig:radial}) could be the result
 of previous scatterings though. The signatures of the strongest earlier bursts to the total angular profile should appear as small-amplitude peaks 
 over a diffuse emission component (halo). Interestingly, the appearance of spiky features between the successive major peaks at late times (with respect to $t_0$) could  also be explained as a residual of prior X-ray activity; see e.g. observations {\#10 and \#11} in Fig.~\ref{fig:radial-all}.
 {If this is the case, their evolution in time should also scale as  $\sqrt{\Delta t}$. We finally note that no significant deviations
 from the rings expansion law $\theta \propto \sqrt{\Delta t}$ are expected as long as the respective peaks in the angular profile 
 can be securely distinguished from the bridge emission. }  For the spectral analysis of the X-ray rings we selected part of the bridge emission as the background   in order to minimize the effect of the previously scattered emission.
%  treated the bridge emission between the rings as background (see also Fig.~\ref{fig:overlap_rings}).

% In order to minimize the effects of the previously X-ray scattered emission
% on the analysis of the \swift/XRT rings due to the last major burst, 
% As a background for the latter, we considered the bridge emission between them and the outermost ring. This was also
% used as  background for the analysis of the outermost ring. 
\section{Discussion}
\label{sec-discussion}
{ Our analysis was based on the RG approximation for the scattering cross section, which
turns out to be insufficient for describing the scattering process of $<1$~keV photons by $\,\mu$m-sized grains.
We therefore restricted the fitting of the scattered X-ray flux and the subsequent analysis to energies $\ge 1$~keV.
Although at 1~keV the RG approximation is marginally valid, the scattered flux at this energy was included in the analysis,
since the available data points for the spectral fitting were already sparse; the dust rings are practically not detected above 2.5~keV (see also 
Fig.~\ref{fig:image_rgb}). Our fitting results regarding the dust properties, and especially, the column densities could therefore
differ from those obtained using the Mie expression for the scattering cross section, which adequately describes
the process at sub-keV energies. To investigate the error introduced in our results, we repeated the fitting procedure of $F_{\rm sc}(E,\theta)$ after excluding the $1$~keV energy bin, where the RG approximation becomes questionable. We find that the quality of the global fit remains the same (e.g. for rings 1-2, $\chi^2_{\rm red}=1.95$ for 60 dof) and that the errors of individual parameters increase due to the smaller number of data points. The best-fit values of all parameters are the same as those listed in Tab.~\ref{tab:intensity} within the 1$\sigma$ errors. We also verified that this applies to the ratios of the column densities.
% $N_{\rm H, 12}/N_{\rm H, 5}=0.033\pm0.010$, $N_{\rm H, 12}/N_{\rm H,34}=0.019\pm0.003$, and $N_{\rm H, 34}/N_{\rm H, 5}=1.825\pm0.605$.

In addition, the difference between the RG approximation and the Mie solution at $1$~keV for $\theta >100$\arcsec \ is expected to be small \citep[see e.g.][]{Smith1998, Corrales2015}. Inspection of Fig.~6 in \citep{Smith1998} shows that for $\theta=100$\arcsec \ and  $E=1$~keV, $I_{\rm sc}^{(\rm RG)}/I_{\rm sc}^{(\rm Mie)} \sim 2$ (3) for grains with sizes $\alpha=0.1\,\mu$m (0.4$\,\mu$m). Taking also into account  that (i) the smallest angular size observed in the \swift/XRT data is $\theta =2$\arcmin \ and that (ii) the average size of dust grains we derived is $\sim 0.1\,\mu$m, we argue that our results would not differ significantly, if we were to perform the analysis using the Mie theory.}

{
We showed that the \swift/XRT observations of the rings can be sufficiently explained by a generalized MRN model for the dust. Our results indicate that the largest grains in all clouds located along the LOS have similar sizes ($\sim0.17-0.24\,\mu$m), while the average grain size, as defined by eq.~(\ref{average}) is  $\sim 0.07-0.1\,\mu$m. 
The best-fit values for the maximum grain size suggest a dust composition of silicate \citep{MRN1977, WD2001}.
However, a composition with $\alpha_{\max} \sim 1\,\mu$m  (e.g. graphite) cannot be excluded for the dust clouds 1-2 and 5, since $\alpha_{\max}$ is not limited from above.  Regarding the minimum size of the grains, our results are compatible with $\alpha_{\min} \lesssim 0.004\,\mu$m (dust clouds 1-2)
and $\lesssim 0.01\,\mu$m (dust clouds 3-4). Interestingly, the scattered flux profile of the fifth cloud showed hints of an exponential cutoff at large angles ($\theta \gtrsim 10$\arcmin). This, in turn, suggests that the minimum grain size in cloud 5 is typically larger than that inferred
for the other clouds (see Fig.~\ref{fig:contours}).  The steepest grain size distribution is obtained for
the most distant clouds, where $q$ lies between 4.35 and 4.5 (at a 3$\sigma$ CL). 
On the contrary, the power-law index of the size distribution in the intermediate clouds lies within the interval $3.4-3.7$, with the best-fit value being close to that of the MRN model \citep{MRN1977}. The obtained size distribution in the fifth cloud is described by $3.6\lesssim q\le 4.2$, namely it is softer than the distribution in the intermediate clouds, yet harder than that in the most distant ones. These results suggest the presence of  gradient in the dust properties along the LOS in the direction of \src that requires further investigation.
}

{
By fitting the profiles of the X-ray scattered flux we were also able to determine the hydrogen column density ratios for different clouds (see eq.~(\ref{norm})). We showed that the intermediate clouds (at distance $\sim 1.6$~kpc) have the highest column density, which is by a factor of $\sim 3.5$ higher than that of the closest to us cloud. In contrast, the dust clouds that are located closer to \src have the lowest column density, namely $\sim 6\%$ of $N_{\rm H,34}$. The above calculation was based on the assumption that the number of grains per hydrogen atom, $A_{\rm i}$, is the same in all clouds. 
We showed that the derived ratios are in qualitative agreement with those inferred from the extinction map in the direction of \src (see Fig.~\ref{fig:dust_extinction}). When viewed in a quantitative way, though, our results overestimate $N_{\rm H,34}$ by a factor or $\sim 3$. This incosistency could easily resolved, if the dust-to-hydrogen ratio in the intermediate clouds was larger than in the other clouds by the same factor. 

In the more general case, where the dust-to-hydrogen ratio differs among the clouds, $A_{\rm i}$ can be deduced by eq.~(\ref{norm}). This requires  knowledge of the burst X-ray fluence $\Phi_{\rm X}(E)$ over the period of the major outburst (see Fig.~\ref{fig:expansion_rings}) as well as of the hydrogen column density of each dust cloud $N_{\rm H, i}$. Both quantities can be roughly estimated as follows. The fluence in the \swift/XRT energy band was 
derived by assuming a correlation between soft (0.3-10~keV) and hard (25-60~keV) X-rays and by using the fluence measured by IBIS/ISGRI (see Sect.~\ref{sec-activity}). Since the 3D extinction map in the direction of \src is known, and the distances of the clouds were determined, the $N_{\rm H,i}$ can be derived using the relation between extinction $A_{\rm V}$ and hydrogen column density \citep[e.g.][]{PredehlSchmitt1995}, i.e. $N_{\rm H}=(2.21\pm0.09)\times 10^{21}\, A_{\rm V}$~cm$^{-2}$ \citep{guver2009}. The derived values for the number of grains per hydrogen atom in the clouds are $A_{12}=10^{-13.9\pm0.2}$, $A_{34}=10^{-13.2\pm0.4}$, and $A_5=10^{-13.9\pm0.5}$.}
{The structure of a dust grain is complex as it is typically composed by many atoms or molecules \citep[e.g.][]{WD2001, draine03}. Yet, a zero-order estimation of total number of atoms contained in a dust grain can be made by comparing the characteristic sizes of atoms and grains.
% Moreover taking into account that a dust grain consists of many atoms we can make a rough estimation about the number of atoms contained within the dust grains. 
The average grain size as determined by our analysis is 0.1 $\mu$m, while the atomic sizes of C and Fe atoms are respectively
$\sim7\times10^{-5}$ $\mu$m and $\sim14\times10^{-5}$ $\mu$m. Roughly speaking, a dust grain can be can be composed by $\sim3.7\times10^{9}$ C atoms or 
 $\sim2.9\times10^{8}$ Fe atoms. Taking into account the $A_{\rm i}$ values we derived for the clouds, we estimate  for the three dust clouds $\sim1.2\times10^{16}$ C atoms cm$^{-2}$ or $\sim9.4\times10^{16}$ Fe atoms cm$^{-2}$. We remark that these values refer to the amount of metals contained in dust grains within the three clouds and not the gas in the ISM.}

{It is interesting that the tracers of atomic and molecular hydrogen become maximum at distances $\sim 0.85$~kpc, where no dust cloud is inferred from the \swift/XRT data (see Fig.~\ref{fig:dust_extinction}). A fiducial dust cloud
at the distance of $\sim 0.85$~kpc would produce an X-ray ring with angular size $\sim 9$\arcmin \ on MJD~57205.5. As this would fall well within the \swift/XRT FOV (see Fig.~\ref{fig:image_rgb}), its absence should be related to the dust properties of the respective cloud. We estimated therefore that 
the maximum grain size should be $\gg 0.2\,\mu$m. A dust cloud composed by large grains ($\sim 1\,\mu$m) would suppress the scattered
intensity even at 1~keV, while it would significantly attenuate X-rays at $\le 0.5$~keV \citep[e.g.][]{Smith1998, Corrales2015}.  Alternatively, the absence of the X-ray ring might indicates a much smaller  number of grains per hydrogen atom at $\sim 0.85$~kpc compared to the other clouds. 
}

% {\bf In Sect.~\ref{sec-dustprop} we calculated the column densities of individual clouds using the best-fit values of the normalization $\mathcal{F}_{\rm X}^{(\rm i)}$, assuming a constant value of $A$ among the clouds, and setting $N_{\rm H,34}$ equal to a fraction of the total $N_{\rm H}$ along the LOS. The latter was derived from the modelling of the source spectrum.  
% If the X-ray fluence of the major outburst (MJD~57199.2-57199.9) was known in the \swift/XRT energy band, then $A$ could be deduced from eq.~(\ref{norm}). 
% % As only two  \swift/XRT snapshots, of 1~ks each, are available in this time interval, a direct measurement of the X-ray fluence cannot be made (see e.g. Fig.~\ref{fig:expansion_rings}). This could be, however, estimated either by assuming that the flux in the \swift/XRT band remains constant over the major outburst or by using the fluence in harder X-rays (25-60~keV) (for more details, see Sect.~\ref{sec-activity}). 
% }

\begin{figure*}
 \centering
\resizebox{\hsize}{!}{\includegraphics[angle=0]{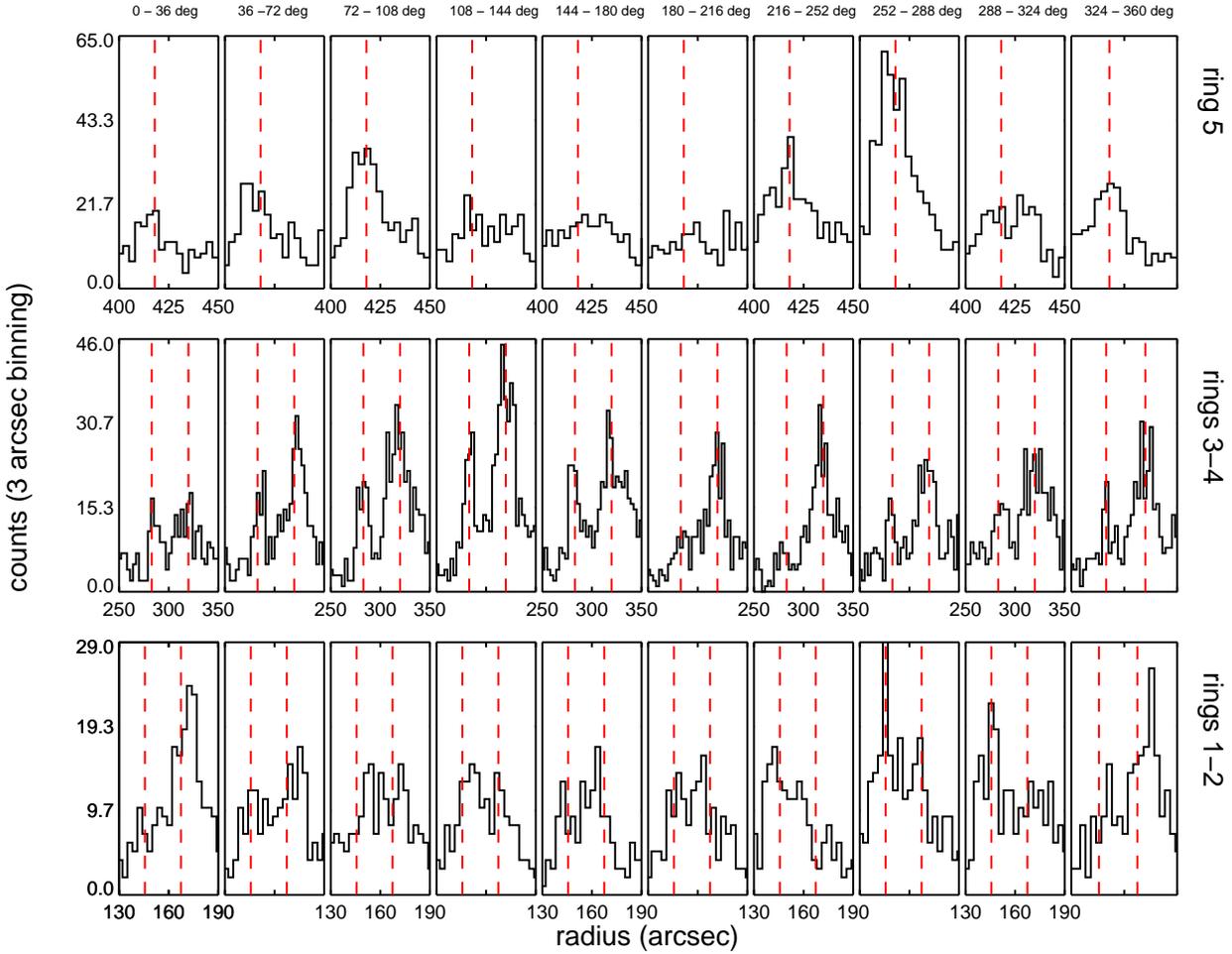}}
\caption{
% Azimuthal variation of the radial profiles of the detected dust scattered rings grouped for a 36\textdegree grid (obsid: 00033861006). 
% From bottom to top are plotted the radial profiles of the inner middle and outer dust scattered rings. The azimuthal variation is presented from left to right. Each panel represent data collected from a 36\textdegree region starting from three o'clock of Fig.~\ref{fig:image_rgb}.
% The dashed red vertical lines represent the positions of the rings as derived from the angular profile of the whole image.
Variation of the radial profile of counts along the azimuthal direction for the \swift/XRT observation { \#2}. 
The profiles have been created using a 36\textdegree \ bin in the azimuthal direction. The azimuth is measured
counterclockwise, starting from the three o'clock ray (east direction) in Fig.~\ref{fig:image_rgb}. The profiles of the innermost, intermediate and outermost
rings are plotted from bottom to top. The dashed red vertical lines denote the peak positions as derived by the analysis
of the azimuthal integrated radial profiles. 
}
\label{fig:tings_azimouth}
\end{figure*}
The analysis presented so far, as in most studies, does not take into account possible
variations of the radial profiles in the azimuthal direction. We remind that the profiles shown in Figs.~\ref{fig:radial} and \ref{fig:radial-all}
were created by summing the counts of each annulus, i.e. integrating over the azimuthal angle. 
Although this method is sufficient for probing the average properties of the dust clouds,  such as position and average column density, it cannot
provide information about the spatial inhomogeneity of the dust clouds and/or their inclination with respect to the LOS.
The photon statistics of the \swift/XRT observations  are sufficient for quantifying the azimuthal variations of the ring intensity (see also Fig.~\ref{fig:image_rgb}). Figure~\ref{fig:tings_azimouth} shows the azimuthal variation of the scattered photons for \swift/XRT 
observation {\#2}. There are three important features that need to be mentioned: 
(i) the radial profile of ring 5 is highly variable along the azimuthal direction, 
with the peak number of counts changing by a factor of $\sim$5
above the background level (see sub-panels for 180\textdegree-216\textdegree and 252\textdegree-288\textdegree); (ii) azimuthal variations are also present in the rings 3 and 4, yet the variability amplitude is lower
when compared to the ring 5. The peak number of counts of the fourth ring
is for all azimuth angles larger compared to that of the third ring; (iii) the azimuthal behavior of the rings 1 and 2 is the most intriguing, as the second peak (corresponding to the ring 2) becomes prominent for azimuth angles 0\textdegree-36\textdegree \ and 324\textdegree-360\textdegree \, while the first peak dominates
for azimuth angles of 216\textdegree-324\textdegree. \ Obviously, this information is lost from the azimuthal integrated radial profiles shown in Fig.~\ref{fig:radial}.
Similar features appear in all \swift/XRT observations with their significance varying according to the photon statistics of the observation.

% The azimuthal variations suggest a non-homogeneous distribution, at sub-pc scales, of matter within
% the clouds\footnote{These variations are present in scales similar to the bin size of the azimuthal profiles, i.e. a few arcmin, yielding
% a projected distance of less than a pc, for the location of the most distant clouds ($\sim2$~kpc).}.  
% 

The azimuthal variations are present in scales similar to the bin size of the azimuthal profiles, i.e. a few arcmin, yielding
a projected distance of less than one pc for the location of the most distant clouds ($\sim2$~kpc).
These results imply a non-homogeneous matter distribution at sub-pc scales  within
the clouds, which could also lead to deviations from the theoretical prediction for the intensity profile $I_{\rm sc}(E,
\theta)$. 
With a time-dependent study of the azimuthal profiles of the X-ray dust scattered rings, one could perform an X-ray tomography of the dust clouds at sub-pc scales. 
A more detailed study lies, however, outside the scope of the present paper.

% In the same concept, since the shape of the best fitted model towards larger angular distances is characterized by the value of the minimum dust grain size, we should expect that the determined minimum dust grain size should be more sensitive (compared to the other parameters of the fit) to the intensity measured for the rings with the largest observed angle.

\section{Summary}
\label{sec-summary}
% \label{conclusion}   
The scattered emission from bright X-ray point sources manifests itself often as an
X-ray halo, while multiple  (typically, one or two) X-ray rings  can be detected
if the X-ray source is variable, exhibiting short duration outbursts, and the dust 
along the LOS is not uniformly distributed but concentrated in discrete dust clouds.
The recent \swift/XRT observations of multiple dust
scattered X-ray rings from the LMXB \src fall into the second category 
and offer a unique opportunity for studying the properties of dust
along the LOS in the direction of \src, for a number of reasons. Most importantly,
these are the first detections of dust scattered emission from the particular LMXB, whose
distance is well constrained. Furthermore, the \swift/XRT observations
span over a period of $\sim 23$~d ($\sim36$~d) for the outermost (innermost) ring(s)
and allow for a time-dependent study of the X-ray scattered emission.

Using a simple but well proven theoretical framework, i.e.
single X-ray scattering of a short X-ray impulse on discrete dust clouds
along the LOS, we modeled the expansion of the dust scattered
rings. We showed that these are produced 
by scattering of the last major outburst of \src, on MJD~57199.7, that lasted
$\sim 0.9$~d. The derived distances of the dust layers
were found to be 2.12, 2.05, 1.63, 1.50
and 1.18~kpc  from the observer. 
These positions coincide roughly with locations of enhanced extinction $A_{\lambda}$ as determined by infrared photometry
and of higher column densities of molecular hydrogen as derived 
by CO radio measurements. 
The spatial proximity of the most distant dust layers, as well as its azimuthal variation, implies the presence of one 
cloud with a finite width of $\sim 100$~pc located at $\sim$2~kpc; 
similar conclusions apply to the intermediate dust layers. 

Assuming that the grain size distribution in all dust layers follows a power-law (i.e., a generalized Mathis-Rumpl-Nordsieck model),
we fitted the angular profiles of the specific intensity at four different energies (1.0, 1.5, 2.0 and 2.5~keV) and 
derived the best-fit model parameters for the dust. In particular, we showed that the size distribution
in the most distant cloud is significantly steeper ($q\sim 4.4$) than the distributions of the other clouds, which
are described by $q\sim 3.4-3.7$. For the two intermediate dust layers we constrained, at a $3\sigma$ level, the maximum grain size in the range $0.16-0.20\,\mu$m, 
and set a lower limit at $\sim 0.18-0.2\,\mu$m for the other clouds. 
% Although the minimum grain size of the dust layers cannot be uniquely determined, we can still set upper limits to the minimum sizes of the dust grain distribution. 
% Thus, we can accept at a $3\sigma$ level the presence of small grains in the dust clouds 1-2, 3-4 and 5 
% with sizes $\alpha_{\min} \lesssim 0.001\,\mu$m, $0.015\,\mu$m, and $0.03\,\mu$m, respectively.
We can also exclude at a $3\sigma$ level that the smallest grains in the dust clouds 1-2, 3-4 and 5 
have sizes $\alpha_{\min} \gtrsim 0.001\,\mu$m, $0.015\,\mu$m, and $0.03\,\mu$m, respectively. 
It is noteworthy that the angular intensity profile of the outermost X-ray
ring showed hints of an exponential cutoff at large angles suggesting the presence
of grains with $0.01\le \alpha_{\min} \le 0.03\,\mu$m (at $1\sigma$ level). 
% Based on the best-fit values we calculated
% the ratios of $N_{\rm H,12}/N_{\rm H, 5}=0.028\pm0.004$, $N_{\rm H,12}/N_{\rm H, 34}=0.016\pm0.001$, 
% and $N_{\rm H,34}/N_{\rm H, 5}=1.765\pm0.072$, 
% showing that the intermediate dust clouds have the highest column density. 
% However, the absolute value of the column densities could not be derived,
% as the fluence of the source outburst in soft X-rays was not directly measured.
% These results are in rough agreement with 
% the extinction histogram shown in Fig.~\ref{fig:dust_extinction}, where
% the lowest extinction values are found at distances $\gtrsim 2$~kpc, i.e. where
% the dust layers 1 and 2 are located.
% As the fluence of the source in soft X-rays was not directly measured during the last outburst,
% an absolute determination of the hydrogen column densities was not possible. Yet, 
The best-fit values of the ratios of the dust column densities $N_{\rm d}$ between different clouds, i.e. $N_{\rm d,12}/N_{\rm d, 5}={0.19\pm0.03}$, $N_{\rm d,12}/N_{\rm d, 34}={0.059\pm0.006}$, and $N_{\rm d,34}/N_{\rm d, 5}={3.5\pm0.6}$, indicate a higher $N_{\rm d}$  value
at a distance $\sim 1.6$~kpc. {Our results suggest the presence of a gradient in the properties of the dust in the direction of \src that requires further investigation.}

\section*{Acknowledgments}
We acknowledge the use of public data from the \swift data archive, which are a result of ToO observations from multiple PIs 
(D. Altamirano,
A.~P. Beardmore,
M. Cadolle Bel,
P. Gandhi,
J.~A. Kennea,
E. Kuulkers,
M. Middleton,
S. Motta,
R.~M. Plotkin,
J. Rodriguez,
G.~R. Sivakoff, 
G. Vasilopoulos),
and we thank the \swift team for accepting and carefully scheduling the ToO observations.  
{ We thank the anonymous referee for useful comments that helped us to clarify subtle points and improve the manuscript.}
We also thank Dr.~P. Predehl for useful discussions, Dr.~S.~Dimitrakoudis for providing Fig.~10 and 
Dr. R. Xue for helpful discussions on the use of CO radio maps.
G.\,V. acknowledges support from the BMWi/DLR grants FKZ 50 OR 1208.
M. P. acknowledges support from NASA through the Einstein Postdoctoral 
Fellowship grant number PF3~140113 awarded by the Chandra X-ray 
Center, which is operated by the Smithsonian Astrophysical Observatory
for NASA under contract NAS8-03060.
% Last but not least, we wish to thank P.~Koliopano for our stimulating discussion.
% \end{acknowledgements}

\bibliographystyle{mn2e}
\bibliography{V404cyg}

\begin{thebibliography}{66}
\expandafter\ifx\csname natexlab\endcsname\relax\def\natexlab#1{#1}\fi

\bibitem[{{Arnaud}(1996)}]{Arnaud1996}
{Arnaud} K.~A., 1996, in Astronomical Society of the Pacific Conference Series,
  Vol. 101, Astronomical Data Analysis Software and Systems V, {Jacoby} G.~H.,
  {Barnes} J., eds., p.~17

\bibitem[{{Beardmore} {et~al}\mbox{.}(2015){Beardmore}, {Altamirano},
  {Kuulkers}, {Motta}, {Osborne}, {Page}, {Sivakoff}, \&
  {Vaughan}}]{Beardmore2015}
{Beardmore} A.~P., {Altamirano} D., {Kuulkers} E., {Motta} S.~E., {Osborne}
  J.~P., {Page} K.~L., {Sivakoff} G.~R., {Vaughan} S.~A., 2015, The
  Astronomer's Telegram, 7736, 1

\bibitem[{{Bhattacharjee} {et~al}\mbox{.}(2014){Bhattacharjee}, {Chaudhury}, \&
  {Kundu}}]{Bhattacharjee2014}
{Bhattacharjee} P., {Chaudhury} S., {Kundu} S., 2014, \apj, 785, 63

\bibitem[{{Brand} \& {Blitz}(1993)}]{BrandBlitz1993}
{Brand} J., {Blitz} L., 1993, \aap, 275, 67

\bibitem[{{Casares} \& {Charles}(1994)}]{CasaresCharles1994}
{Casares} J., {Charles} P.~A., 1994, \mnras, 271, L5

\bibitem[{{Casares} {et~al}\mbox{.}(1992){Casares}, {Charles}, \&
  {Naylor}}]{Casares1992}
{Casares} J., {Charles} P.~A., {Naylor} T., 1992, \nat, 355, 614

\bibitem[{{Casares} {et~al}\mbox{.}(1993){Casares}, {Charles}, {Naylor}, \&
  {Pavlenko}}]{Casares1993}
{Casares} J., {Charles} P.~A., {Naylor} T., {Pavlenko} E.~P., 1993, \mnras,
  265, 834

\bibitem[{{Cash}(1979)}]{Cash1979}
{Cash} W., 1979, \apj, 228, 939

\bibitem[{{Chatzopoulos} {et~al}\mbox{.}(2015){Chatzopoulos}, {Fritz},
  {Gerhard}, {Gillessen}, {Wegg}, {Genzel}, \& {Pfuhl}}]{Chatzopoulos2015}
{Chatzopoulos} S., {Fritz} T.~K., {Gerhard} O., {Gillessen} S., {Wegg} C.,
  {Genzel} R., {Pfuhl} O., 2015, \mnras, 447, 948

\bibitem[{{Clemens}(1985)}]{Clemens1985}
{Clemens} D.~P., 1985, \apj, 295, 422

\bibitem[{{Corrales} \& {Paerels}(2015)}]{Corrales2015}
{Corrales} L.~R., {Paerels} F., 2015, \mnras, 453, 1121

\bibitem[{{Costantini} {et~al}\mbox{.}(2005){Costantini}, {Freyberg}, \&
  {Predehl}}]{Costantini2005}
{Costantini} E., {Freyberg} M.~J., {Predehl} P., 2005, \aap, 444, 187

\bibitem[{{Dame}(2011)}]{Dame2011}
{Dame} T.~M., 2011, ArXiv e-prints:1101.1499

\bibitem[{{Dame} {et~al}\mbox{.}(2001){Dame}, {Hartmann}, \&
  {Thaddeus}}]{Dame2001}
{Dame} T.~M., {Hartmann} D., {Thaddeus} P., 2001, \apj, 547, 792

\bibitem[{Dewey {et~al}\mbox{.}(1969)Dewey, Mapes, \& Reynolds}]{Dewey1969}
Dewey R., Mapes R., Reynolds T., 1969, Handbook of X-ray and Microprobe Data,
  Analytical chemistry. Pergamon Press

\bibitem[{{Draine}(2003)}]{draine03}
{Draine} B.~T., 2003, \apj, 598, 1026

\bibitem[{{Draine}(2011)}]{Draine2011}
{Draine} B.~T., 2011, {Physics of the Interstellar and Intergalactic Medium}

\bibitem[{{Draine} \& {Bond}(2004)}]{Draine2004}
{Draine} B.~T., {Bond} N.~A., 2004, \apj, 617, 987

\bibitem[{{Draine} \& {Tan}(2003)}]{Draine2003}
{Draine} B.~T., {Tan} J.~C., 2003, \apj, 594, 347

\bibitem[{{Ferrigno} {et~al}\mbox{.}(2015){Ferrigno}, {Fotopoulou}, {Domingo},
  {Alfonso-Garz{\'o}n}, {Rodriguez}, {Motta}, {Kuulkers}, {Sanchez-Fernandez},
  \& {Cadolle Bel}}]{FerrignoAtel2015}
{Ferrigno} C. {et~al.}, 2015, The Astronomer's Telegram, 7662, 1

\bibitem[{{Garner} {et~al}\mbox{.}(2015){Garner}, {Eikenberry}, {Stelter},
  {Raines}, {Charcos}, {Edwards}, {Lasso-Cabrera}, {Marin-Franch}, {Cenarro},
  {Bennett}, {Mullin}, {Chinn}, {Ackley}, {Varosi}, {Warner}, {Frommeyer},
  {Herlevich}, {Miller}, {Murphey}, {Donoso}, {Vega}, {Packham}, {Dallilar},
  {Scarpa}, {Gerarts}, {Martin}, {Calero}, {Sanchez}, {Siegel}, {Losada},
  {Perez}, {Sendra}, \& {Acosta}}]{Garner2015}
{Garner} A. {et~al.}, 2015, The Astronomer's Telegram, 7663, 1

\bibitem[{{Gazeas} {et~al}\mbox{.}(2015){Gazeas}, {Vasilopoulos},
  {Petropoulou}, \& {Sapountzis}}]{Gazeas2015}
{Gazeas} K., {Vasilopoulos} G., {Petropoulou} M., {Sapountzis} K., 2015, The
  Astronomer's Telegram, 7650, 1

\bibitem[{{Greiner} {et~al}\mbox{.}(1996){Greiner}, {Dennerl}, \&
  {Predehl}}]{GreinerDennerl1996}
{Greiner} J., {Dennerl} K., {Predehl} P., 1996, \aap, 314, L21

\bibitem[{{Greiner} {et~al}\mbox{.}(1995){Greiner}, {Predehl}, \&
  {Pohl}}]{GreinerPredehl1995}
{Greiner} J., {Predehl} P., {Pohl} M., 1995, \aap, 297, L67

\bibitem[{{G{\"u}ver} \& {{\"O}zel}(2009)}]{guver2009}
{G{\"u}ver} T., {{\"O}zel} F., 2009, \mnras, 400, 2050

\bibitem[{{Henke}(1981)}]{Henke1981}
{Henke} B.~L., 1981, in American Institute of Physics Conference Series,
  Vol.~75, Low Energy X-ray Diagnostics, {Attwood} D.~T., {Henke} B.~L., eds.,
  pp. 146--155

\bibitem[{{King} {et~al}\mbox{.}(2015){King}, {Miller}, {Raymond}, {Reynolds},
  \& {Morningstar}}]{King2015}
{King} A.~L., {Miller} J.~M., {Raymond} J., {Reynolds} M.~T., {Morningstar} W.,
  2015, ArXiv e-prints

\bibitem[{{Klose}(1994)}]{Klose1994}
{Klose} S., 1994, \apjl, 423, L23

\bibitem[{{Kouveliotou} {et~al}\mbox{.}(2001){Kouveliotou}, {Tennant}, {Woods},
  {Weisskopf}, {Hurley}, {Fender}, {Garrington}, {Patel}, \& {G{\"o}{\v
  g}{\"u}{\c s}}}]{Kouveliotou2001}
{Kouveliotou} C. {et~al.}, 2001, \apjl, 558, L47

\bibitem[{{Krimm} {et~al}\mbox{.}(2013){Krimm}, {Holland}, {Corbet},
  {Pearlman}, {Romano}, {Kennea}, {Bloom}, {Barthelmy}, {Baumgartner},
  {Cummings}, {Gehrels}, {Lien}, {Markwardt}, {Palmer}, {Sakamoto},
  {Stamatikos}, \& {Ukwatta}}]{Krimm2013}
{Krimm} H.~A. {et~al.}, 2013, \apjs, 209, 14

\bibitem[{{Kuulkers}(2015)}]{Kuulkers2015}
{Kuulkers} E., 2015, The Astronomer's Telegram, 7758, 1

\bibitem[{{Makino} {et~al}\mbox{.}(1989){Makino}, {Wagner}, {Starrfield},
  {Buie}, {Bond}, {Johnson}, {Harrison}, \& {Gehrz}}]{Makino1989}
{Makino} F., {Wagner} R.~M., {Starrfield} S., {Buie} M.~W., {Bond} H.~E.,
  {Johnson} J., {Harrison} T., {Gehrz} R.~D., 1989, \iaucirc, 4786, 1

\bibitem[{{Mathis} \& {Lee}(1991)}]{MathisLee1991}
{Mathis} J.~S., {Lee} C.-W., 1991, \apj, 376, 490

\bibitem[{{Mathis} {et~al}\mbox{.}(1977){Mathis}, {Rumpl}, \&
  {Nordsieck}}]{MRN1977}
{Mathis} J.~S., {Rumpl} W., {Nordsieck} K.~H., 1977, \apj, 217, 425

\bibitem[{{Mauche} \& {Gorenstein}(1986)}]{Mauche1986}
{Mauche} C.~W., {Gorenstein} P., 1986, \apj, 302, 371

\bibitem[{{Miller-Jones} {et~al}\mbox{.}(2009){Miller-Jones}, {Jonker},
  {Dhawan}, {Brisken}, {Rupen}, {Nelemans}, \& {Gallo}}]{Miller-Jones2009}
{Miller-Jones} J.~C.~A., {Jonker} P.~G., {Dhawan} V., {Brisken} W., {Rupen}
  M.~P., {Nelemans} G., {Gallo} E., 2009, \apjl, 706, L230

\bibitem[{{Mooley} {et~al}\mbox{.}(2015){Mooley}, {Fender}, {Anderson},
  {Staley}, {Kuulkers}, \& {Rumsey}}]{Mooley2015}
{Mooley} K., {Fender} R., {Anderson} G., {Staley} T., {Kuulkers} E., {Rumsey}
  C., 2015, The Astronomer's Telegram, 7658, 1

\bibitem[{{Moretti} {et~al}\mbox{.}(2005){Moretti}, {Campana}, {Mineo},
  {Romano}, {Abbey}, {Angelini}, {Beardmore}, {Burkert}, {Burrows}, {Capalbi},
  {Chincarini}, {Citterio}, {Cusumano}, {Freyberg}, {Giommi}, {Goad}, {Godet},
  {Hartner}, {Hill}, {Kennea}, {La Parola}, {Mangano}, {Morris}, {Nousek},
  {Osborne}, {Page}, {Pagani}, {Perri}, {Tagliaferri}, {Tamburelli}, \&
  {Wells}}]{Moretti2005}
{Moretti} A. {et~al.}, 2005, in Society of Photo-Optical Instrumentation
  Engineers (SPIE) Conference Series, Vol. 5898, UV, X-Ray, and Gamma-Ray Space
  Instrumentation for Astronomy XIV, {Siegmund} O.~H.~W., ed., pp. 360--368

\bibitem[{{Motta} {et~al}\mbox{.}(2015){Motta}, {Beardmore}, {Oates}, {Sanna},
  {Kuulkers}, {Kajava}, \& {Sanchez-Fernanedz}}]{MottaAtel2015}
{Motta} S., {Beardmore} A., {Oates} S., {Sanna} N.~P.~M.~K.~A., {Kuulkers} E.,
  {Kajava} J., {Sanchez-Fernanedz} C., 2015, The Astronomer's Telegram, 7665, 1

\bibitem[{{Negoro} {et~al}\mbox{.}(2015){Negoro}, {Matsumitsu}, {Mihara},
  {Serino}, {Matsuoka}, {Nakahira}, {Ueno}, {Tomida}, {Kimura}, {Ishikawa},
  {Nakagawa}, {Sugizaki}, {Shidatsu}, {Sugimoto}, {Takagi}, {Kawai}, {Yoshii},
  {Tachibana}, {Yoshida}, {Sakamoto}, {Kawakubo}, {Ohtsuki}, {Tsunemi},
  {Imatani}, {Nakajima}, {Tanaka}, {Ueda}, {Kawamuro}, {Hori}, {Tsuboi},
  {Kanetou}, {Yamauchi}, {Itoh}, {Yamaoka}, \& {Morii}}]{MAXI2015}
{Negoro} H. {et~al.}, 2015, The Astronomer's Telegram, 7646, 1

\bibitem[{{Overbeck}(1965)}]{Overbeck1965}
{Overbeck} J.~W., 1965, \apj, 141, 864

\bibitem[{{Predehl} {et~al}\mbox{.}(2000){Predehl}, {Burwitz}, {Paerels}, \&
  {Tr{\"u}mper}}]{Predehl2000}
{Predehl} P., {Burwitz} V., {Paerels} F., {Tr{\"u}mper} J., 2000, \aap, 357,
  L25

\bibitem[{{Predehl} \& {Klose}(1996)}]{PredehlKlose1996}
{Predehl} P., {Klose} S., 1996, \aap, 306, 283

\bibitem[{{Predehl} \& {Schmitt}(1995)}]{PredehlSchmitt1995}
{Predehl} P., {Schmitt} J.~H.~M.~M., 1995, \aap, 293, 889

\bibitem[{{Richter}(1989)}]{Richter1989}
{Richter} G.~A., 1989, Information Bulletin on Variable Stars, 3362, 1

\bibitem[{{Rodriguez} {et~al}\mbox{.}(2015{\natexlab{a}}){Rodriguez}, {Cadolle
  Bel}, {Alfonso-Garz{\'o}n}, {Siegert}, {Zhang}, {Grinberg}, {Savchenko},
  {Tomsick}, {Chenevez}, {Clavel}, {Corbel}, {Diehl}, {Domingo},
  {Gouiff{\`e}s}, {Greiner}, {Krause}, {Laurent}, {Loh}, {Markoff},
  {Mas-Hesse}, {Miller-Jones}, {Russell}, \& {Wilms}}]{Rodriguez2015}
{Rodriguez} J. {et~al.}, 2015{\natexlab{a}}, \aap, 581, L9

\bibitem[{{Rodriguez} {et~al}\mbox{.}(2015{\natexlab{b}}){Rodriguez},
  {Ferrigno}, {Cadolle Bel}, {Clavel}, {Loh}, {Corbel}, {Laurent}, {Markoff},
  {Miller-Jones}, {Russell}, {Tomsick}, \& {Wilms}}]{RodriguezAtel2015}
{Rodriguez} J. {et~al.}, 2015{\natexlab{b}}, The Astronomer's Telegram, 7702, 1

\bibitem[{{Rolf}(1983)}]{Rolf1983}
{Rolf} D.~P., 1983, \nat, 302, 46

\bibitem[{{Sale} {et~al}\mbox{.}(2014){Sale}, {Drew}, {Barentsen}, {Farnhill},
  {Raddi}, {Barlow}, {Eisl{\"o}ffel}, {Vink}, {Rodr{\'{\i}}guez-Gil}, \&
  {Wright}}]{Sale2014}
{Sale} S.~E. {et~al.}, 2014, \mnras, 443, 2907

\bibitem[{{Segreto} {et~al}\mbox{.}(2015){Segreto}, {Del Santo}, {D'A{\'{\i}}},
  {La Parola}, {Cusumano}, {Mineo}, \& {Malzac}}]{SegretoAtel2015}
{Segreto} A., {Del Santo} M., {D'A{\'{\i}}} A., {La Parola} V., {Cusumano} G.,
  {Mineo} T., {Malzac} J., 2015, The Astronomer's Telegram, 7755, 1

\bibitem[{{Shahbaz} {et~al}\mbox{.}(1996){Shahbaz}, {Bandyopadhyay}, {Charles},
  \& {Naylor}}]{Shahbaz1996}
{Shahbaz} T., {Bandyopadhyay} R., {Charles} P.~A., {Naylor} T., 1996, \mnras,
  282, 977

\bibitem[{{Smith} {et~al}\mbox{.}(2006){Smith}, {Dame}, {Costantini}, \&
  {Predehl}}]{SmithDame2006}
{Smith} R.~K., {Dame} T.~M., {Costantini} E., {Predehl} P., 2006, \apj, 648,
  452

\bibitem[{{Smith} \& {Dwek}(1998)}]{Smith1998}
{Smith} R.~K., {Dwek} E., 1998, \apj, 503, 831

\bibitem[{{Sofue} {et~al}\mbox{.}(2009){Sofue}, {Honma}, \&
  {Omodaka}}]{Sofue2009}
{Sofue} Y., {Honma} M., {Omodaka} T., 2009, \pasj, 61, 227

\bibitem[{{Svirski} {et~al}\mbox{.}(2011){Svirski}, {Nakar}, \&
  {Ofek}}]{Svirski2011}
{Svirski} G., {Nakar} E., {Ofek} E.~O., 2011, \mnras, 415, 2485

\bibitem[{{Tetarenko} {et~al}\mbox{.}(2015){Tetarenko}, {Sivakoff}, {Gurwell},
  {Petitpas}, {Wouterloot}, \& {Miller-Jones}}]{Tetarenko2015}
{Tetarenko} A., {Sivakoff} G.~R., {Gurwell} M.~A., {Petitpas} G., {Wouterloot}
  J.~G.~A., {Miller-Jones} J.~C., 2015, The Astronomer's Telegram, 7661, 1

\bibitem[{{Tiengo} {et~al}\mbox{.}(2010){Tiengo}, {Vianello}, {Esposito},
  {Mereghetti}, {Giuliani}, {Costantini}, {Israel}, {Stella}, {Turolla},
  {Zane}, {Rea}, {G{\"o}tz}, {Bernardini}, {Moretti}, {Romano}, {Ehle}, \&
  {Gehrels}}]{Tiengo2010}
{Tiengo} A. {et~al.}, 2010, \apj, 710, 227

\bibitem[{{Tr{\"u}mper} \& {Sch{\"o}nfelder}(1973)}]{Truemper1973}
{Tr{\"u}mper} J., {Sch{\"o}nfelder} V., 1973, \aap, 25, 445

\bibitem[{{van de Hulst}(1957)}]{vandeHulst1957}
{van de Hulst} H.~C., 1957, {Light Scattering by Small Particles}

\bibitem[{{Vaughan} {et~al}\mbox{.}(2004){Vaughan}, {Willingale}, {O'Brien},
  {Osborne}, {Reeves}, {Levan}, {Watson}, {Tedds}, {Watson}, {Santos-Lle{\'o}},
  {Rodr{\'{\i}}guez-Pascual}, \& {Schartel}}]{Vaughan2004}
{Vaughan} S. {et~al.}, 2004, \apjl, 603, L5

\bibitem[{{Vianello} {et~al}\mbox{.}(2007){Vianello}, {Tiengo}, \&
  {Mereghetti}}]{Vianello2007}
{Vianello} G., {Tiengo} A., {Mereghetti} S., 2007, \aap, 473, 423

\bibitem[{{Wagner} {et~al}\mbox{.}(1992){Wagner}, {Kreidl}, {Howell}, \&
  {Starrfield}}]{Wagner1992}
{Wagner} R.~M., {Kreidl} T.~J., {Howell} S.~B., {Starrfield} S.~G., 1992,
  \apjl, 401, L97

\bibitem[{{Weingartner} \& {Draine}(2001)}]{WD2001}
{Weingartner} J.~C., {Draine} B.~T., 2001, \apj, 548, 296

\bibitem[{{Wilms} {et~al}\mbox{.}(2000){Wilms}, {Allen}, \&
  {McCray}}]{wilms2000}
{Wilms} J., {Allen} A., {McCray} R., 2000, \apj, 542, 914

\bibitem[{{Xiang} {et~al}\mbox{.}(2011){Xiang}, {Lee}, {Nowak}, \&
  {Wilms}}]{Xiang2011}
{Xiang} J., {Lee} J.~C., {Nowak} M.~A., {Wilms} J., 2011, \apj, 738, 78

\bibitem[{{{\.Z}ycki} {et~al}\mbox{.}(1999){{\.Z}ycki}, {Done}, \&
  {Smith}}]{Zycki1999}
{{\.Z}ycki} P.~T., {Done} C., {Smith} D.~A., 1999, \mnras, 309, 561

\end{thebibliography}
%%%%%%%%%%%%%%%%%%%%%%%%%%%%%%%%%%%%%%%%%%%%%%%%%%%%%%%50
 %%%%%%%%%%%%%%%%%LONG TABLE%%%%%%%%%%%%%%%%%%%%%%%%%%
%%%%%%%%%%%%%%%%%%%%%%%%%%%%%%%%%%%%%%%%%%%%%%%%%%%%%%%5
\onecolumn
%\begin{appendix}
\appendix
\section[]{X-ray observation logs}
\label{appenA}

% \FloatBarrier

\begin{table*}
\caption{X-ray observations log of \src.}
\begin{center}
\scalebox{0.9}{
\begin{threeparttable}
\begin{tabular}{cccccccc}
\hline\hline  
\#& OBS-ID  & Instrument (mode)&  T\_start  & T\_stop  & T\_mean$^a$    & Duration$^b$ &exposure$^c$   \\
        &     		   &   MJD [d] & MJD [d] & MJD [d]   &    [h]    &[h]   &[sec]  \\
\hline\noalign{\smallskip}      
1& 00031403071 &  XRT [PC]  &  57203.453    &   57203.465  &   57203.459 &   0.28       &    979   \\   \noalign{\smallskip}
2& 00033861006 &  XRT [PC]  &  57205.452    &   57205.472  &   57205.463 &   0.48       &    1711   \\   \noalign{\smallskip}
3& 00031403072 &  XRT [PC]  &  57205.794    &   57205.805  &   57205.800 &   0.27       &    949   \\   \noalign{\smallskip}
4& 00031403074 &  XRT [PC]  &  57206.660    &   57206.670  &   57206.666 &   0.23       &    819   \\   \noalign{\smallskip}
5& 00033861007 &  XRT [PC]  &  57207.377    &   57207.394  &   57207.386 &   0.42       &    1478   \\   \noalign{\smallskip}
6& 00031403076 &  XRT [PC]  &  57207.521    &   57207.535  &   57207.528 &   0.33       &    1164   \\   \noalign{\smallskip}
7& 00033861008 &  XRT [PC]  &  57208.373    &   57208.390  &   57208.382 &   0.42       &    1471   \\   \noalign{\smallskip}
8& 00031403079 &  XRT [PC]  &  57208.916    &   57208.933  &   57208.925 &   0.41       &    1451   \\   \noalign{\smallskip}
9& 00031403078 &  XRT [PC]  &  57208.986    &   57208.996  &   57208.992 &   0.25       &    874   \\   \noalign{\smallskip}
10& 00031403080 &  XRT [PC]  &  57209.516    &   57209.526  &   57209.522 &   0.26       &    901   \\   \noalign{\smallskip}
11& 00081751001 &  XRT [PC]  &  57209.660    &   57209.783  &   57209.714 &   2.97       &    1763   \\   \noalign{\smallskip}
12& 00031403081 &  XRT [PC]  &  57210.379    &   57210.390  &   57210.385 &   0.26       &    909   \\   \noalign{\smallskip}
13& 00031403083 &  XRT [PC]  &  57210.309    &   57210.319  &   57210.314 &   0.24       &    852   \\   \noalign{\smallskip}
14& 00031403084 &  XRT [PC]  &  57211.509    &   57211.519  &   57211.514 &   0.24       &    854   \\   \noalign{\smallskip}
15& 00031403085 &  XRT [PC]  &  57211.717    &   57211.859  &   57211.786 &   3.41       &    1746   \\   \noalign{\smallskip}
16& 00031403086 &  XRT [PC]  &  57212.589    &   57212.992  &   57212.808 &   9.67       &    1988   \\   \noalign{\smallskip}
17& 00031403087 &  XRT [PC]  &  57213.050    &   57213.058  &   57213.055 &   0.20       &    717   \\   \noalign{\smallskip}
18& 00033861010 &  XRT [PC]  &  57218.177    &   57218.848  &   57218.498 &   16.1       &    5304   \\   \noalign{\smallskip}
19& 00031403107 &  XRT [PC]  &  57223.092    &   57223.962  &   57223.431 &   20.9       &    9168   \\   \noalign{\smallskip}
20& 00031403108 &  XRT [PC]  &  57226.474    &   57226.823  &   57226.639 &   8.37       &    6785   \\   \noalign{\smallskip}
21& 00031403111 &  XRT [PC]  &  57231.002    &   57231.478  &   57231.292 &   11.4       &    11170   \\   \noalign{\smallskip}
22& 00031403113 &  XRT [PC]  &  57235.248    &   57235.794  &   57235.495 &   13.1       &    10186   \\   \noalign{\smallskip}
23& 00031403115 &  XRT [PC]  &  57239.171    &   57239.521  &   57239.343 &   8.37       &    9742   \\   \noalign{\smallskip}
\hline 
\end{tabular}
\tnote{a} The average arrival time of all the photons during the multiple \swift/XRT snapshots. \\
{\tnote{b} Defined as the time interval between the start of the first \swift/XRT snapshot and the end time last snapshot. \\
\tnote{c} Total \swift/XRT exposure time of all snapshots taken within the duration of the obs-id. \\}
% \tnote{b} xxx \\
% \tnote{b} The mean of the Gaussian curve. \\
% \tnote{c} $\sigma$ of the  Gaussian curve. \\
% \tnote{d} The total area of the  Gaussian curve.
\end{threeparttable}
}
\end{center}
\label{tab:xray-obs}
\end{table*}

% \FloatBarrier

\section[]{Angular profile of the dust scattered X-ray emission}
\label{appenB}
\begin{figure*}
\includegraphics[width=0.32\textwidth,angle=0,clip=]{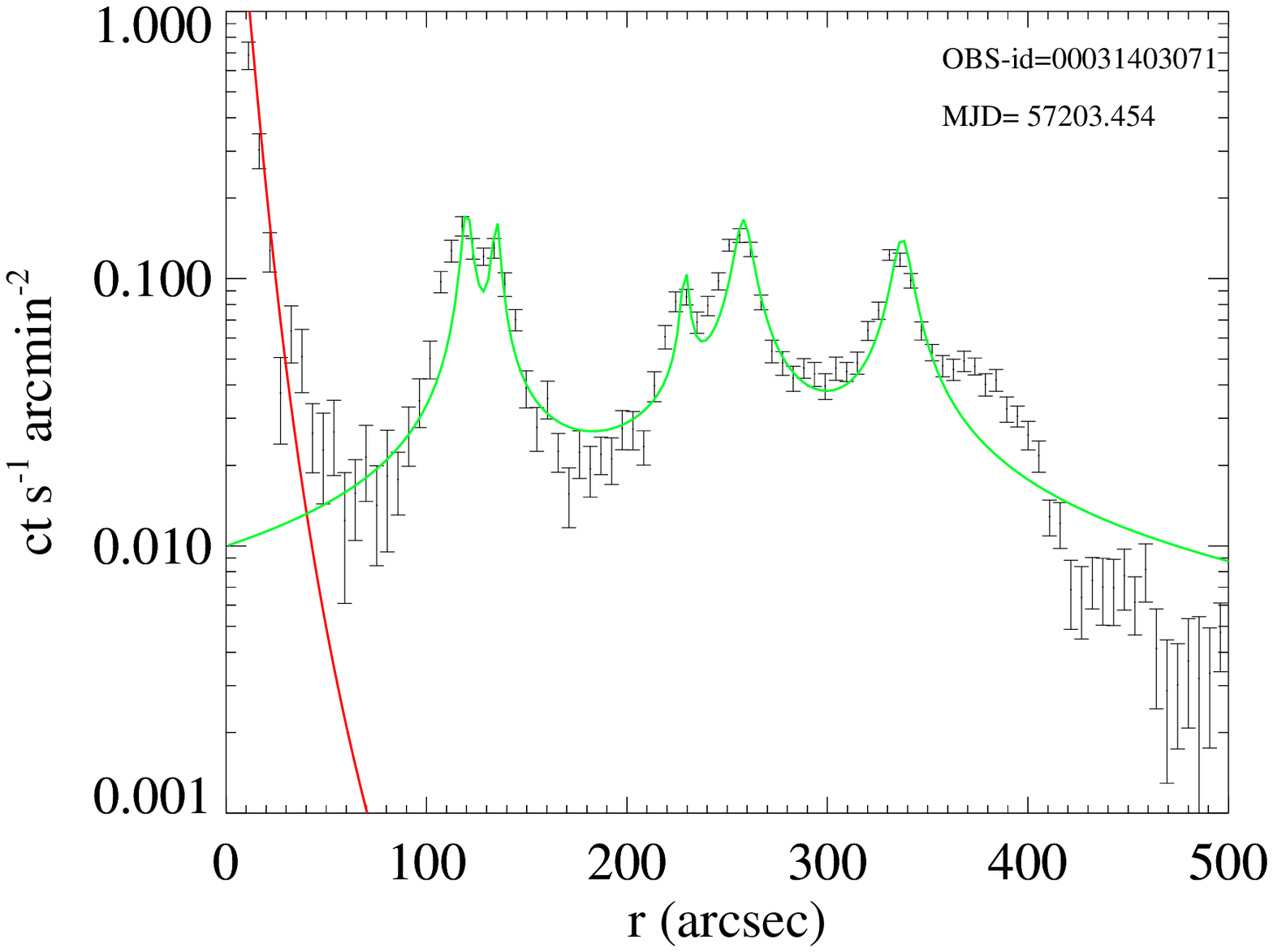}
\includegraphics[width=0.32\textwidth,angle=0,clip=]{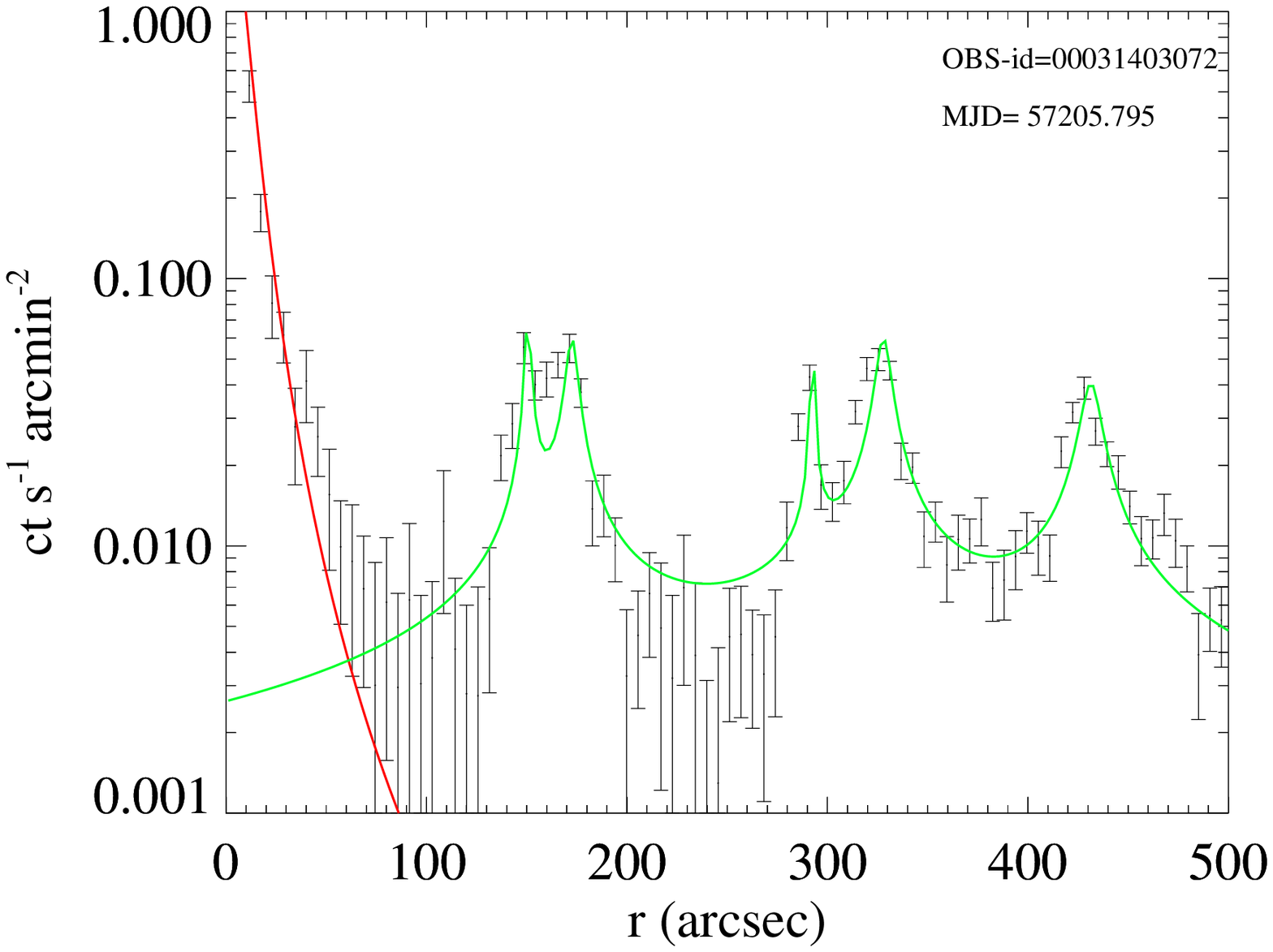}
\includegraphics[width=0.32\textwidth,angle=0,clip=]{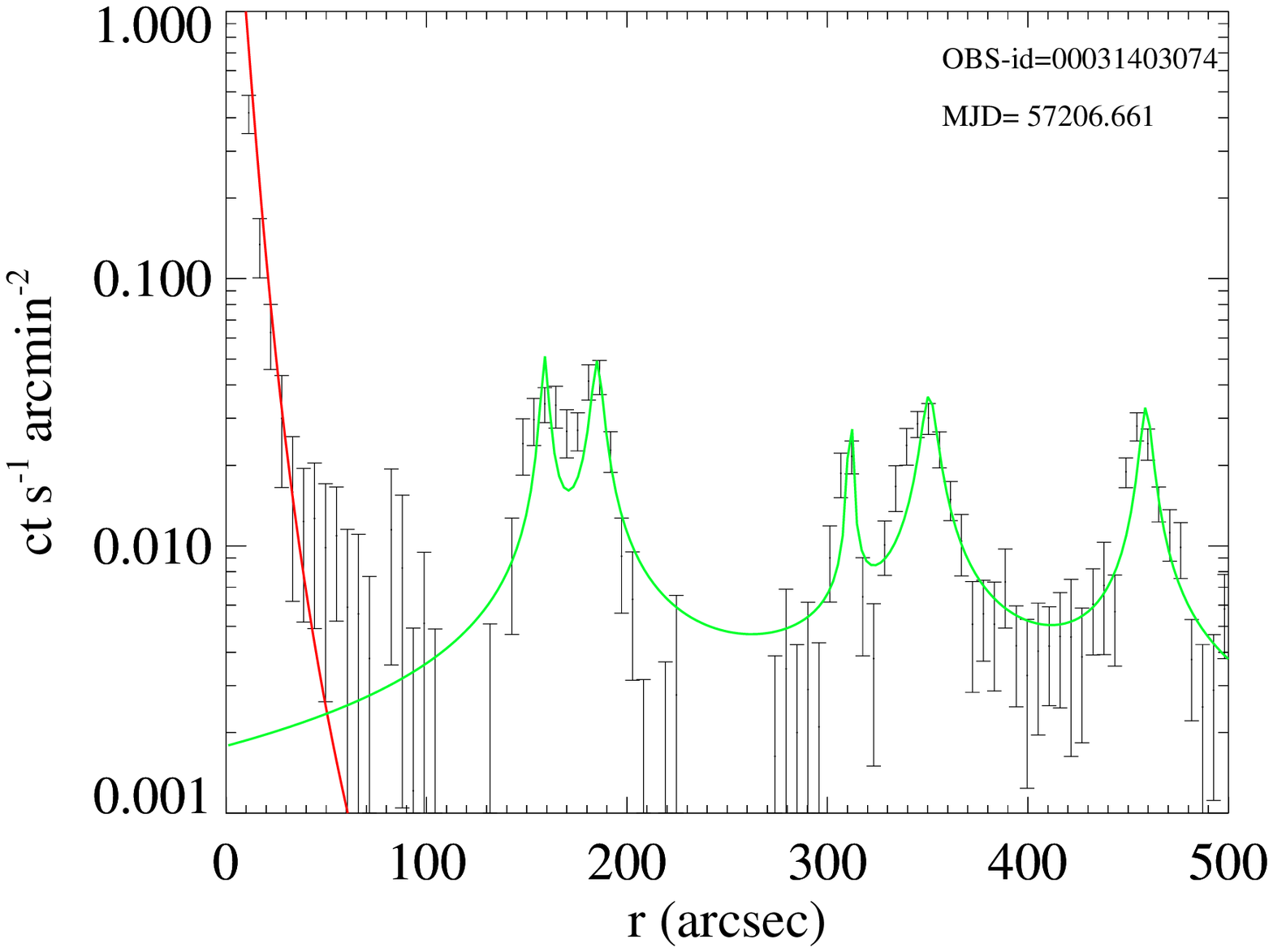}\\
\includegraphics[width=0.32\textwidth,angle=0,clip=]{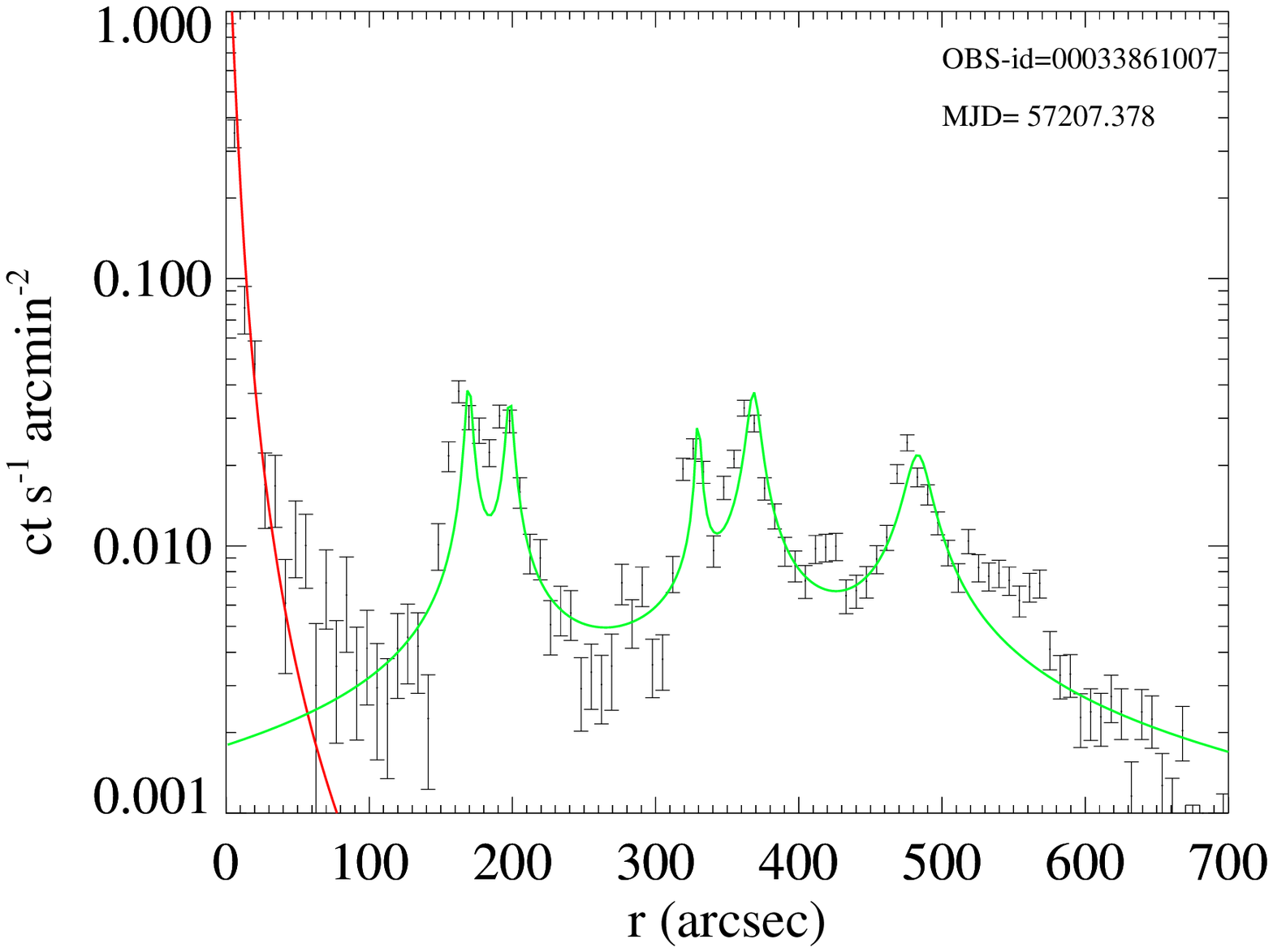}
\includegraphics[width=0.32\textwidth,angle=0,clip=]{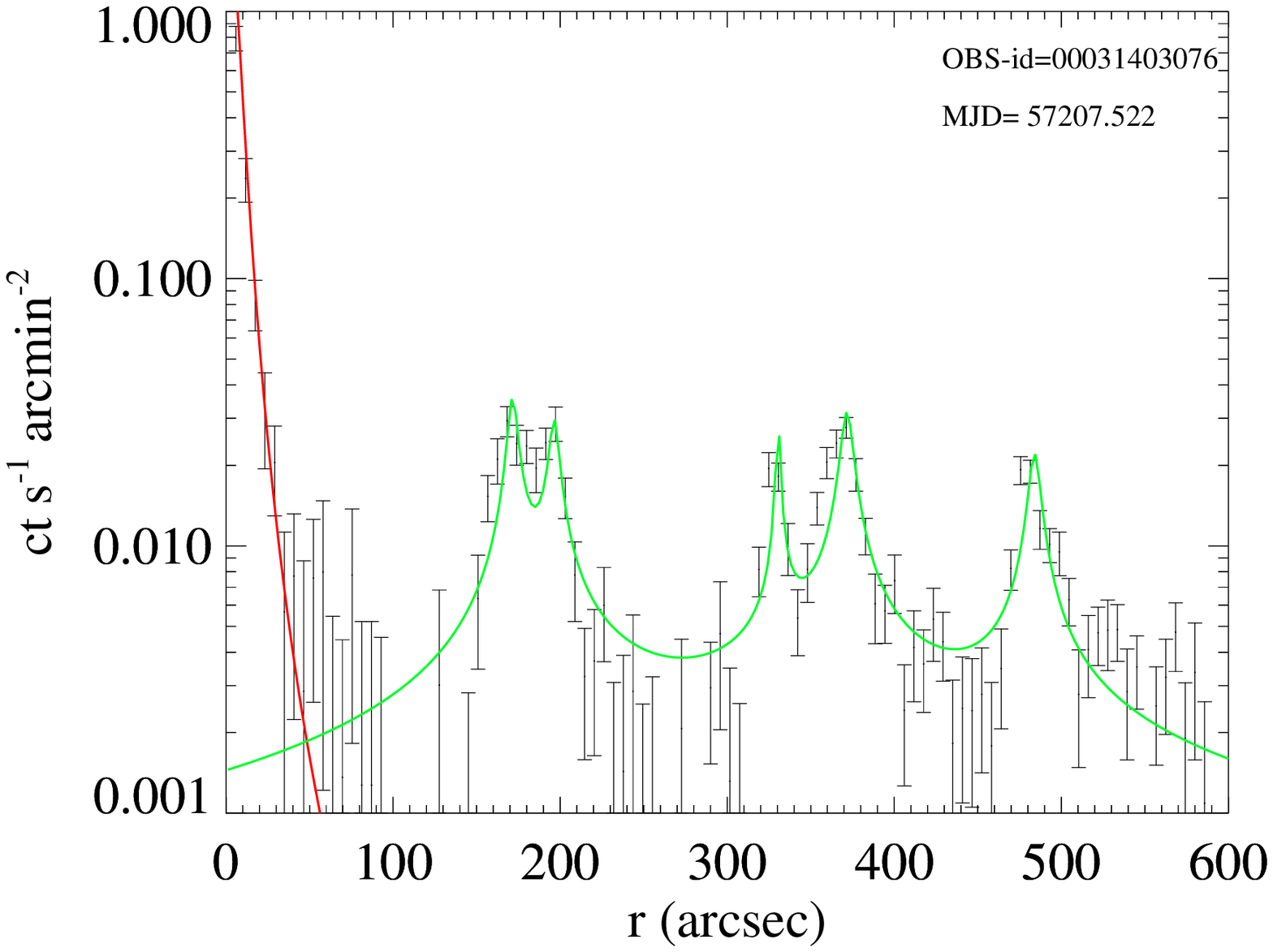}
\includegraphics[width=0.32\textwidth,angle=0,clip=]{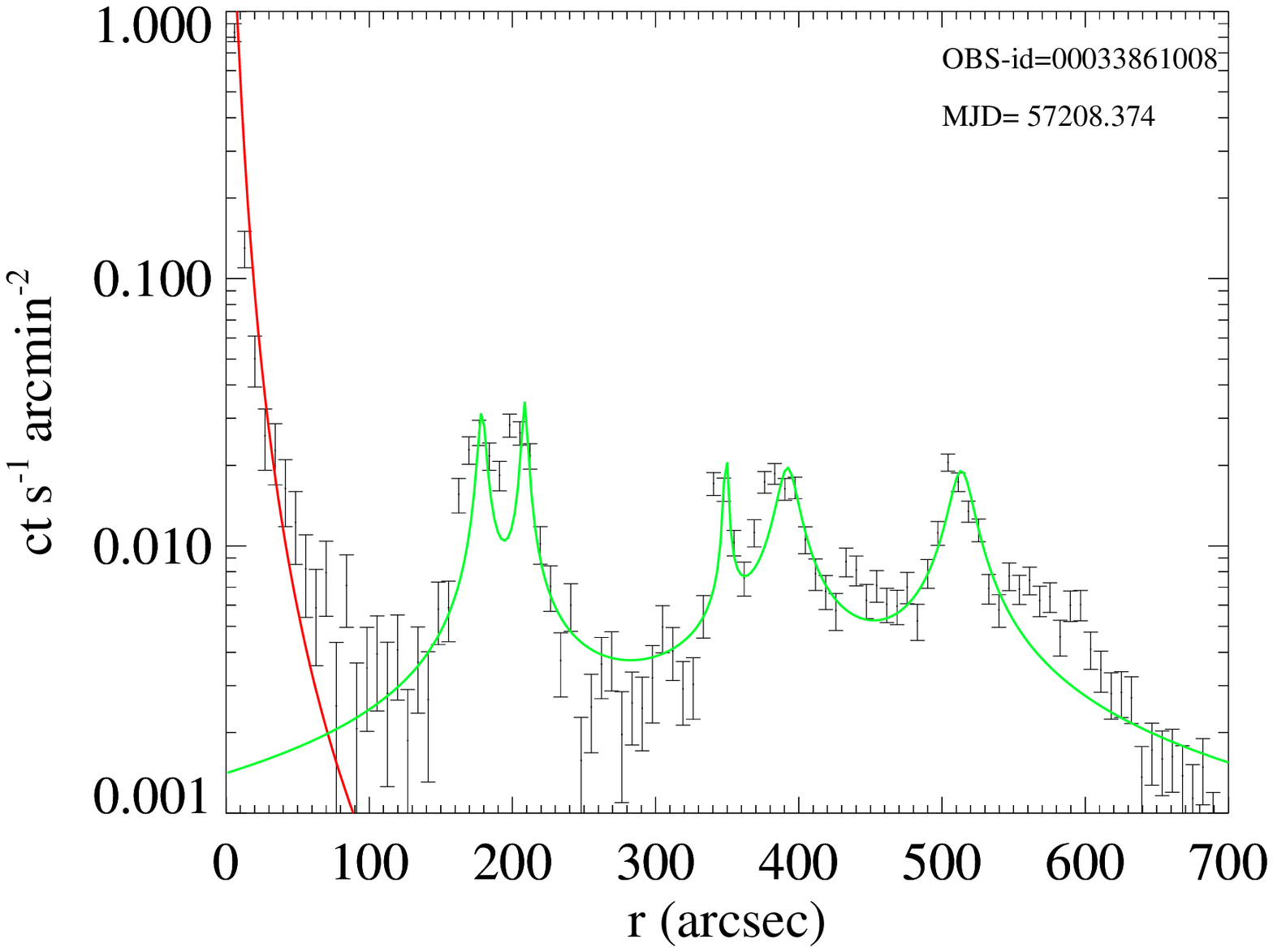}\\
\includegraphics[width=0.32\textwidth,angle=0,clip=]{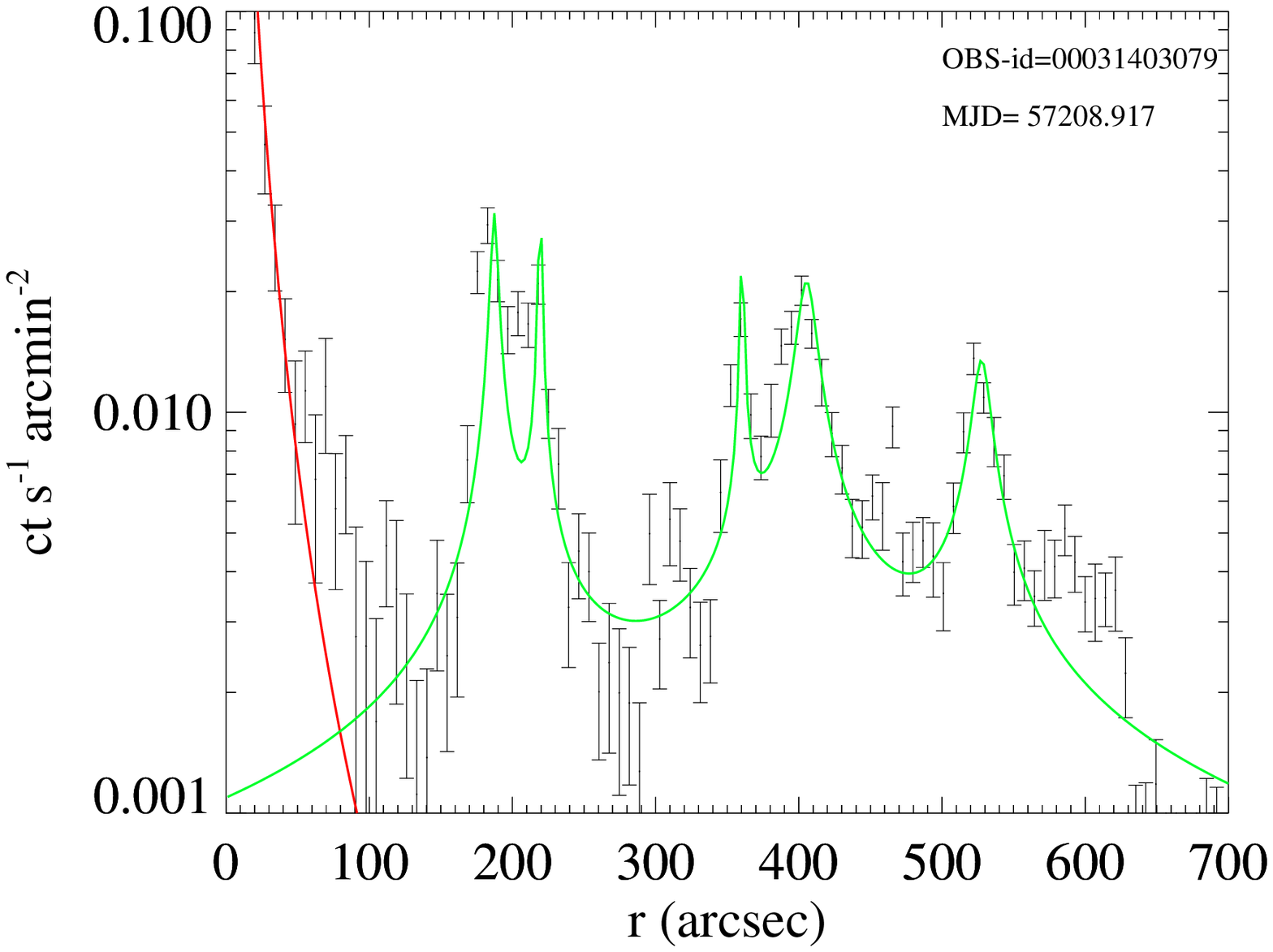}
\includegraphics[width=0.32\textwidth,angle=0,clip=]{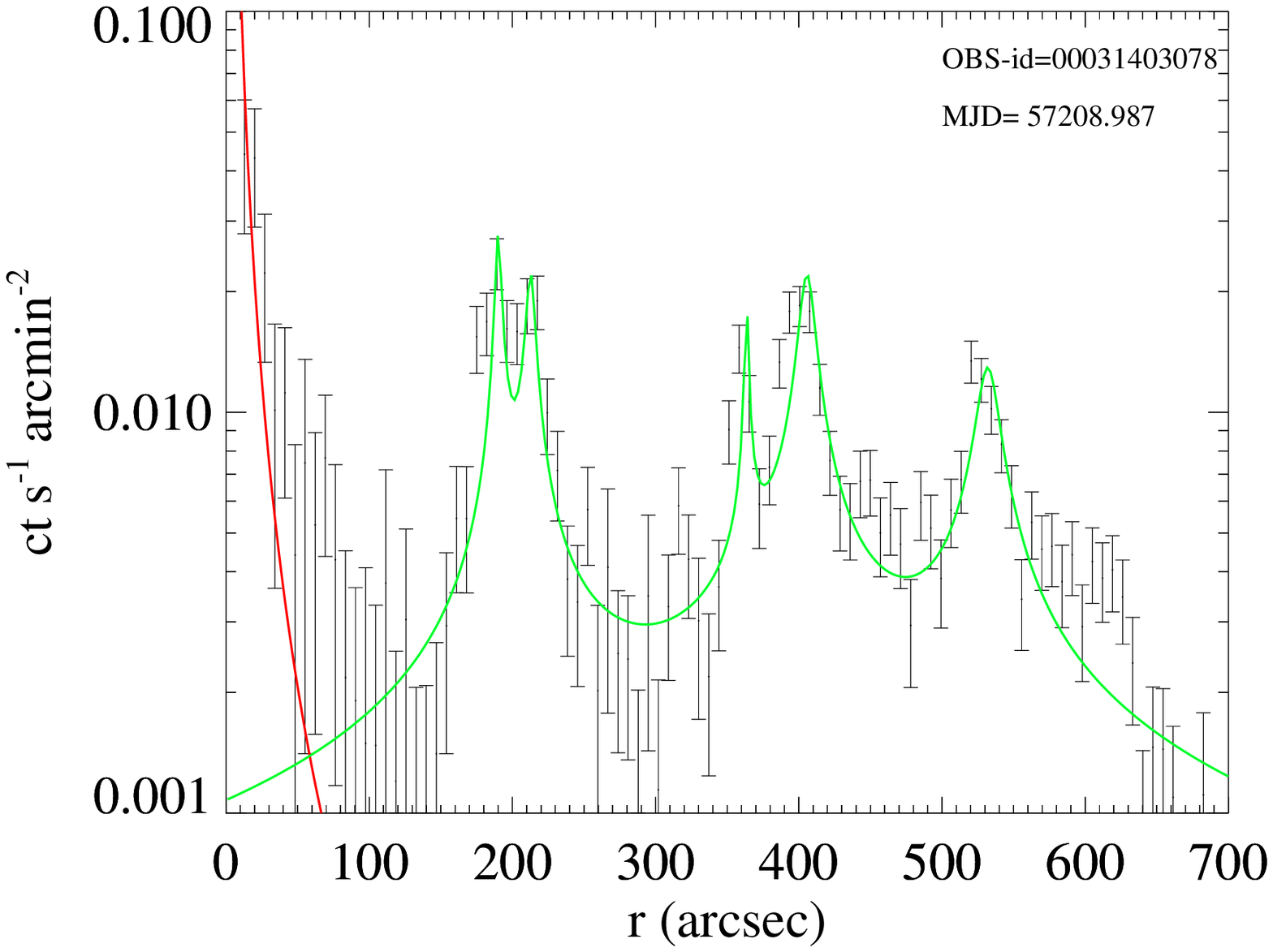}
\includegraphics[width=0.32\textwidth,angle=0,clip=]{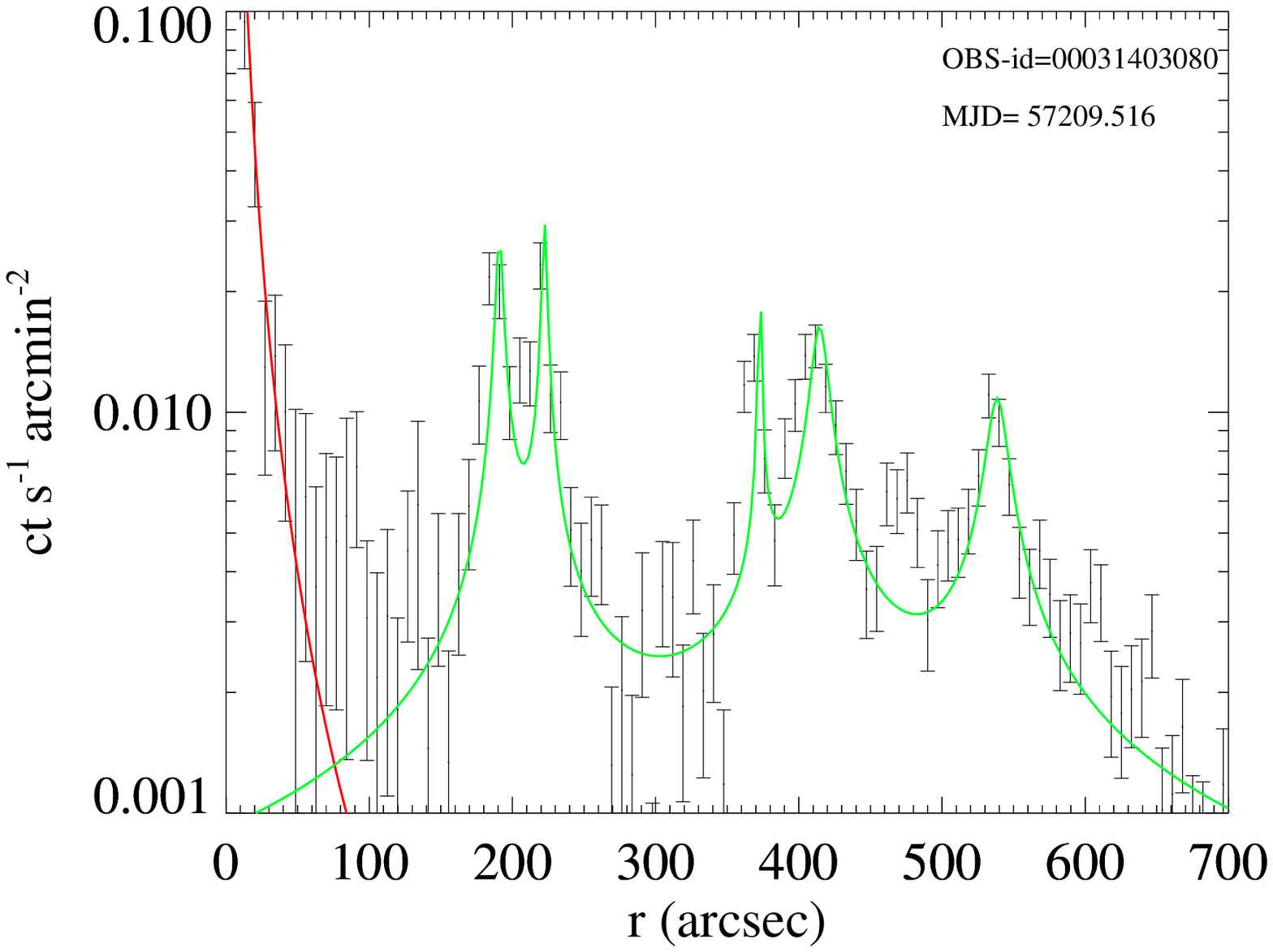}\\
\includegraphics[width=0.32\textwidth,angle=0,clip=]{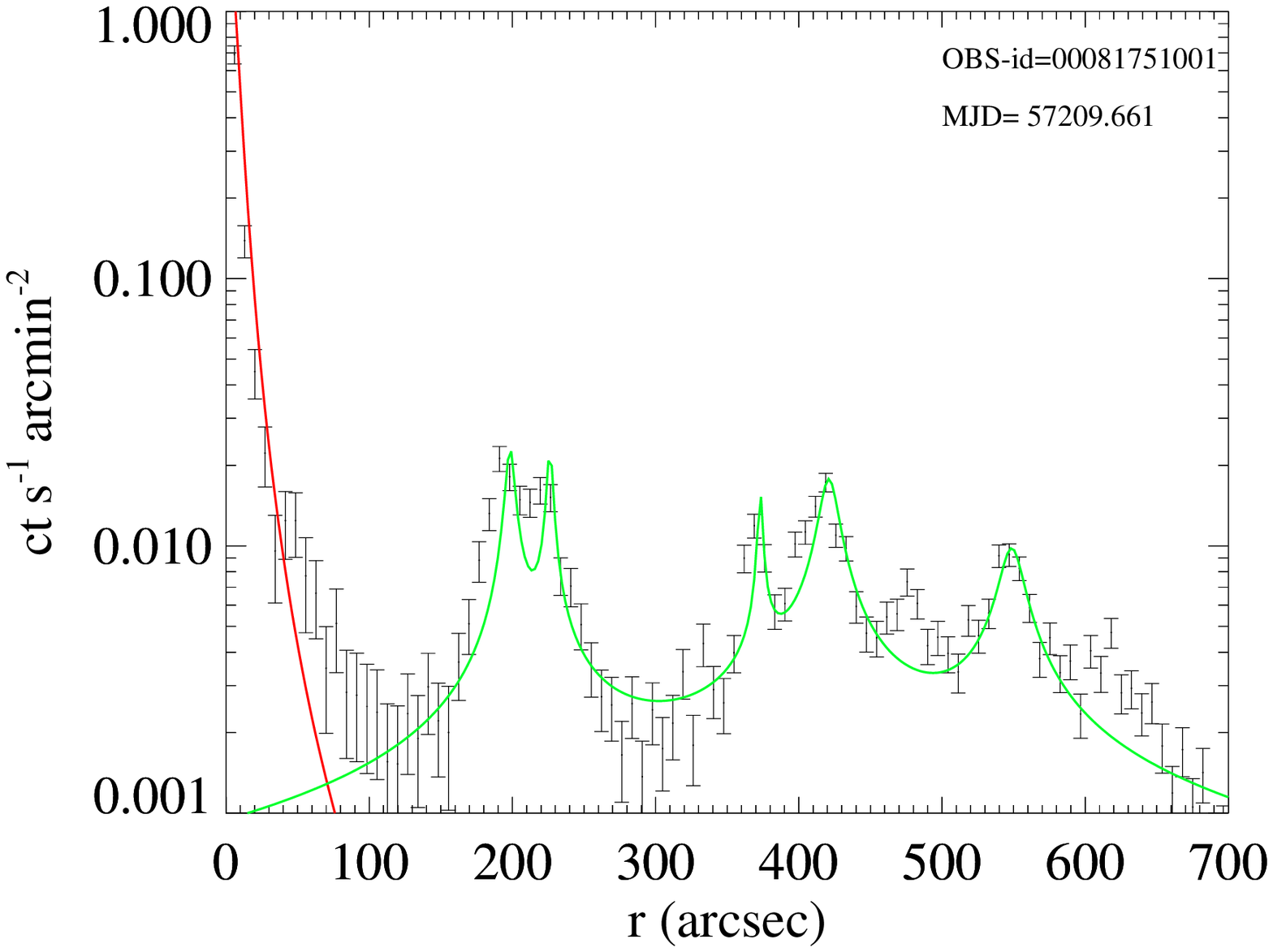}
\includegraphics[width=0.32\textwidth,angle=0,clip=]{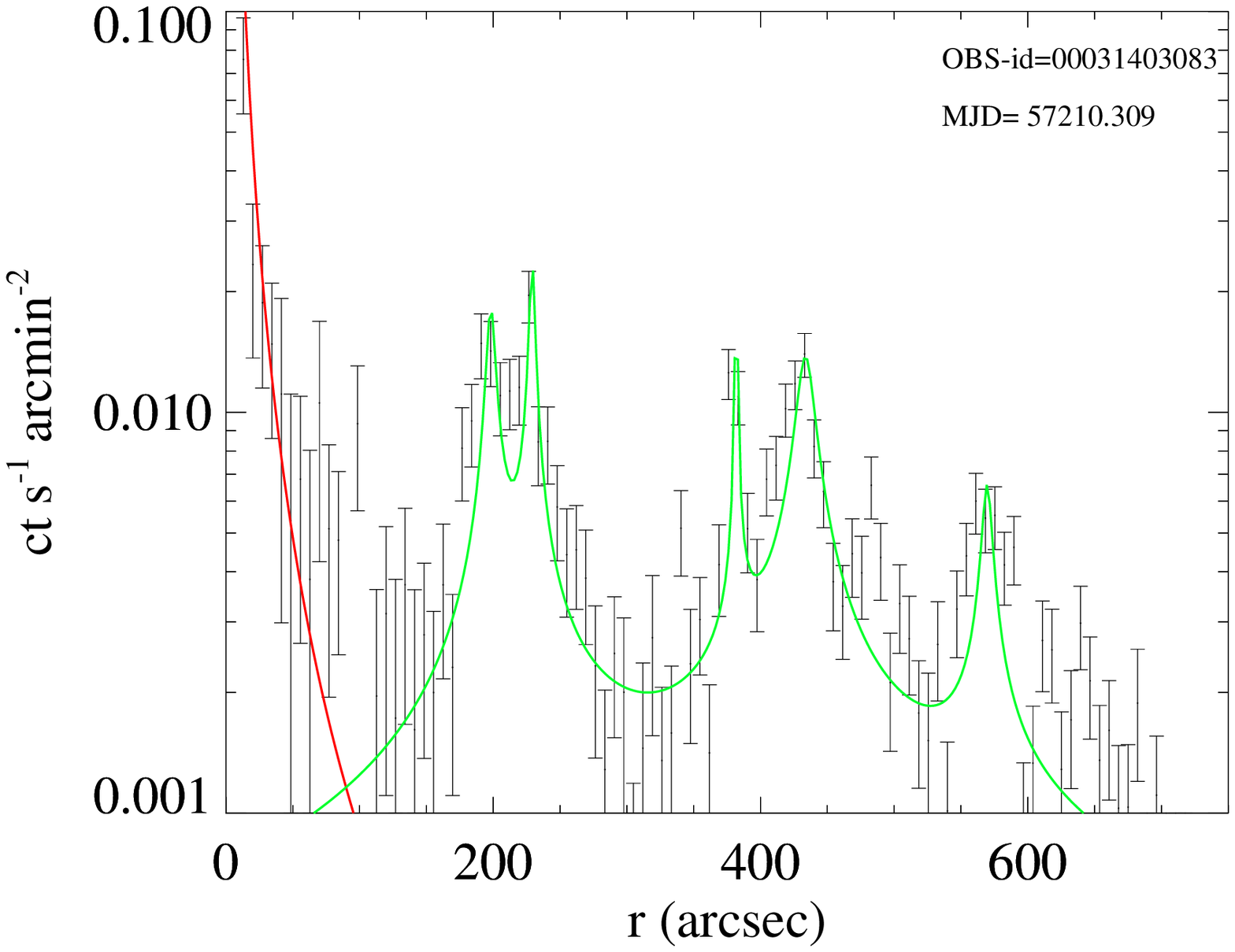}
\includegraphics[width=0.32\textwidth,angle=0,clip=]{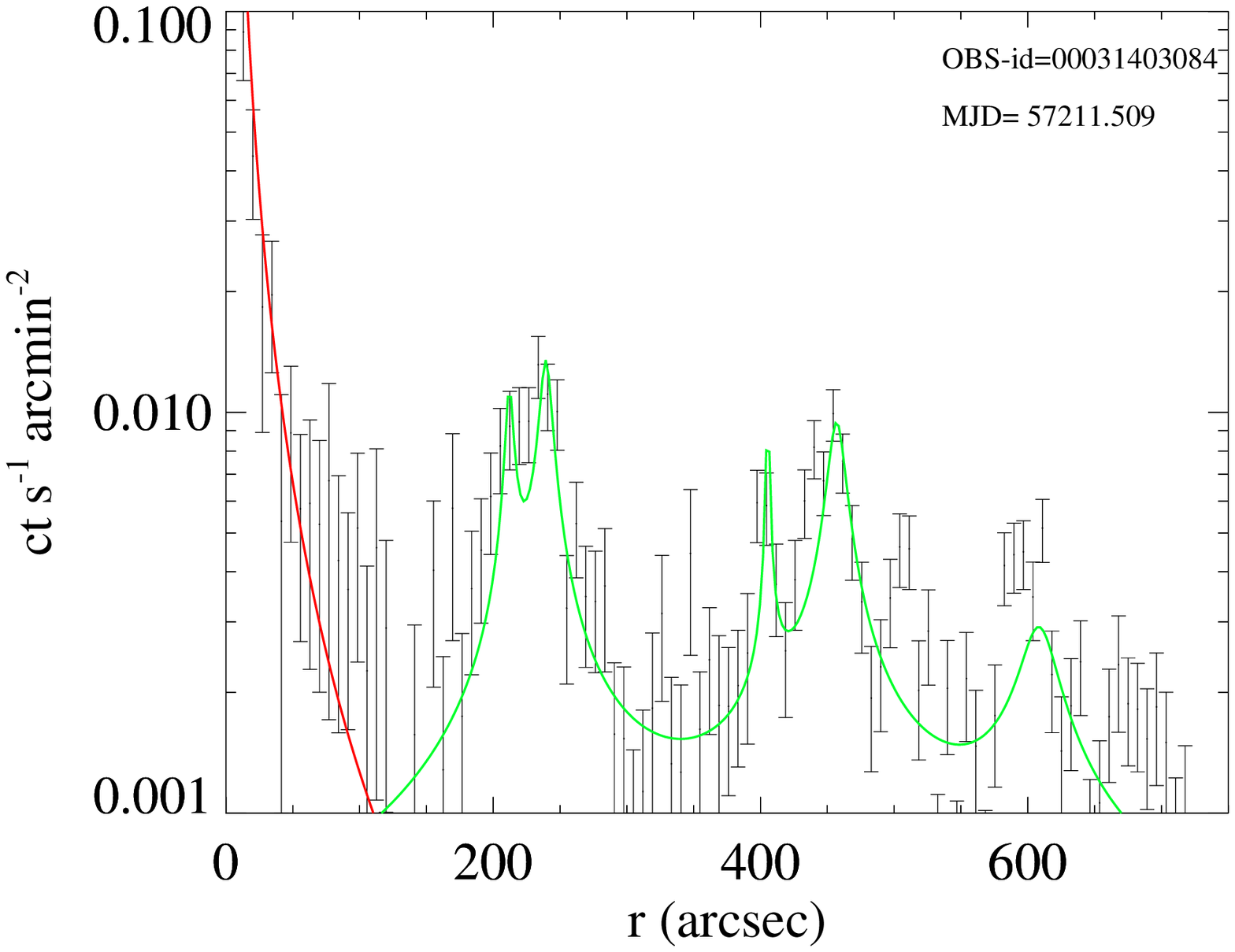}\\
\includegraphics[width=0.32\textwidth,angle=0,clip=]{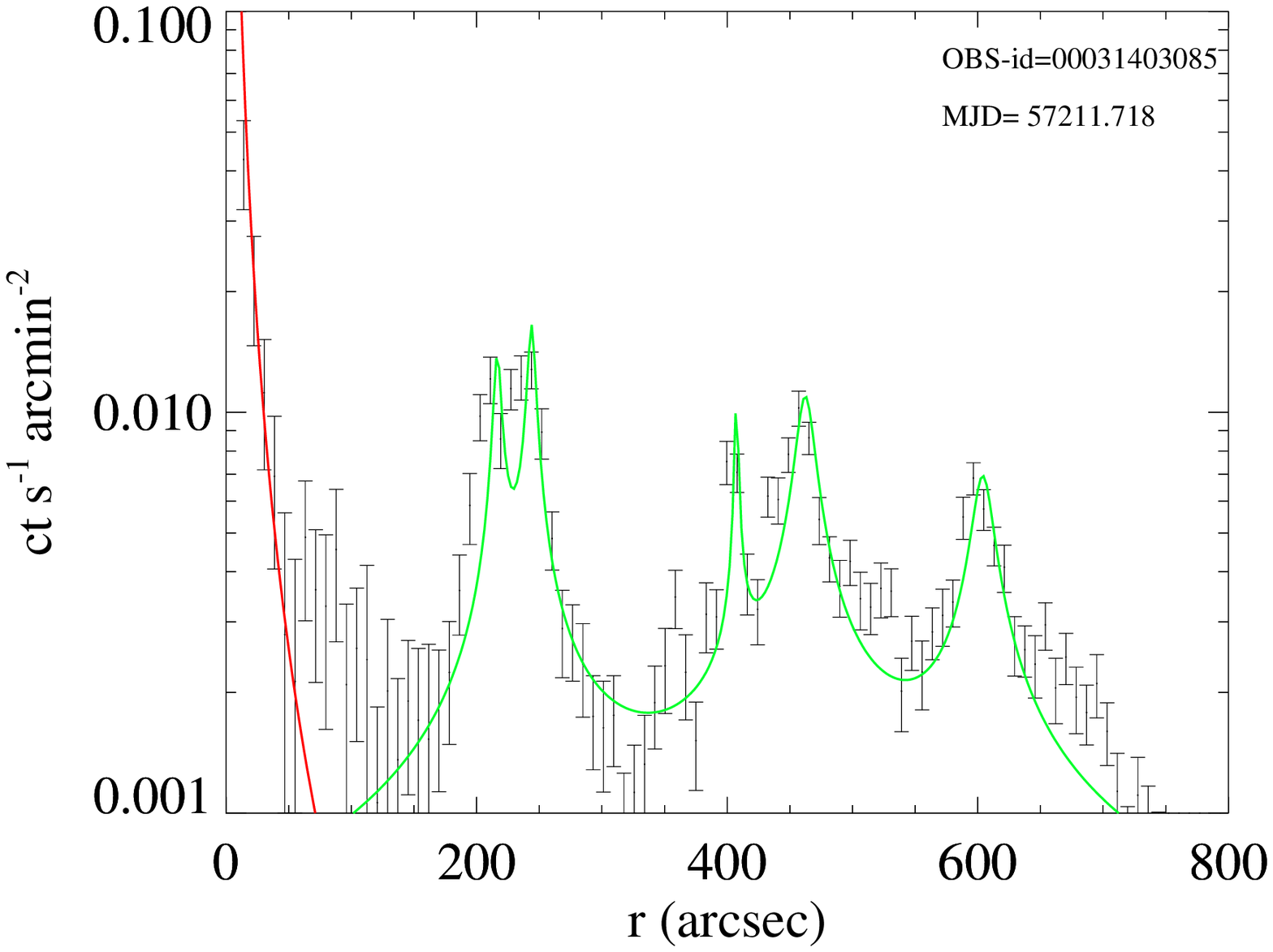}
\includegraphics[width=0.32\textwidth,angle=0,clip=]{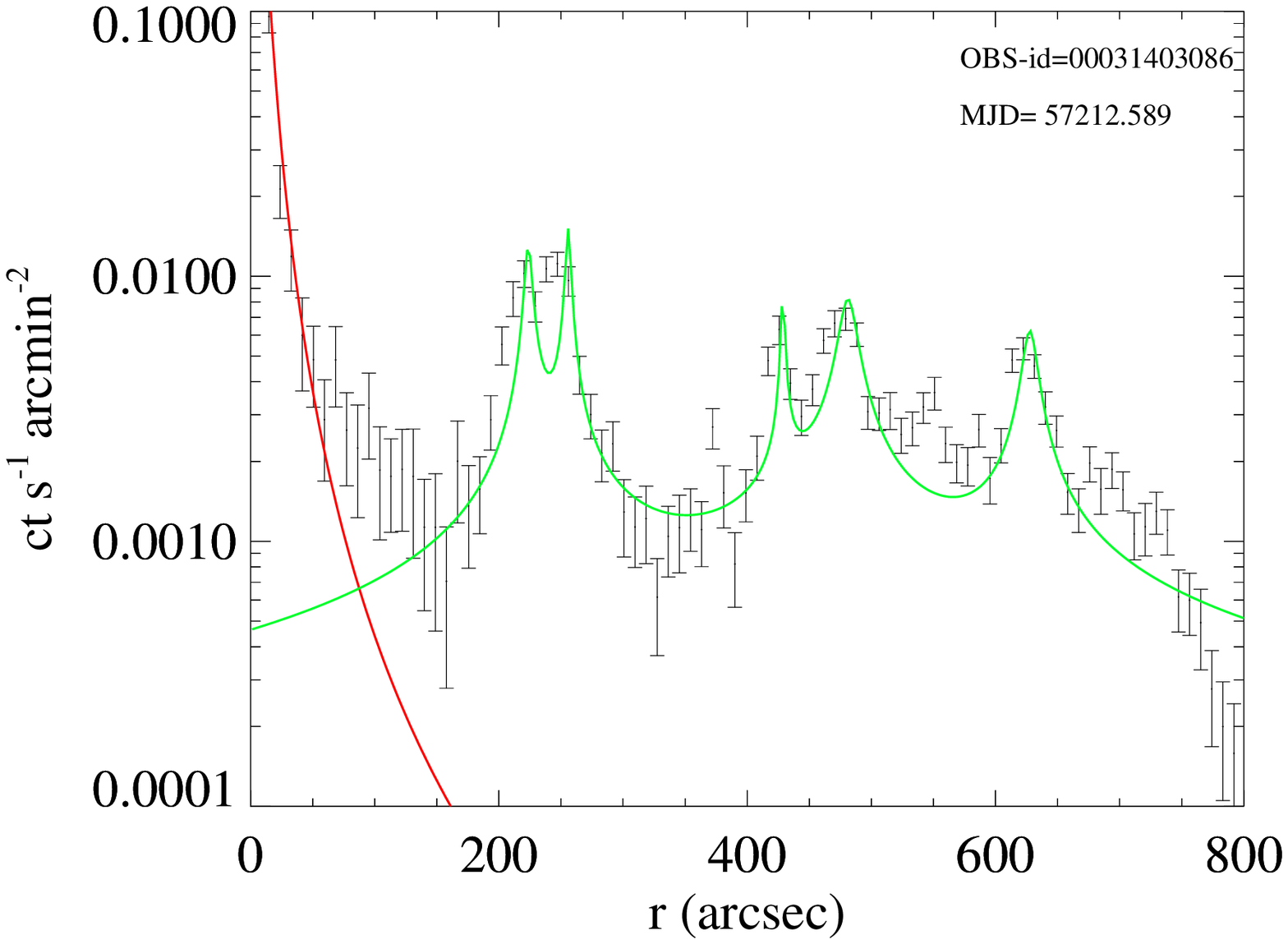}
\includegraphics[width=0.32\textwidth,angle=0,clip=]{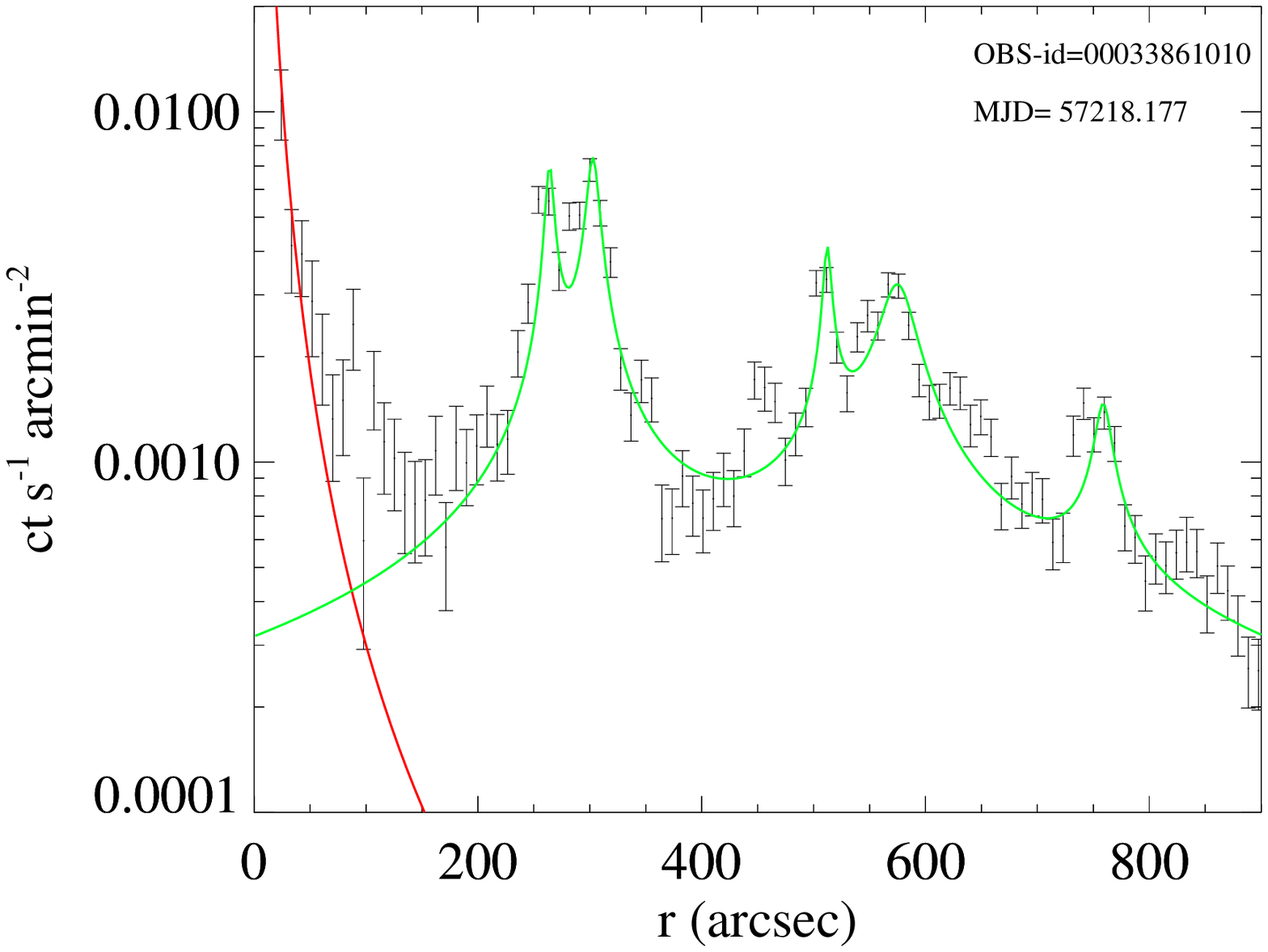}
  \caption{The background subtracted radial profile of the dust rings for a sample of the \swift/XRT observations covering the period
  MJD~57203.4 - 57218.2. The fitted model  is composed by a King profile (red line)  and five Lorentzian functions (green line). 
  } 
  \label{fig:radial-all}
\end{figure*}

\begin{table*}
\caption{Peak position $r_{0i}$ (in arcsec) of the five Lorentzian functions shown in Figs.~\ref{fig:radial} and \ref{fig:radial-all} and the reduced $\chi^2$ value
of the global fit.}
\begin{center}
\scalebox{0.9}{
\begin{threeparttable}
\begin{tabular}{ccc cccc cc}
\hline\hline  
\#&OBS-ID  & MJD\tnote{a} & $r_{01}$ & $r_{02}$  & $r_{03}$ & $r_{04}$&$r_{05}$& $\chi^2$/dof \\
\#&   & [d] & [\arcsec] & [\arcsec]  & [\arcsec] & [\arcsec]&[\arcsec]&   \\
\hline\noalign{\smallskip}
1& 00031403071&57203.459&120.1$\pm$0.8 & 134.9$\pm$0.8 & 229.0$\pm$0.8 & 258.1$\pm$ 0.8 & 337.3$\pm$0.7 & 783/201 \\   \noalign{\smallskip} %bg01
2& 00033861006&57205.463&145.6$\pm$0.8 & 167.4$\pm$1.0 & 282.9$\pm$0.4 & 319.9$\pm$ 0.8 & 418.0$\pm$0.9 & 919/236 \\   \noalign{\smallskip} %bg01
3& 00031403072&57205.800&150.6$\pm$0.5 & 172.5$\pm$0.8 & 292.5$\pm$0.3 & 327.9$\pm$ 0.7 & 431.4$\pm$0.8 & 279/189 \\   \noalign{\smallskip} %bg01
4& 00031403074&57206.660&158.9$\pm$0.8 & 185.1$\pm$0.8 & 311.4$\pm$0.3 & 350.7$\pm$ 0.8 & 458.9$\pm$0.6 & 159/156 \\   \noalign{\smallskip} %bg01
5& 00033861007&57207.386&169.5$\pm$0.9 & 198.1$\pm$0.9 & 329.8$\pm$0.6 & 368.4$\pm$ 0.8 & 483.2$\pm$1.4 & 360/194 \\   \noalign{\smallskip} %bg0015
6& 00031403076&57207.528&171.5$\pm$1.0 & 196.4$\pm$1.1 & 330.5$\pm$0.6 & 371.7$\pm$ 0.8 & 484.0$\pm$0.8 & 250/168 \\   \noalign{\smallskip} %bg01
7& 00033861008&57208.382&178.7$\pm$1.0 & 208.6$\pm$0.8 & 349.1$\pm$0.5 & 392.1$\pm$ 1.6& 513.5$\pm$1.3 & 631/279 \\   \noalign{\smallskip}%bg0015
8& 00031403079&57208.925&187.4$\pm$0.8 & 219.4$\pm$0.4 & 360.4$\pm$0.4 & 405.5$\pm$ 1.2 & 527.7$\pm$1.3 & 549/272 \\   \noalign{\smallskip} %bg0015
9& 00031403078&57208.992&190.1$\pm$1.0 & 212.7$\pm$1.3 & 363.6$\pm$0.8 & 405.7$\pm$ 1.1 & 532.0$\pm$1.6 & 317/253 \\   \noalign{\smallskip} %bg0015
10&00031403080&57209.522&190.9$\pm$0.8 & 222.4$\pm$0.7 & 372.8$\pm$0.5 & 414.6$\pm$ 1.3 & 538.6$\pm$1.5 & 329/267 \\   \noalign{\smallskip}  %bg0005
11&00081751001&57209.661&198.4$\pm$1.1 & 226.2$\pm$0.7 & 373.0$\pm$0.8 & 421.0$\pm$ 1.1 & 548.7$\pm$1.6 & 580/291 \\   \noalign{\smallskip} %bg0005
12&00031403081&57210.385&200.0$\pm$0.6 & 229.6$\pm$1.0 & 384.2$\pm$0.3 & 435.1$\pm$ 1.3 & 566.6$\pm$1.6 & 308/255 \\   \noalign{\smallskip} %bg0005
13&00031403083&57210.314&198.2$\pm$1.4 & 229.1$\pm$0.8 & 381.8$\pm$0.3 & 433.6$\pm$ 1.5 & 569.6$\pm$1.6 & 326/255 \\   \noalign{\smallskip} %bg0005
14&00031403084&57211.514&212.1$\pm$1.5 & 239.1$\pm$1.6 & 405.4$\pm$0.6 & 457.0$\pm$ 1.6 & 608.$\pm$5. & 250/245 \\   \noalign{\smallskip} %bg0005
15&00031403085&57211.786&216.5$\pm$1.1 & 243.9$\pm$1.0 & 407.4$\pm$0.7 & 462.3$\pm$ 1.2 & 604.0$\pm$1.5 & 435/297 \\   \noalign{\smallskip} %bg0001
16&00031403086&57212.808&223.7$\pm$1.1 & 255.8$\pm$0.8 & 428.7$\pm$0.7 & 481.1$\pm$ 1.4 & 627.7$\pm$1.3 & 545/325 \\   \noalign{\smallskip} %bg0001
17&00031403087\tnote{b}&57213.055&233.0$\pm$1.4 & 261.7$\pm$1.0 & 429.2$\pm$0.6 & 488.$\pm$ 2. & 637.4$\pm$1.2 & 234/237 \\   \noalign{\smallskip}%bg0001
18&00033861010&57218.498&264.0$\pm$0.9 & 302.8$\pm$1.0 & 512.2$\pm$1.0 & 575.$\pm$ 2. & 759.$\pm$2. & 526/368 \\   \noalign{\smallskip}%bg0001
19&00031403107\tnote{c}&57223.431&299.3$\pm$1.2 & 343.8$\pm$1.2 & 572.62$\pm$0.9 & 644.7$\pm$ 1.6 & 852.0$\pm$4. & 635/376 \\   \noalign{\smallskip} %bg0001
20&00031403108&57226.475&319.2$\pm$1.2 & 362.9$\pm$1.5 & 614. $\pm$2.  & 686.$\pm$ 4.   & 880.$\pm$20. & 447/361 \\   \noalign{\smallskip} %bg0001
21&00031403111&57231.292&350.0$\pm$2. & 394.8$\pm$1.4 & 654.$\pm$9. & 738.$\pm$ 6. &  $>$ FOV \tnote{d}         & 432/369 \\   \noalign{\smallskip} %bg0001
22&00031403113&57235.495&370.7$\pm$1.5 & 418.3$\pm$1.6 &  not resolved \tnote{e} &  not resolved \tnote{e}   & $>$ FOV \tnote{d}           & 406/361 \\   \noalign{\smallskip} %bg0001
23&00031403115&57239.343&388.4$\pm$2. & 448.7$\pm$1.5 &  not resolved \tnote{e} &  not resolved \tnote{e}   & $>$ FOV  \tnote{d}  & 473/359    \\   \noalign{\smallskip} %bg0001
\hline 
\end{tabular} 
\tnote{a} The average time of the respective exposure (see also Table~\ref{Tab:modelfit}).  \\
\tnote{b} { Being an} observation with  small exposure time {and} low signal to noise, it will not be included in the analysis of the intensity profiles. \\
\tnote{c} The fifth ring was modeled by a Gaussian function for the fit. \\
\tnote{d} The fifth X-ray ring is  outside the FOV  and cannot be constrained.          \\
\tnote{e} The peak count rate of the ring is too small to allow for a reliable determination of its peak position.
\end{threeparttable}
}
\end{center}
\label{Tab:modelfit}
\end{table*}

\end{document}